\documentclass[10pt]{article}

\usepackage{amsmath}

\usepackage{mathrsfs}

\usepackage{hyperref}

\usepackage{color}

\usepackage{graphicx}

\usepackage{caption}

\usepackage{subcaption}

\usepackage{cancel}

\newcommand{\be}{\begin{equation}}

\newcommand{\ee}{\end{equation}}

\newcommand{\beq}{\begin{equation}}

\newcommand{\eeq}{\end{equation}}

\newcommand{\bea}{\begin{eqnarray}}

\newcommand{\eea}{\end{eqnarray}}

\begin{document}


\begin{titlepage}
\vskip  .7in

\begin{center}
{\large \bf Compactifications of the Klebanov-Witten CFT and new $AdS_3$ backgrounds.}

\vskip 0.4in

{\bf Yago Bea}${}^{a}\,$\footnote{{\tt yago.bea@fpaxp1.usc.es}},\phantom{x}{\bf Jos\'e D. Edelstein}${}^{a}\,$\footnote{{\tt jose.edelstein@usc.es}},\phantom{x}
{\bf Georgios Itsios}${}^{a}\,$\footnote{{\tt georgios.itsios@usc.es}},\phantom{x}{\bf Karta S. Kooner}${}^{b}\,$\footnote{{\tt K.S.Kooner.745700@swansea.ac.uk}},\\[5pt]
{\bf Carlos N\'u\~nez}${}^{b}\,$\footnote{{\tt c.nunez@swansea.ac.uk}. Also at CP3-Origins, Odense, Denmark.},\phantom{x}{\bf Daniel Schofield}${}^{c,d}\,$\footnote{{\tt djschofield@phys.uoa.gr}},\phantom{x}and {\bf J. An\'{\i}bal Sierra-Garc\'ia }${}^{a}\,$\footnote{{\tt jesusanibal.sierra@usc.es}}
\vskip 0.1in
{\em
\vskip .15in
${}^a$Departamento de F\'isica de Part\'iculas and Instituto Galego de F\'isica de Altas Enerx\'ias\\
Universidade de Santiago de Compostela\\
E-15782 Santiago de Compostela, Spain\\
\vskip 0.1in
${}^b$Swansea University, School of Physical Sciences,\\
Singleton Park, Swansea, SA2 8PP,  UK
\vskip 0.1 in
${}^c$Department of Mathematics, University of Patras,
26110 Patras, Greece
\vskip 0.1in
${}^d$Department of Nuclear and Particle Physics,
Faculty of Physics,\\
University of Athens,
15771 Athens, Greece
}

\vskip .4in
\end{center}

\centerline{\bf Abstract}

\noindent In this paper we find various new backgrounds in Type IIB, IIA and M-theory with an $AdS_3$-factor. The solutions are smooth and preserve small amounts of SUSY. These new backgrounds are found by application of non-Abelian T-duality (sometimes combined with T-duality)
on the supergravity solution dual to the Klebanov-Witten CFT compactified to two dimensions.
The field theory aspects encoded by these backgrounds are studied. We give a detailed  account of conserved charges, central charges, entanglement entropy and Wilson loops. Further,
we present a possible field theory interpretation for our backgrounds.

\end{titlepage}

\newpage


\tableofcontents

\baselineskip=16pt

\pagestyle{plain}

\setcounter{page}{1}



\newcommand{\startappendix}{
\setcounter{section}{0}
\renewcommand{\thesection}{\Alph{section}}}
\newcommand{\Appendix}[1]{
\refstepcounter{section}
\begin{flushleft}
{\large\bf Appendix \thesection: #1}
\end{flushleft}}



\def\del{{\partial}}

\def\vev#1{\left\langle #1 \right\rangle}

\def\cn{{\cal N}}

\def\co{{\cal O}}

\newfont{\Bbb}{msbm10 scaled 1200}     

\newcommand{\mathbb}[1]{\mbox{\Bbb #1}}

\def\IC{{\mathbb C}}

\def\IR{{\mathbb R}}

\def\IZ{{\mathbb Z}}

\def\RP{{\bf RP}}

\def\CP{{\bf CP}}

\def\Poincare{{Poincar\'e }}

\def\tr{{\rm tr}}

\def\tp{{\tilde \Phi}}

\def\TL{\hfil$\displaystyle{##}$}

\def\TR{$\displaystyle{{}##}$\hfil}

\def\TC{\hfil$\displaystyle{##}$\hfil}

\def\TT{\hbox{##}}

\def\HLINE{\noalign{\vskip1\jot}\hline\noalign{\vskip1\jot}}

\def\seqalign#1#2{\vcenter{\openup1\jot

  \halign{\strut #1\cr #2 \cr}}}

\def\lbldef#1#2{\expandafter\gdef\csname #1\endcsname {#2}}

\def\eqn#1#2{\lbldef{#1}{(\ref{#1})}%

\begin{equation} #2 \label{#1} \end{equation}}

\def\eqalign#1{\vcenter{\openup1\jot

    \halign{\strut\span\TL & \span\TR\cr #1 \cr

   }}}

\def\eno#1{(\ref{#1})}

\def\half{{1 \over 2}}



\def\ads{{\it AdS}}

\def\adsp{{\it AdS}$_{p+2}$}

\def\cft{{\it CFT}}

\newcommand{\ber}{\begin{eqnarray}}

\newcommand{\eer}{\end{eqnarray}}

\newcommand{\beqar}{\begin{eqnarray}}

\newcommand{\cN}{{\cal N}}

\newcommand{\cO}{{\cal O}}

\newcommand{\cA}{{\cal A}}

\newcommand{\cT}{{\cal T}}

\newcommand{\cF}{{\cal F}}

\newcommand{\cC}{{\cal C}}

\newcommand{\cR}{{\cal R}}

\newcommand{\cW}{{\cal W}}

\newcommand{\eeqar}{\end{eqnarray}}

\newcommand{\tht}{\thteta}

\newcommand{\lm}{\lambda}
\newcommand{\Lm}{\Lambda}

\newcommand{\eps}{\epsilon}


\newcommand{\nonu}{\nonumber}

\newcommand{\oh}{\displaystyle{\frac{1}{2}}}

\newcommand{\dsl}


\newcommand{\id}{i\!\!\not\!\partial}

\newcommand{\as}{\not\!\! A}

\newcommand{\ps}{\not\! p}

\newcommand{\ks}{\not\! k}

\newcommand{\D}{{\cal{D}}}

\newcommand{\dv}{d^2x}

\newcommand{\Z}{{\cal Z}}

\newcommand{\N}{{\cal N}}

\newcommand{\Dsl}{\not\!\! D}

\newcommand{\Bsl}{\not\!\! B}

\newcommand{\Psl}{\not\!\! P}

\newcommand{\eeqarr}{\end{eqnarray}}

\newcommand{\ZZ}{{\rm \kern 0.275em Z \kern -0.92em Z}\;}



\def\del{{\delta^{\hbox{\sevenrm B}}}} \def\ex{{\hbox{\rm e}}}

\def\azb{A_{\bar z}} \def\az{A_z} \def\bzb{B_{\bar z}} \def\bz{B_z}

\def\czb{C_{\bar z}} \def\cz{C_z} \def\dzb{D_{\bar z}} \def\dz{D_z}

\def\im{{\hbox{\rm Im}}} \def\mod{{\hbox{\rm mod}}} \def\tr{{\hbox{\rm Tr}}}

\def\ch{{\hbox{\rm ch}}} \def\imp{{\hbox{\sevenrm Im}}}

\def\trp{{\hbox{\sevenrm Tr}}} \def\vol{{\hbox{\rm Vol}}}

\def\rl{\Lambda_{\hbox{\sevenrm R}}} \def\wl{\Lambda_{\hbox{\sevenrm W}}}

\def\fc{{\cal F}_{k+\cox}} \def\vev{vacuum expectation value}

\def\nodiv{\mid{\hbox{\hskip-7.8pt/}}}

\def\ie{{\em i.e.}}

\def\ie{\hbox{\it i.e.}}

\def\CC{{\mathchoice

{\rm C\mkern-8mu\vrule height1.45ex depth-.05ex

width.05em\mkern9mu\kern-.05em}

{\rm C\mkern-8mu\vrule height1.45ex depth-.05ex

width.05em\mkern9mu\kern-.05em}

{\rm C\mkern-8mu\vrule height1ex depth-.07ex

width.035em\mkern9mu\kern-.035em}

{\rm C\mkern-8mu\vrule height.65ex depth-.1ex

width.025em\mkern8mu\kern-.025em}}}

\def\RR{{\rm I\kern-1.6pt {\rm R}}}

\def\NN{{\rm I\!N}}

\def\ZZ{{\rm Z}\kern-3.8pt {\rm Z} \kern2pt}

\def\IB{\relax{\rm I\kern-.18em B}}

\def\ID{\relax{\rm I\kern-.18em D}}

\def\II{\relax{\rm I\kern-.18em I}}

\def\IP{\relax{\rm I\kern-.18em P}}

\newcommand{\CS}{{\scriptstyle {\rm CS}}}

\newcommand{\CSs}{{\scriptscriptstyle {\rm CS}}}

\newcommand{\rc}{\nonumber\\}

\newcommand{\bear}{\begin{eqnarray}}

\newcommand{\eear}{\end{eqnarray}}

\newcommand{\W}{{\cal W}}

\newcommand{\F}{{\cal F}}

\newcommand{\x}{{\cal O}}

\newcommand{\LL}{{\cal L}}

\def\mani{{\cal M}}

\def\calo{{\cal O}}

\def\calb{{\cal B}}

\def\calw{{\cal W}}

\def\calz{{\cal Z}}

\def\cald{{\cal D}}

\def\calc{{\cal C}}

\def\to{\rightarrow}

\def\ele{{\hbox{\sevenrm L}}}

\def\ere{{\hbox{\sevenrm R}}}

\def\zb{{\bar z}}

\def\wb{{\bar w}}

\def\nodiv{\mid{\hbox{\hskip-7.8pt/}}}

\def\menos{\hbox{\hskip-2.9pt}}

\def\dr{\dot R_}

\def\drr{\dot r_}

\def\ds{\dot s_}

\def\da{\dot A_}

\def\dga{\dot \gamma_}

\def\ga{\gamma_}

\def\dal{\dot\alpha_}

\def\al{\alpha_}

\def\cl{{closed}}

\def\cls{{closing}}

\def\vev{vacuum expectation value}

\def\tr{{\rm Tr}}

\def\to{\rightarrow}

\def\too{\longrightarrow}


\def\a{\alpha}

\def\b{\beta}

\def\c{\gamma}

\def\d{\delta}

\def\e{\epsilon}           

\def\f{\phi}               

\def\vf{\varphi}  \def\tvf{\tilde{\varphi}}

\def\vp{\varphi}

\def\g{\gamma}

\def\h{\eta}

\def\i{\iota}

\def\j{\psi}

\def\k{\kappa}                    

\def\l{\lambda}

\def\m{\mu}

\def\n{\nu}

\def\o{\omega}  \def\w{\omega}

\def\q{\theta}  \def\th{\theta}                  

\def\r{\rho}                                     

\def\s{\sigma}                                   

\def\t{\tau}

\def\u{\upsilon}

\def\x{\xi}

\def\z{\zeta}

\def\pt{\tilde{\varphi}}

\def\tt{\tilde{\theta}}

\def\lab{\label}  

\def\6{\partial}

\def\wg{\wedge}

\def\atanh{{\rm arctanh}}

\def\bpsi{\bar{\psi}}

\def\bt{\bar{\theta}}

\def\bvf{\bar{\varphi}}

%



\newfont{\namefont}{cmr10}


\newfont{\addfont}{cmti7 scaled 1440}

\newfont{\boldmathfont}{cmbx10}


\newfont{\headfontb}{cmbx10 scaled 1728}

\numberwithin{equation}{section}

\section{Introduction and General Idea}

\def\magenta{\textcolor{magenta}}
\def\Vol{\textrm{Vol}}

While perturbation theory is a central tool for any physicist, there
are no general equivalent methods to study quantum field theories (QFTs)
at large coupling. In principle, using the Path Integral
and the Renormalization Group (RG) proposed by Wilson one could calculate
any observable for any QFT, 
by discretizing the theory on a lattice and
having access to a large enough computer. As it is well
known,  there are limitations to this Lattice Field theory-project
and the practicalities of it are quite involved (but progress has been 
continuously ongoing for the past few decades).

Another prominent tool--{\it duality}-- emerged in different examples
almost seventy years ago. In the past quarter-century it has become
a central idea in Mathematical Physics.

The two dualities that guide the present investigation are
non-Abelian T-duality (NATD) and the duality between gauge 
theories and quantum gravity theories proposed by Maldacena 
 in \cite{Maldacena:1997re}. 

The non-Abelian T-duality
originally presented in 
\cite{de la Ossa:1992vc}
was further developed 
and carefully inspected in \cite{Giveon:1993ai}-
\cite{Borlaf:1996na}. See the lectures \cite{Quevedo:1997jb} for 
a nice account of some of the dualities that follow from a Buscher procedure.

The interaction between the Maldacena duality and NATD started in the
paper \cite{Sfetsos:2010uq}. Indeed, Sfetsos and Thompson applied NATD to the
maximally symmetric example of $AdS_5\times S^5$, finding a 
metric and RR-fields that preserved ${\cal N}=2$ SUSY. When lifted 
to eleven dimensions, the background fits (not surprisingly) into 
the classification of
\cite{Lin:2004nb}. What is interesting 
is that Sfetsos and Thompson \cite{Sfetsos:2010uq} 
{\it generated} a new solution to the 
Gaiotto-Maldacena differential equation
\cite{Gaiotto:2009gz}, describing
${\cal N}=2$ SUSY CFTs of the Gaiotto-type \cite{Gaiotto:2009we}, \cite{Tachikawa:2013kta}.
This logic was profusely applied to less supersymmetric cases in 
\cite{Itsios:2012zv}- 
\cite{Itsios:2012dc}; finding new metrics and defining new QFTs by
the calculation of their observables.

Nevertheless, various puzzles associated with NATD remain.
One of them, which has been around since the early days of the topic, refers to global aspects
of the metrics generated. In more concrete terms, the periodicity (if any) 
of the
dual coordinates (the Lagrange multipliers in the sigma-model) is not known.
In the same vein, the precise dual field theory to the 
backgrounds generated by NATD is not clear 
at the moment of writing this article.

In this paper we present possible solutions to the two puzzles above---
at least in particular examples.
Our study begins with Type IIB backgrounds dual to a compactification of the
Klebanov-Witten CFT \cite{Klebanov:1998hh} to a 2-d CFT. We will present
these backgrounds for different compactifications and perform a NATD
transformation on them, hence generating new smooth and SUSY 
solutions with $AdS_3$-factors
in Type IIA and M-theory. Application of a further T-duality
generates new backgrounds in Type IIB with an $AdS_3$-factor which are also smooth and preserve the same amount of SUSY. We will
make a proposal for the dual QFT and interpret the range of the 
dual coordinates in terms of a field theoretic operation.

The picture that emerges is that our geometries describe QFTs 
that become conformal 
at low energies. These CFTs live on the intersection of D2 and D6 branes 
suspended
between $NS_5$ branes. While crossing the $NS_5$ branes, 
charge for D4 branes is induced and new nodes of the quiver appear.

We will present the calculation of different observables of the 
associated QFTs that support the proposal made above. 
These calculations are performed in smooth 
supergravity solutions, hence they are trustable and capture the strong
dynamics of the associated 2-d CFTs.

The logic of this paper is a continuation of  that in the previous works of Sfetsos-Thompson
\cite{Sfetsos:2010uq} and Itsios-Nunez-Sfetsos-Thompson (INST) \cite{Itsios:2012zv},
\cite{Itsios:2013wd}. The connection between the material in this work
and those papers is depicted in Fig. \ref{RoadMap} below.
\begin{figure}[h]
	\centering
	\includegraphics[scale=0.6]{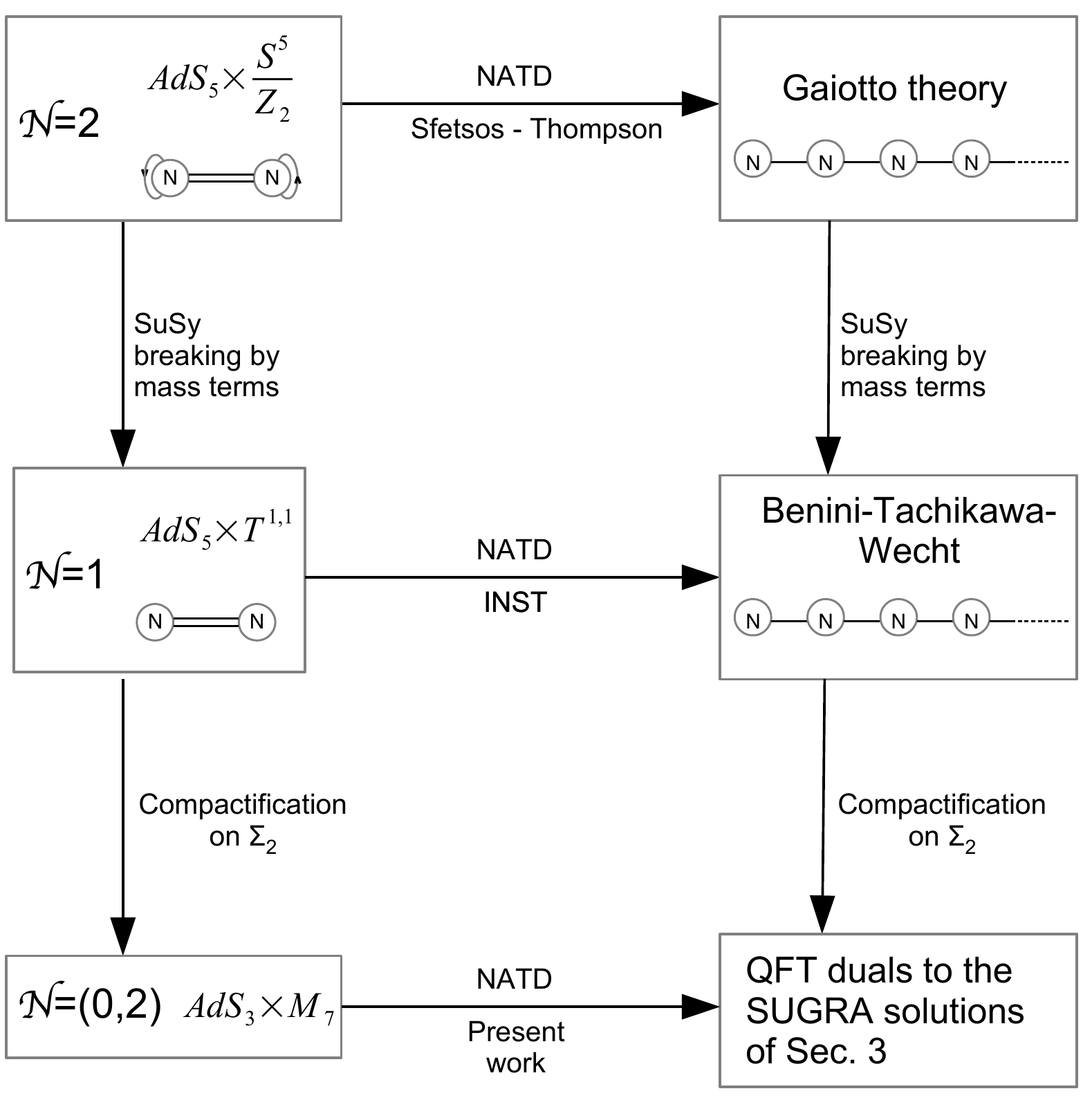}
	\caption{On the left: known solutions on which NATD is performed. On the right: QFT's that correspond to the NATD SUGRA solutions.}
\label{RoadMap}
\end{figure}

The organization of this paper is as follows. In 
Part \ref{geometries} 
of this work, covered in Sections \ref{section1.1} to \ref{seccionlift}, we present  Type IIB backgrounds that are already 
known and the new ones (in Type IIA, IIB and M-theory)
that we construct. The Table 1 summarizes these solutions. In Part \ref{qftxaxa}, starting in Section \ref{comentarios},
 we begin the study of  the
field theoretical aspects encoded by the backgrounds of Part I.
Section \ref{chargessectionxx}  
deals with the quantized charges, defining ranks of the gauge groups.
Section \ref{centralchargessectionxx} 
studies the central charge computed holographically, either at the 
fixed points or along the anisotropic flows 
(where a proposal for a c-function is analyzed).
This observable  presents many clues
towards the understanding of the associated  QFTs. 
Section \ref{sectionEEWilson}
presents a detailed study of Wilson loops and entanglement entropy of the QFT
at the fixed points and along the flow. Summary, conclusions and future directions of research
are spelt out in Section \ref{conclusiones}. The presentation is 
complemented by generous appendices where many useful technical points
have been relegated.

{\it \underline{Note added}:} While this paper was in the final writing stages, the work
\cite{Araujo:2015npa} appeared. 
It superposes with the material in our Section \ref{natdDG}.

\vskip 20pt

\begin{table}[h]
\begin{tabular}{|c|c|c|c|c|}
\hline
\textrm{\textbf{Solutions}} & \textrm{\textbf{IIB}} &  \textrm{\textbf{NATD}} & \textrm{\textbf{NATD-T}} &\textrm{\textbf{Uplift}}
 \\
\hline
$
\begin{array}{l}
            \textrm{Flow from} \; AdS_5 \times T^{1,1} \; \textrm{to} \; AdS_3 \times H_2 \times T^{1,1}  \; \big ( \mathcal{N} = 1\big)
            \\[5pt]
            \textrm{Fixed points} \; AdS_3 \times \Sigma_2 \times T^{1,1}: \;\; \Sigma_2 = S^2, T^2, H_2 \; (\textrm{non-SUSY})
\end{array}
$
                                          &  \ref{section211xx} & \ref{s2h2natd} & \ref{NATD-T} & \ref{uplifteds2h2natd} 
\\
\hline
\textrm{The Donos-Gauntlett solution} &  \ref{sectionDG} &  \ref{natdDG} & \textrm{-}  &  \ref{upliftnatdDG}\\
\hline
\end{tabular}
\caption{Summary of solutions contained in Part \ref{geometries}. The numbers indicate  the section where they are discussed.}
\label{tabla1}
\end{table}

\part{Geometry}\label{geometries}

In this Part, we will exhaustively present a large set  of backgrounds solving
the Type IIB  or Type IIA Supergravity equations. Most of them are new, but some are already present in the bibliography.
New solutions in eleven-dimensional
Supergravity will also 
be discussed. These geometries, for the most part,
preserve some amount of SUSY.

The common denominator of these backgrounds will be the presence of 
an $AdS_3$ sub-manifold in the
ten or eleven dimensional metric. This will be interpreted as 
the dual description of strongly coupled two dimensional 
conformal dynamics. In most of the cases, there is also a flow, 
connecting from an $AdS_3$ fixed point to an $AdS_5$, 
with boundary $R^{1,1}\times \Sigma_2$.
The manifold $\Sigma_2$ will be a constant curvature Riemann surface. 
As a consequence
we conjecture that the full geometry is describing the strongly 
coupled dynamics of a four dimensional QFT,
that is conformal at high energies and gets 
compactified on $\Sigma_2$ (typically preserving some amount of SUSY).
The QFT flows at low energies to a  2-d CFT that is also strongly coupled.

The solutions {that we are going to present} in this section, can be found by inspection of 
the Type II equations. This requires a quite inspired ansatz. 
More practical is to search for solutions of this kind in 
five dimensional gauged supergravity, see the papers \cite{Buchel:2006gb,Gauntlett:2009zw}
for a detailed account of the Lagrangians. 
Some other solutions are efficiently obtained by the use 
of generating techniques, for example a combination of
Abelian and  non-Abelian T-duality, 
that are applied to known (or new) backgrounds, as we show below. 

We now present in detail,  the solutions we will work with in this paper. 
 
\section{Simple flows from $AdS_5 \times T^{1,1}$ to $AdS_3\times M_7$ in Type IIB}
\label{section1.1}

We start this section by proposing a 
simple background in Type IIB. In the sense of the Maldacena duality,
this describes the strongly coupled dynamics of an   
${\cal N}=1$ SUSY QFT in four dimensions, that 
is compactified to two dimensions on a manifold $\Sigma_2$.
In order to allow such a compactification we turn on a 1-form field, $A_1$, on the Riemann surface
$\Sigma_2$. Motivated by the works \cite{Buchel:2006gb,Gauntlett:2009zw}, where the authors consider dimensional reductions to five dimensions of Type IIB supergravity backgrounds on any Sasaki-Einstein manifold, we propose the following ansatz,
\begin{equation}
 \begin{aligned}
 & \frac{ds^{2}}{L^2}=e^{2 A} \left(- dy_0^2+d y_1^2\right)+e^{2 B} ds^2_{\Sigma_{{2}}}+ dr^2+e^{2U}ds_{KE} ^2+e^{2V} \left( \eta + z A_1 \right)^2, 
 \\[5pt]
 & \frac{F_5}{L^4}=4 e^{-4U-V} \textrm{Vol}_5+2 J \wedge J \wedge \left( \eta + z A_1 \right) - z \textrm{Vol}_{\Sigma_2} \wedge J \wedge \left( \eta + z A_1 \right)
 \\[5pt]
&   ~~~~~ - ze^{-2B-V} \textrm{Vol}_{AdS_3} \wedge J, \;\;\; 
 \\[5pt]
&    \Phi=0, ~~~~~~ C_0=0, ~~~~~~ F_3=0, ~~~~~~ B_2=0.
 \end{aligned}
\label{NN02}
\end{equation}
%
%
 We will focus on the case in which the Sasaki-Einstein space is $T^{1,1}$, hence the K\"ahler-Einstein manifold is
 \bea
 ds_{KE}^2=\frac{1}{6}(\sigma_{1}^2+\sigma_{2}^2+\omega_1^2+\omega_2^2) \ ,
 \label{zazar}
\eea
where we have defined,
\begin{equation}
\label{gaga}
 \begin{array}{lll}
  \sigma_{1}= d \theta_1,  & \sigma_{2}= \sin \theta_1 d\phi_1,   & \sigma_{3}= \cos \theta_1 d\phi_1,
 \\[10pt]
  \omega_1=\cos\psi \sin\theta_2 d\phi_2 -\sin\psi d\theta_2  , & \omega_2=\sin\psi \sin\theta_2 d\phi_2  + \cos\psi d\theta_2 , & \omega_3=d\psi +\cos\theta_2 d\phi_2  , 
  \\[10pt]
  \textrm{Vol}_{AdS_3}=e^{2A} dy_0 \wedge dy_1 \wedge dr, & \textrm{Vol}_5= e^{2B} \textrm{Vol}_{AdS_3} \wedge \textrm{Vol}_{\Sigma_{{2}}} , & z \in \mathbb{R} ,
  \\[10pt]
  \eta =\frac{1}{3}\left( d\psi + \cos \theta_1 d \phi_1 + \cos \theta_2 d \phi_2 \right), &
  J=-\frac{1}{6}\left( \sin \theta_1 d \theta_1 \wedge d \phi_1 + \sin \theta_2 d \theta_2\wedge d \phi_2  \right). & \textrm{}
 \end{array}
\end{equation}
%
%
The forms $\eta$ and $J$ verify the relation $d\eta = 2 \ J$.  The 
range of the angles in the $\sigma_{i}'s$ and the $\omega_i\textrm{'}s$ 
---the left invariant forms of $SU(2)$---  
is given by $0\leq\theta_{1,2}<\pi$, 
$0\leq\phi_{1,2}<2\pi$ and $0\leq\psi<4\pi$. {The $\omega_i$
satisfy $d\omega_i = \frac{1}{2} \ \epsilon_{ijk} \ \omega_j \wedge \omega_k$.}
 We also defined a one form $A_1$, that verifies $dA_1= \textrm{Vol}_{\Sigma_2}$. As usual $ds_{\Sigma_2}^2$ is the metric of the two dimensional surface of curvature $\kappa=(1,-1,0)$, denoting a sphere, hyperbolic plane\footnote{To be precise, we do not consider the hyperbolic plane $H_2$, as it has infinite volume. What we consider is a compact space $H_2/\Gamma$ obtained by quotient by a proper Fuchsian group \cite{kehagiasrusso}, and its volume is given by $4\pi (g-1)$, where $g$ is the genus of $H_2/\Gamma$.} or a torus 
respectively. In local coordinates these read,
\begin{equation}
\label{NN0}
 \begin{array}{llll}
  A_1= - \cos \alpha \ d \beta,  &  \textrm{Vol}_{\Sigma_{{2}}}= \sin \alpha \ d\alpha \wedge d \beta,     &  ds^2_{\Sigma_{{2}}}= d\alpha^2 + \sin^2 \alpha d \beta^2, & (\kappa=1) \ ,
 \\[10pt]
 A_1= \cosh \alpha \ d \beta,  & \textrm{Vol}_{\Sigma_{{2}}}= \sinh \alpha \ d\alpha 
\wedge d \beta,  &  ds^2_{\Sigma_{{2}}}= d\alpha^2 + \sinh^2 \alpha 
d \beta^2, & (\kappa=-1) \ ,
 \\[10pt]
   A_1 =  \alpha \ d \beta,  & \textrm{Vol}_{\Sigma_{{2}}}= d\alpha \wedge d \beta,  &  ds^2_{\Sigma_{{2}}}= d\alpha^2 + d \beta^2, & (\kappa=0) \ .
 \end{array}
\end{equation}
%
%
A natural vielbein for the metric (\ref{NN02}) is,
\begin{equation}
\begin{aligned}
& e^{y_0}=L e^A dy_0 \ ,   ~~~~~~  e^{y_1}=L e^A dy_1 \ ,  ~~~~~~  e^{\alpha}=L e^B d\alpha \ ,~~~~~~~~  e^{\beta}=L e^B A_0 d\beta \ ,  ~~~~~~~~  e^{r}=L dr \ , ~~~~~~~~~~~~
\\[5pt]
& e^{\sigma_{1}}= L \frac{e^{U}}{\sqrt{6}} \sigma_{1} \ ,  ~~~~~  e^{\sigma_{2}}= L \frac{e^{U}}{\sqrt{6}} \sigma_{2} \ ,    ~~~~~  e^{1}= L \frac{e^{U}}{\sqrt{6}} \omega_{1} \ ,  ~~~~~ e^{2}=  L \frac{e^{U}}{\sqrt{6}} \omega_{2} \ ,   ~~~~~ e^{3}=L e^V (\eta + z A_1) \ ,
\label{vielbein00}
\end{aligned}
\end{equation}
with $A_0=\sinh \alpha$ for $H_2$, $A_0=\sin \alpha$ for $S^2$  and $A_0=1$ for $T^2$.

As anticipated, the background above describes 
the strong dynamics for a compactification of a 
four dimensional QFT to two dimensions. 
{In the case that we are interested in this work} --- in which the 
K\"ahler-Einstein manifold is the one in eq. \eqref{zazar}--- 
the four dimensional QFT at high energy asymptotes to  
the Klebanov-Witten quiver 
\cite{Klebanov:1998hh} on $R^{1,1}\times \Sigma_2$. 
As it will be clear, most of our results will be valid
for the case of a general $Y^{p,q}$ or any other Sasaki-Einstein 
manifold and their associated QFT. Indeed, these solutions 
can be obtained by lifting to Type IIB, simpler backgrounds of
the five-dimensional supergravity in \cite{Gauntlett:2009zw}.
In fact, the 5-d supergravity Lagrangian was written for any Sasaki-Einstein 
internal space.

Assuming that the functions $A,B,U,V$ depend only on the radial 
coordinate $r$, we can calculate the BPS equations describing 
the SUSY preserving flow from $AdS_5\times T^{1,1}$ 
at large values of the radial coordinate to $AdS_3\times M_7$. 
The end-point of the flow will be dual to a  
2-d CFT obtained after taking the low energy limit of
a twisted KK compactification of the Klebanov-Witten QFT
on $\Sigma_2$. Imposing a set of projections on the 
SUSY spinors of Type IIB---see Appendix \ref{appendixsusy} for details--- 
we find,
\bea
& & A'-e^{-V-4U}\pm \frac{z}{2}e^{-2B-2U-V}=0,\nonumber\\[5pt]
& & B'-e^{-V-4U}\mp \frac{z}{2}e^{-2B-2U-V} \mp \frac{z}{2}e^{-2B+V}=0,\nonumber\\[5pt]
& & U'+e^{-V-4U}-e^{V-2U}=0,\label{O17}
\\[5pt]
& & V'-3e^{-V}+2e^{V-2U}+e^{-V-4U} \mp \frac{z}{2}e^{-2B-2U-V} \pm \frac{z}{2}e^{-2B+V}=0 \ ,\nonumber
\eea
where the upper signs are for $H_2$, the lower signs for $S^2$,  and $z=-\frac{1}{3}$ for both cases.
In the case of the torus the variation of the gravitino will force $z=0$, obtaining $A'=B'$, which does not permit an $AdS_3$ solution.
We now attempt {to} find simple solutions to the eqs.(\ref{O17}).
\subsection{Solution of the form $AdS_3\times H_2$, $AdS_3\times S^2$ and $AdS_3\times T^2$ with 'twisting'}
\label{section211xx}
{At this point we are going to} construct a flow between $AdS_5\times T^{1,1}$ 
and $AdS_3\times \Sigma_2\times M_5$.
To simplify the task, we  propose that the functions 
$U,V$ are constant, then the BPS equations imply that $U=V=0$, 
leaving us---in the case 
of $H_2$ with,
\beq
A'= 1 + \frac{e^{-2B}}{6},\;\;\; B'=1-\frac{e^{-2B}}{3} \ ,
\label{BPSH2}
\eeq
that can be immediately integrated,
 \begin{equation}
A=\frac{3}{2}r-\frac{1}{4} \ln \left( 1+e^{2r} \right) +a_0 \ ,    ~~~~~~ B=\ln\frac{1}{\sqrt{3}} + \frac{1}{2} \ln \left( 1+e^{2r} \right).
 \label{O19}
 \end{equation}
One of the  integration constants associated with these solutions 
corresponds to a choice of the origin 
of the holographic variable and the other constant, 
$e^{2a_0}$, sets the size of the 
three dimensional space ($y_0,y_1,r$). We will choose $a_0=0$ 
in what follows.

In the limit $r\rightarrow \infty$ (capturing the UV-dynamics of the QFT) we recover a Klebanov-Witten metric
     \begin{equation}
   A\sim r -\frac{1}{4}e^{-2r} +\frac{1}{8}e^{-4r} \ ,   ~~~~~ 
B \sim r -\frac{1}{2}\ln 3 +\frac{1}{2}e^{-2r}-\frac{1}{4}e^{-4r} \ ,
     \label{O20x}
     \end{equation}
whilst in the limit $r\rightarrow -\infty$ (that is dual to  the IR 
in the dual QFT) we obtain a supersymmetric solution of 
the form $AdS_3 \times H_2$ ,
     \begin{equation}
   A\sim\frac{3}{2}r -\frac{1}{4}e^{2r}+\frac{1}{8}e^{4r} \ ,  ~~~~~ 
B\sim\ln\frac{1}{\sqrt{3}} +\frac{1}{2}e^{2r} -\frac{1}{4}e^{4r}.
     \label{O21}
     \end{equation}
Let us write explicitly this background using eq. \eqref{zazar},
\bea
& &      \frac{ds_{10}^2}{L^2}=\frac{e^{3r}}{\sqrt{1+e^{2r}}}\left( -dy_0^2+dy_1^2 \right)+\frac{1+e^{2r}}{3}\left( d\alpha^2+\sinh^2\alpha d\beta^2 \right)+dr^2+ {ds^2_{KE}} +\left( \eta - \frac{1}{3} A_1 \right)^2 \ ,\nonumber\\[5pt]
& & \frac{F_5}{L^4}=\frac{4}{3}e^{3r}\sqrt{1+e^{2r}}\sinh \alpha \ dy_0 \wedge dy_1 \wedge d\alpha \wedge d\beta \wedge dr +2 J \wedge J \wedge \left( \eta  - \frac{1}{3} A_1 \right) \nonumber\\[5pt]
& &  
     \qquad +\frac{1}{3} \sinh \alpha \ d \alpha \wedge d \beta \wedge J \wedge \left( \eta - \frac{1}{3}  A_1 \right) + \frac{e^{3r}}{\left(1+e^{2r}\right)^{\frac{3}{2}}} dy_0 \wedge dy_1 \wedge dr  \wedge J \ ,
     \\[5pt]
& &       A_1= \cosh \alpha \ d \beta,~~~~~~\Phi=0, ~~~~~~ C_0=0 ,~~~~~~ F_3=0, ~~~~~~ B_2=0 \ .
\nonumber
     \label{O20}
\eea
The solution above was originally presented in \cite{Gauntlett:2006af}.

We consider now the case of $S^2$. With the same assumptions 
about the functions $U,V$, the BPS equations (\ref{O17}) read,
\beq
A'=1-\frac{e^{-2B}}{6},\;\;\;     B'=1+\frac{e^{-2B}}{3}.
     \label{O25}
     \end{equation}
These can also be immediately integrated (with a suitable 
choice of integration constants), 
     \begin{equation}
   A=\frac{3}{2}r-\frac{1}{4} \ln \left( e^{2r} -1 \right) \ ,   ~~~~~~ B=\ln\frac{1}{\sqrt{3}} + \frac{1}{2} \ln \left( e^{2r}-1 \right) \ , ~~~~~~ r>0 \ .
     \label{O26}
     \end{equation}
This solution seems to be problematic close to $r=0$. 
Indeed, if we compute the Ricci scalar we obtain $R=0$, 
nevertheless, $R_{\mu\nu}R^{\mu\nu}\sim \frac{3}{32 L^4 r^4}+.... $ 
close to $r=0$. The solution is singular and we will not
study it further. 

It is interesting to notice that  a family of {\it non-SUSY} 
fixed point solutions exists. Indeed, we can consider 
the situation where $B,U,V$  and $A'(r)=a_1$ are constant. 
For $S^2$ and $H_2$, we find then that the Einstein equations impose $U=V=0$ and 
 \begin{equation}
     8+e^{-4B}z^2-4 \ a_1^2=0 \ ,  ~~~~~~ 4-z^2 e^{-4B}+\kappa \ e^{-2B}=0.
     \label{O12}
     \end{equation}
Where $\kappa=+1$ for $S^2$ and $\kappa=-1$ for $H_2$. The solution is,
  \begin{equation}
     z^2= e^{2B} \left( 4 e^{2B}  +  \kappa \right) \ ,   ~~~~~~  a_1^2=3  +  \kappa \ \frac{e^{-2B}}{4}
     \label{O13}
     \end{equation}
%
 For the $S^2$ case the range of parameters is,
     \begin{equation}
     B \in \mathbb{R} \ , ~~~~~~ z \in \mathbb{R}-\{0\} \ , ~~~~~~ a_1 \in \left( -\infty,\sqrt{3}\right) \cup \left( \sqrt{3}, +\infty \right) \ ,
     \label{O14}
     \end{equation}
while for ${H_2}$ we find,
  \begin{equation}
     B \in \Big[-\ln 2, +\infty \Big) \ , ~~~~~~ z \in \mathbb{R} \ , ~~~~~~ a_1 \in \Big( -\sqrt{3},-\sqrt{2} \ \Big] \cup \Big[ \ \sqrt{2}, \sqrt{3} \ \Big) \ .
     \label{O15}
     \end{equation}
Notice that in the case of $H_2$, there is a non-SUSY solution with $z=0$, hence no fibration
between the hyperbolic plane and the Reeb 
vector $\eta$. The SUSY fixed point  
in eq. \eqref{O20} is part of the family 
in eq. \eqref{O15}, with $z=-\frac{1}{3}$.

For the $T^2$ we also find an $AdS_3$ solution,
\begin{equation}
     a_1^2=3 \ , ~~~~~~ z ^2=4e^{4B} \ , ~~~~~~ U=0 \ , ~~~~~~ V=0 \ .
     \label{T7}
\end{equation}


%
     Notice that $z \in \mathbb{R}-\{0\} $.
This completes our presentation of what we will refer as 'twisted' solutions. 
By twisted we mean solutions where a gauge field is switched on the Riemann surface 
generating a fibration  between $\Sigma_2$ and the R-symmetry 
direction. In the following, we will present a background that 
also contains an $AdS_3$ factor, it 
flows in the UV to an $AdS_5$, 
but the field content and the mechanism of SUSY preservation are different
from the ones above.

\subsection{The Donos-Gauntlett-(Kim) background}
\label{sectionDG}

In this section we revisit a beautiful solution  written in 
\cite{Donos:2014eua}---this type of solution
was first studied in \cite{Donos:2008ug}. Due to the more
detailed study of \cite{Donos:2014eua}, we will refer to it as
the Donos-Gauntlett solution in the rest of this paper.  

The background in \cite{Donos:2014eua} 
describes a flow  in the radial coordinate, from 
$AdS_5\times T^{1,1}$ to $AdS_3\times M_7$. 
The solution is very original. While the boundary of $AdS_5$ 
is of the form $R^{1,1}\times T^2$, the compactification on the
Riemann surface (a torus) does not use a 'twist'
of the 4d-QFT. This is reflected by the absence of a 
fibration of the Riemann surface on the R-symmetry direction $\eta$. 
Still, the background  preserves SUSY 
\footnote{The solutions in Eqs.(63)-(80) of the paper \cite{Nunez:2001pt}, 
can be thought
as an 'ugly' ancestor of the Donos-Gauntlett background.}. 
The {solution} contains 
an active NS-three form $H_3$ that together with 
the RR five form $F_5$ implies the presence of a 
RR-three form $F_3$. The authors of \cite{Donos:2014eua} found this configuration
by using a very inspired ansatz. 
We review this solution below, adding new information to complement that in
\cite{Donos:2014eua}.

The metric ansatz is given by,
\bea
& &  \frac{ds^2}{L^2}   = e^{2A}\left(-dy_0^2+dy_1^2\right)+e^{2B}\left(d\alpha^2+d\beta^2\right)+dr^2+ e^{2U}ds_{KE}^{2}+e^{2V} \eta^2 \ , 
\label{metric-bef}
\eea
where $A,B,U,V$ are functions of the radial coordinate $r$ only. 
The line element
 $ds_{KE}^{2}$ is defined in eq. \eqref{zazar} 
and $\sigma_{i}$, $\omega_{i}$, $\eta$ are given in eq. \eqref{gaga}.
%
The natural vielbein is,
\begin{equation}
\begin{aligned}
& e^{y_0}=L e^{A} dy_0, \quad e^{y_1}=L e^{A} dy_1, \quad e^{\alpha}= Le^{B} d\alpha, \quad e^{\beta}= L e^{B} d\beta, \quad e^{r} = L dr,		
\\[5pt]
& e^{\sigma_{1}}= L \frac{e^{U}}{\sqrt{6}}\sigma_{1},  
\quad e^{\sigma_{2}}=L \frac{e^{U}}{\sqrt{6}}\sigma_{2},   \quad e^{1}= L \frac{e^{U}}{\sqrt{6}} \omega_1, \quad e^{2}= L \frac{e^{U}}{\sqrt{6}} \omega_2, \quad e^{3}=L e^V \eta\ .
\end{aligned}
\label{veil-bef}
\end{equation}
%
Note that compared to \cite{Donos:2014eua} 
we have relabelled $\phi_i \rightarrow -\phi_i$.
To complete the definition of the background, we also need 
\bea
 {{\textrm{Vol}}_1= - \sigma_{1} \wedge \sigma_{2}, 
\qquad  {\textrm{Vol}}_2=\omega_1\wedge\omega_2}, 
\label{bdefn-bef}
\eea
and the fluxes,
\begin{equation}
\begin{aligned}
& \frac{1}{L^{4}}F_{5} =  4e^{2A+2B-V-4U}dy_0 \wedge dy_1 \wedge d\alpha \wedge d\beta \wedge d r+
\frac{1}{9}\eta \wedge {\textrm{Vol}}_1 \wedge {\textrm{Vol}}_2		
 \\[5pt]
& \qquad \quad  {+\frac{\lambda^{2}}{12} \Big[ d\alpha \wedge d\beta \wedge \eta \wedge ( \textrm{Vol}_1+{\textrm{Vol}}_2) + e^{2A-2B-V} dy_0 \wedge dy_1 \wedge d r \wedge ({\textrm{Vol}}_1+{\textrm{Vol}}_2)\Big]}, 
\\[5pt]
&  {\frac{1}{L^{2}} F_3=\frac{\lambda}{6} d\beta \wedge ({\textrm{Vol}}_1-{\textrm{Vol}}_2), \qquad \frac{1}{L^{2}}  B_2 = -\frac{\lambda }{6}\alpha\ ({\textrm{Vol}}_1-{\textrm{Vol}}_2)},
\end{aligned}
\label{fluxes-bef}
\end{equation}
where $\lambda$ is a constant that encodes the deformation of the space
(and the corresponding operator in the 4d CFT). 
The BPS equations for the above system are given by,
\begin{equation}
\label{BPS}
\begin{aligned}
& A' =	\frac{1}{4} \lambda ^2 e^{-2 B-2 U-V}+e^{-4 U-V}, 
\qquad B' =		e^{-4 U-V}-\frac{1}{4} \lambda ^2 e^{-2 B-2 U-V},	 
\\[5pt]
& U' =	e^{V-2 U}-e^{-4 U-V}, \qquad V' = 	-\frac{1}{4} \lambda ^2 e^{-2 B-2 U-V}-e^{-4 U-V}-2 e^{V-2 U}+3 e^{-V}.
\end{aligned}
\end{equation}
It is possible to recover the $AdS_5 \times T^{1,1}$ solution by setting $\lambda=0$ and
\begin{equation}
A=B=r, \qquad U=V=0.
\label{AdST11sol}
\end{equation}
In Appendix \ref{appendixDG}, we will write an asymptotic expansion
showing how the $AdS_5$ fixed point is deformed by $\lambda$.
Further we can recover the $AdS_3$ solution 
by setting $\lambda=2$ (this is just a conventional value adopted in
\cite{Donos:2014eua}) {and}
\begin{equation}
A=\frac{3^{3/4}}{\sqrt{2}}r, \qquad B=\frac{1}{4}\ln\left(\frac{4}{3}\right), \qquad U=\frac{1}{4}\ln\left(\frac{4}{3}\right), \qquad V=-\frac{1}{4}\ln\left(\frac{4}{3}\right) .
\label{AdS3sol}
\end{equation}
{A flow
(triggered by the deformation parameterized by  $\lambda$) between the asymptotic $AdS_5$ and
the $AdS_3$ fixed point}
can be {found numerically}.  

It is also possible to find 
an analytic 
approximation for the solution of the BPS system which makes contact with aspects of  
thermodynamics. This is well explained in Appendix \ref{appendixDG}, where a similar analysis for the geometry of the $H_2$ flow is also considered. We approximate the various functions that appear in the metric of the Donos-Gauntlett geometry using the following ansatz,
\begin{equation}
\label{AnalyticApproxAnsatz}
\begin{aligned}
& A  = \frac{r}{R_{DG}}+(1-\frac{1}{R_{DG}})\mu_{A}\ln\left[e^{\frac{r}{\mu_{A}}}+e^{\frac{r_{A}}{\mu_{A}}}\right] \ ,
\qquad
U  = \frac{1}{2}\frac{\ln\frac{2}{\sqrt{3}}}{1+e^{\frac{r-r_{U}}{\mu_{U}}}}  \ ,
\\[10pt]
& B  = r+r_{B} + \m_B \ln\Big[e^{-\frac{r}{\mu_{B}}}+e^{-\frac{r_{B}}{\mu_{B}}}\Big]  ,
\qquad\qquad\quad \;\;
V  =\frac{1}{2} \frac{\ln\frac{\sqrt{3}}{2}}{1+e^{\frac{r-r_{V}}{\mu_{V}}}} \ .
\end{aligned}
\end{equation}
%
%
The comparison with the numerical solutions is illustrated in Figure \ref{Fig:Flow}. \\
%
%

Up to this point, we have set the stage for our study, 
but most of the solutions discussed above have already been stated in the literature. {Here} we just added some new backgrounds and technical elaborations on the known ones.
Below, we will present genuinely new Type IIA/B solutions. 
The technical tool {that we have used is} non-Abelian T-duality. 
This technique, when applied to an $SU(2)$
isometry of the previously discussed solutions will generate new Type IIA configurations. {In our examples these new solutions are nonsingular}. Their lift to eleven dimensions will 
produce new, smooth, $AdS_3$ configurations in M-theory. {Moreover, performing an additional Abelian T-duality transformation
we generate new Type IIB  backgrounds} with all fluxes turned on and an $AdS_3$ fixed point at the IR.
We move on to describe these.

\section{New backgrounds in Type IIA: use of  non-Abelian T-duality}

In this section we apply the technique of non-Abelian T-duality on the Type IIB backgrounds that we presented above. As a result we obtain new solutions of the Type IIA supergravity.

\subsection{The non-Abelian T-dual of the twisted solutions}
\label{s2h2natd}

We will start by applying non-Abelian T-duality (NATD) to the 
background obtained via a twisted compactification
in Section \ref{section1.1}. The configuration we will focus on
 is a particular case of that in eq. \eqref{NN02}. 
Specifically, in what follows we consider $U = V = 0$. These values for the functions $U$ and $V$ are compatible with the BPS system \eqref{O17}. In this case the background \eqref{NN02} simplifies to, 
\begin{equation}
 \label{NN01}
 \begin{aligned}
   & \frac{ds^{2}}{L^2}=e^{2 A} \left(- dy_0^2+d y_1^2\right)+e^{2 B} ds^2_{\Sigma_{{2}}}+ dr^2+\frac{1}{6} \left(\sigma_{1}^2+\sigma_{2}^2\right)+\frac{1}{6} \left(\omega_1^2+\omega_2^2\right) + \left( \eta + z A_1 \right)^2 \ ,
\\[5pt]
   & \frac{F_5}{L^4}=4 \textrm{Vol}_5+2 J \wedge J \wedge \left( \eta + z A_1 \right) - z \textrm{Vol}_{\Sigma_2} \wedge J \wedge \left( \eta + z A_1 \right) - ze^{-2B} \textrm{Vol}_{AdS_3} \wedge J \ .
 \end{aligned}
\end{equation}
%
%
As before, all other RR and NS fields are taken to vanish. Also, the 1-form $A_1$, the line element of the Riemann surface $\Sigma_2$ and the corresponding volume form, for each of the three cases that we consider here, are given in eq. \eqref{NN0}.

We will now present the {details} for the background after NATD
has been applied on the $SU(2)$ {isometry} described by the coordinates $(\theta_2,\phi_2,\psi)$. As is well-known
a gauge fixing has to be implemented {during the NATD procedure}. This leads to a  choice of three 'new coordinates' among the 
Lagrange multipliers {$(x_1,x_2,x_3)$} used in the NATD procedure  
and the 'old coordinates' ($\theta_2,\phi_2,\psi$) 
\footnote{The process of NATD and the needed gauge fixing was 
described in detail in \cite{Itsios:2013wd}, 
\cite{Macpherson:2014eza}.}. 
In all of our examples we consider a gauge fixing of the form,
\begin{equation}
 \theta_2 = \phi_2 = \psi = 0.
\label{gaugefixing012}
\end{equation}
 As a result, the Lagrange multipliers $x_1, x_2$ and $x_3$ play the r\^ole of the dual coordinates in the new background.
To display the natural symmetries of the background, 
we will quote the results using  spherical coordinates. 
This makes the expressions compact. We refer the 
interested reader to Appendix \ref{cylcart}, 
for the expressions after NATD in cartesian and cylindrical coordinates.
We define,
\begin{equation}
 \begin{aligned}
  & x_1=\rho \cos \xi \sin \chi \ , ~~~~~~ x_2=\rho \sin \xi \sin \chi \ , ~~~~~~ x_3=\rho \cos \chi \ ,
  \\[5pt]
  & \Delta =L^4 +54 \alpha'^2 \rho^2 \sin^2 \chi +36 \alpha'^2 \rho^2 \cos^2 \chi \ ,  ~~~~~ \tilde{\sigma}_{3}= \cos \theta_1 d \phi_1 + 3 z A_1.
    \label{NN10}
 \end{aligned}
\end{equation}
%
%
The NSNS sector of the transformed IIA background reads,
\begin{equation}
\label{NN11}
 \begin{aligned}
  & e^{-2 \widehat{\Phi}}=\frac{L^2}{324 \alpha'^3} \Delta, 
  \\
  & \frac{d\hat{s}^{2}}{L^2}=e^{2 A} \left(- dy_0^2+d y_1^2\right)+e^{2 B} ds^2_{\Sigma_2}+ dr^2+\frac{1}{6} \left(\sigma_{1}^2+\sigma_{2}^2\right) + \frac{\alpha'^2}{\Delta} \Bigg[ 6 \Big( \sin^2 \chi \big( d \rho^2 + \rho^2 (d \xi + \tilde{\sigma}_{3})^2 \big) 
  \\
   & \qquad + \rho \sin (2 \chi) d\rho d\chi+\rho^2 \cos^2 \chi d \chi^2 \Big) + 9 \Big( \cos \chi d \rho - \rho \sin \chi d \chi \Big)^2 +\frac{324 \alpha'^2}{L^4} \rho^2 d\rho^2 \Bigg] ,
  \\
  & \widehat{B}_2=\frac{\alpha'^3}{\Delta} \Bigg[ 36 \rho \cos \chi \Big(\rho \tilde{\sigma}_{3}\wedge d\rho +\rho \sin \chi d\xi \wedge \big(\sin \chi d\rho+\rho \cos \chi d \chi \big)\Big)
  \\
  & \qquad + \Big(  \frac{L^4}{\alpha'^2} \tilde{\sigma}_{3} - 54 \rho^2 \sin^2 \chi d \xi  \Big)\wedge \big( \cos \chi d \rho - \rho \sin \chi d \chi \big) \Bigg] \ ,
 \end{aligned}
\end{equation}
%
 %
while the RR sector is, 
\footnote{
 According to the democratic formalism, the higher rank RR forms are related to those of lower rank through the relation $ F_p = (-1)^{[\frac{p}{2}]} * F_{10-p}$.
}
 \begin{align}
 \label{NN12}
  \widehat{F}_0 &=0, \qquad \widehat{F}_2=\frac{L^4}{54\alpha'^{\frac{3}{2}}} \left( 2 {\sigma_{1}\wedge \sigma_{2}} + 3 z \textrm{Vol}_{\Sigma_2} \right) \ ,
  \nonumber\\[5pt]
  \widehat{F}_4 &=\frac{L^4}{18 \sqrt{\alpha'}} \Bigg[   3 z e^{-2B}  \big(d\rho \cos \chi - \rho \sin \chi d\chi \big)  \wedge \textrm{Vol}_{AdS_3} + z \ \rho \ \cos \chi \textrm{Vol}_{\Sigma_2} \wedge {\sigma_{1} \wedge \sigma_{2}}
  \nonumber\\[5pt]
   & -\frac{18 \alpha'^2 }{\Delta} \rho^2 \sin  \chi \Big( z \textrm{Vol}_{\Sigma_2}  + \frac{2}{3} {\sigma_{1} \wedge \sigma_{2}} \Big) \wedge \Big( 2  \cos \chi \big(\sin \chi d\rho+\rho \cos \chi d \chi \big)
  \\
  &  +3   \sin  \chi \big(\rho \sin \chi d \chi -\cos \chi d \rho \big) \Big)\wedge \big(d \xi +  \tilde{\sigma}_{3} \big) \Bigg]. 
 \nonumber
\nonumber
 \end{align}
%
%
%
The SUSY preserved by this background is discussed in 
Appendix \ref{appendixsusy}. We have checked that the equations of motion are solved by this background.

\subsection{The non-Abelian T-dual of the Donos-Gauntlett solution}
\label{natdDG}

{In this section } we will briefly present the result of applying NATD to the Donos-Gauntlett solution \cite{Donos:2014eua}
that we described in detail in Section \ref{sectionDG} and Appendix \ref{appendixDG}.

Like above, we will perform the NATD choosing a gauge such that the new coordinates are $(x_1,x_2,x_3)$.
We will quote the result in spherical coordinates, but the expressions in cylindrical and cartesian coordinates will be written in the Appendix \ref{cylcart}.
The expressions below are naturally more involved than those in Section \ref{s2h2natd}.
{This is due to the fact that now there is a non-trivial NS 2-form that enters in the procedure (explicitly, in the string sigma model) 
which makes things more complicated.}
 As above, we start with some definitions,
\begin{equation}
  \mathcal{B}_{\pm}=\rho \cos \chi\pm\frac{L^2 \lambda}{6 \alpha'} \alpha \ , \qquad \mathcal{B}=\mathcal{B}_{+} \ ,  \qquad
  \Delta =L^4 e^{4 U+2 V}+54 \alpha'^2 e^{2 U} \rho^2 \sin^2 \chi +36 \alpha'^2 \mathcal{B}^2 e^{2V} \ .
    \label{s1}
\end{equation}
The NSNS sector is given by,
%
 \begin{align}
  e^{-2 \widehat{\Phi}} & =\frac{L^2}{324 \ \alpha'^3} \Delta,   \label{s5} 
\nonumber\\[5pt]
  \frac{d\hat{s}^{2}}{L^2} & =e^{2 A} \left(- dy_0^2+d y_1^2\right)+e^{2 B} \left( d \alpha^2+ d \beta^2\right)+ dr^2+\frac{e^{2 U}}{6} \left(\sigma_{1}^2+\sigma_{2}^2\right)+
\nonumber\\[5pt]
  &  +\frac{\alpha'^2}{\Delta} \Bigg[ 6 e^{2 U+2 V} \Bigg( \sin^2 \chi \Big( d \rho^2 + \rho^2 \big(d \xi + \sigma_{3} \big)^2 \Big)+ \rho \sin (2 \chi) d\rho d\chi+\rho^2 \cos^2 \chi d \chi^2 \Bigg) 
\nonumber\\[5pt]
  &  + 9 e^{4 U} \big( \cos \chi d \rho - \rho \sin \chi d \chi \big)^2 + \frac{324 \alpha'^2}{L^4} \Big( \big(\mathcal{B} \cos \chi+ \rho \sin^2 \chi \big) d\rho+ \rho \big(\rho \cos \chi - \mathcal{B} \big)\sin \chi d\chi \Big)^2 \Bigg],
\\[5pt]
  \widehat{B}_2 & =L^2\frac{\lambda}{6} \ \alpha \ {\sigma_{1}\wedge \sigma_{2}}+\frac{\alpha'^3}{\Delta} \Bigg[ 36 \ \mathcal{B} \  e^{2V} \Bigg(\sigma_{3}\wedge \Big( \big(\mathcal{B} \cos \chi+ \rho \sin^2 \chi \big) d\rho+ \rho \big(\rho \cos \chi - \mathcal{B} \big) \sin \chi d\chi \Big) 
\nonumber\\[5pt]
  &  +\rho \sin \chi d\xi \wedge \big(\sin \chi d\rho+\rho \cos \chi d \chi \big) \Bigg)+ e^{2U} \Big( e^{2V+2U} \frac{L^4}{\alpha'^2} \sigma_{3} - 54 \rho^2 \sin^2 \chi d \xi  \Big)\wedge \big( \cos \chi d \rho - \rho \sin \chi d \chi \big) \Bigg].
\nonumber
 \end{align}
%
%
The RR sector reads,
\begin{equation}
\label{s11}
 \begin{aligned}
  & \widehat{F}_0=0, \qquad
\widehat{F}_2=\frac{L^2}{6\alpha'^{\frac{3}{2}}} \Big( \lambda \alpha' d\mathcal{B}_{-}\wedge d\beta + \frac{2}{9} L^2 {\sigma_{1}\wedge \sigma_{2}} \Big), 
\\[5pt]
  & \widehat{F}_4=\frac{L^4 \lambda}{36 \sqrt{\alpha'}} \Bigg[ e^{V}\Big( \frac{2 L^2}{\alpha'}d\alpha -3 e^{-2B-2V} \lambda \big( \cos \chi d \rho - \rho \sin \chi d \chi \big) \Big) \wedge \textrm{Vol}_{AdS_3}
\\[5pt]
  &  \quad \; -\frac{6 \alpha'}{L^2} \Big( \rho \sin \chi \big( \sin \chi d\rho + \rho \cos \chi d \chi \big) \mathcal{B} d \mathcal{B}\Big) \wedge d \beta \wedge {\sigma_{1} \wedge \sigma_{2}}
\\[5pt]
  & \quad \; +\frac{36 \alpha'^2 }{\Delta} \rho \sin  \chi \Big( \frac{3 \alpha'}{L^2} d \beta \wedge d\mathcal{B}_{-}  -\frac{2}{3 \lambda} {\sigma_{1} \wedge \sigma_{2}} \Big) \wedge \Big( 2 e^{2V}\mathcal{B}  \big(\sin \chi d\rho+\rho \cos \chi d \chi \big)+ 
\\[5pt]
  & \quad \; +3 e^{2U}  \rho \sin  \chi \big(\rho \sin \chi d \chi -\cos \chi d \rho \big) \Big)\wedge \big(d \xi +  \sigma_{3} \big) \Bigg] \ .
 \end{aligned}
\end{equation}
%
Here again, aspects of the SUSY preserved by this background are
relegated to Appendix \ref{appendixsusy}. We have checked that the equations of motion are solved by this background.

\section{T-dualizing back from Type IIA to IIB }
\label{NATD-T}
In this section, we will construct {\it new} Type IIB 
Supergravity backgrounds with an $AdS_3$ factor at the IR. The idea is to obtain
these new solutions by performing an (Abelian) T-duality
on the configurations  described 
by eqs. \eqref{NN11}-\eqref{NN12}; which in turn,
were obtained by performing NATD on the backgrounds 
of eq. \eqref{NN01}.
The full chain of dualities is NATD -T. 
The new solutions present an $AdS_3$ fixed point and
all RR and NS fields are switched on. 

It should be interesting to study if the $AdS_3$ fixed point of this geometry falls  within known classifications \cite{Gauntlett:2006ux}. If not,
to use them as inspiring ansatz to extend these taxonomic efforts.

In order to perform the T-duality, we will choose a Killing vector that has no fixed points, in such a way that the dual geometry has no singularities. An adapted system of coordinates for that Killing vector is obtained through the change of variables,

\begin{equation}
\alpha = \textrm{arccosh} \left( \cosh a \cosh b \right) \ , \qquad \beta = \arctan \left(  \frac{\sinh b}{\tanh a}  \right) \ ,
\label{NN2TTwewewe1}
\end{equation}
obtaining \footnote{
In fact using the coordinate transformation eq. \eqref{NN2TTwewewe1} we obtain $A_1 = \sinh a \ db + \textrm{total derivative}. $
},
\begin{equation}
 A_1= \sinh a \ db \ ,  \qquad  \textrm{Vol}_{\Sigma_2}= \cosh a\  da \wedge db \ , \qquad  ds^2_{\Sigma_2}=da^2 + \cosh^2 a \ db^2 \ .
\label{NN0000021}
\end{equation}

The Killing vector that we choose is the one given by translations along the $b$ direction. Its modulus is proportional to the quantity,
\begin{equation}
    \Delta_T = 54 \alpha'^2 z^2 \rho^2 \sin^2 \chi  \sinh^2 a + e^{2B} \cosh^2 a \  \Delta \ ,
\label{moduluskillingvector} 
\end{equation}
where $\Delta$ is defined in eq.(\ref{NN10}). Since $\Delta_T$  is never vanishing the isometry has no fixed points. 

To describe these new configurations, we define,
\begin{equation}
 A_3=3 z \sinh a \ \big( \cos \chi d\rho - 
\rho \sin \chi d \chi \big) - db \ .
  \label{NN1TT}
\end{equation}
 Then, we will have a NSNS sector,
%
\begin{align}
\label{NN2TT}
    e^{-2 \widetilde{\Phi}} & =\frac{L^4}{324 \alpha'^4} \Delta_T \ , 
\nonumber\\[5pt]
    \frac{d\tilde{s}^{2}}{L^2} & =e^{2 A} \left(- dy_0^2+d y_1^2\right)+e^{2 B} da^2+ dr^2+\frac{1}{6} \left(\sigma_{1}^2+\sigma_{2}^2\right)
\nonumber\\[5pt]
    & +\frac{\alpha'^2}{\Delta_T} \Bigg[  \frac{\Delta}{L^4} db^2 + 6 e^{2B} \cosh^2 a  \Big(  \rho^2 \sin^2 \chi      \big(  \sigma_{3}+ d\xi \big)^2 + \big(d \rho \  \sin \chi +\rho \ \cos \chi \  d \chi \big)^2 \Big)
\nonumber\\[5pt]
    &  +9 \big( z^2 \sinh^2 a+e^{2B} \cosh^2 a  \big) \Big( \big(d\rho \ \cos \chi -\rho \  \sin \chi \ d\chi \big)^2 + \frac{36 \alpha'^2}{L^4} \rho^2 d\rho^2 \Big) 
\nonumber\\[5pt]
    &   - 6 z \sinh a \ d b  \Bigg(  \Big(\frac{36 \alpha'^2}{L^4} \rho^2+1\Big)  \cos \chi \  d\rho - \rho \ \sin \chi \ d \chi \Bigg) \Bigg] \ ,
\\[5pt]
    \widetilde{B}_2 & =\frac{\alpha'^3}{\Delta_T}     \Bigg[  e^{2B} \cosh^2 a   \Bigg( 36 \rho \cos \chi \Big[ \rho \sigma_{3}\wedge d\rho + \rho  \sin \chi  d\xi \wedge \Big(\sin \chi   d\rho +\rho  \cos \chi   d\chi \Big)\Big]
\nonumber\\[5pt]
    & + \Big( \frac{L^4}{\alpha'^2} \sigma_{3} - 54 \rho ^2 d\xi  \sin ^2 \chi  \Big)\wedge \Big(d \rho  \cos \chi - \rho  \sin \chi d \chi  \Big) \Bigg)  
 - 18 \ z \ \rho^2 \ \sinh a \sin ^2 \chi \Big(      d \xi  \wedge A_3 + d b \wedge \sigma_{3}   \Big) \Bigg] \ ,
\nonumber
\end{align}
%
and a RR sector that reads,
%
\begin{align}
\label{NN3TT}
& \widetilde{F}_1 = \frac{z L^4}{18 \alpha'^2} \cosh a \ d a \ ,
\nonumber\\[5pt]
& \widetilde{F}_3 = \frac{L^4}{54\alpha'} \big(  2 A_3 + 3 z \rho \cos \chi \cosh a \ d a  \big)    \wedge         \sigma_{1}\wedge \sigma_{2}
\nonumber\\[5pt]
& \qquad +\frac{ z L^4 \alpha'  \cosh a }{\Delta_T} \rho^2 \sin \chi \big(  \sigma_{3} +  d \xi \big) \wedge \Big[    e^{2B} \cosh^2 a  \Big( \big ( 2 + \sin^2 \chi \big) d\chi - \sin \chi \cos \chi d \rho \Big) 
\nonumber\\[5pt]
& \qquad - z \sinh a \sin \chi A_3 \Big] \wedge d a \ ,
\nonumber\\[5pt]
& \widetilde{F}_5 = \frac{L^4}{6} e^{-2B}  \Big( 24 e^{4B} \rho \cosh a  da \wedge d \rho + z \big( d\rho  \cos \chi -\rho d\chi \sin \chi \big) \wedge d b \Big) \wedge \textrm{Vol}_{AdS_3} 
\\[5pt]
& \qquad + \frac{L^4 \alpha'^2 \cosh a}{18 \Delta_T} \Bigg[   e^{2B} \cosh a \rho  \sin \chi  d \xi \wedge  \Bigg(  \big( d\rho  \sin \chi +\rho d \chi  \cos \chi \big)  \wedge \Big[ 24 \rho  \cos \chi A_3
\nonumber\\[5pt]
&  \qquad - z \Big( \frac{L^4}{\alpha'^2} + 54 \rho ^2 \sin ^2 \chi  \Big) \cosh a da \Big]
 + 18 \ \rho \  \sin  \chi  \big( 3 z \rho \cos \chi \cosh a d a -2 d\beta \big) \wedge \big(d \rho  \cos \chi -\rho  d \chi  \sin \chi \big) \Bigg)
\nonumber\\[5pt]
& \qquad - 18 z^2 \  \rho ^3 \sinh a \ \sin ^2 \chi \  d \xi \wedge \Big( 3 z \sinh a \big(\sin^2 \chi d \rho + \rho \sin \chi \cos \chi d \chi \big) + \cos \chi A_3 \Big)  \wedge d a  \Bigg] \wedge  \sigma_{1}\wedge \sigma_{2} \ .    
\nonumber
\end{align}
%
We have checked that the equations of motion are solved by this background, either assuming the BPS equations (\ref{BPSH2}) or the non-SUSY solution (\ref{O15}). We have also checked that the Kosmann derivative vanishes without the need to impose further projections. These facts point to the conclusion that this new and smooth solution is also SUSY preserving.
 
 We will now present new backgrounds of eleven-dimensional Supergravity.


\section{ Lift to M - theory }\label{seccionlift}

Here, we lift the solutions of  
Sections \ref{s2h2natd} and \ref{natdDG} to eleven dimensions. 
This constitutes another original contribution of this paper, 
presenting new and smooth backgrounds of M-theory that describe 
the strong dynamics of a SUSY 2d CFT.

It is well known that given a solution of the 
Type IIA SUGRA the metric of the uplifted to eleven dimensions 
solution, has the following form,
\begin{equation}
ds^2_{11} = e^{-\frac{2}{3} \widehat{\Phi}} \ ds^2_{IIA} \ + \ e^{\frac{4}{3} \widehat{\Phi}} \ \big(  dx_{11} + \widehat{C}_1  \big)^2 \ ,
\end{equation}
where $\widehat{\Phi}$ is the dilaton of the 10-dimensional 
solution of the Type IIA SUGRA and $\widehat{C}_1$ is 
the 1-form potential that corresponds to the 
RR 2-form of the Type IIA background. Also, by
$x_{11}$ we denote the $11^{\textrm{th}}$ coordinate which corresponds 
to a $U(1)$ isometry as neither the metric tensor or flux  
explicitly depend on it.

The 11-dimensional geometry is supported by a 3-form potential $C^M_3$ 
which gives rise to a 4-form $F^M_4 = dC^M_3$. 
This 3-form potential can be written in terms 
of the 10-dimensional forms and the differential 
of the $11^{\textrm{th}}$ coordinate as,
\begin{equation}
 C^M_3 = \widehat{C}_3 + \widehat{B}_2 \wedge dx_{11} \ .
\end{equation}
The 3-form $\widehat{C}_3$ corresponds to the closed part of 
the 10-dimensional RR form 
$\widehat{F}_4 = d\widehat{C}_3 -\widehat{H}_3 \wedge \widehat{C}_1$.
Here, $\widehat{B}_2$ is the NS 2-form of the 10-dimensional 
type-IIA theory and $\widehat{H}_3 = d\widehat{B}_2$.
Hence we see that in order to describe the 11-dimensional 
solution we need the following ingredients,
\begin{equation}
 ds^2_{IIA} \ , \quad \widehat{\Phi} \ , \quad \widehat{B}_2 \ , \quad \widehat{C}_1 \ , \quad\widehat{C}_3 \,\,\,\, \textrm{or} \,\,\,\, \widehat{F}_4 \ .
\end{equation}
Let us now present these quantities for the cases of interest.
\subsection{Uplift of the NATD of the twisted solutions}
\label{uplifteds2h2natd}
As we mentioned above, in order to specify 
the M-theory background we need to read the field 
content of the 10-dimensional solution. For the case at hand we 
wrote the metric $ds^2_{IIA}$, 
the dilaton $\widehat{\Phi}$
and the NS 2-form $\widehat{B}_2$ in eq. \eqref{NN11}. 
Moreover, from the expression of the RR 2-form 
in eq. \eqref{NN12} we can immediately extract the 
1-form potential $\widehat{C}_1$, 
\begin{equation}
 \widehat{C}_1 = \frac{L^4}{54 \ a'^{\frac{3}{2}}} \ 
\big(  3 \ z \ A_1 - 2 \ \s_3  \big) \ .
\end{equation}
The 3-form potential--$\widehat{C}_3$--can be obtained 
from the RR 4-form of eq. \eqref{NN12}. After some algebra one finds,
\begin{equation}
\begin{aligned}
  \widehat{C}_3 & = \frac{L^4 \ \ e^{-2B}}{6 \ a'^{\frac{1}{2}}} \ z \ \r \ \cos\chi \ \textrm{Vol}_{AdS_3} + \frac{L^4 \ z \ \big(L^4 + 36 a'^2 \r^2\big) \ \cos\chi}{6 \ \a'^{\frac{1}{2}} \ \Delta} A_1 \wedge \sigma_{3} \wedge d\r
  \\[5pt]
                         & + \frac{L^4 \ \a'^{\frac{3}{2}} \ \r^2 \sin\chi}{\Delta} \Big( z \ A_1 - \frac{2}{3} \sigma_{3}  \Big) \wedge d\xi \wedge \Big( \r \ \big( 2 + \sin^2\chi  \big) \ d\chi -\sin\chi \ \cos\chi \ d\r  \Big)
   \\[5pt]
                         & + \frac{L^4 \ z \ \r}{18 \ \a'^{\frac{1}{2}} \ \Delta} \Big( \Delta \ \cos\chi \ \big(  \sigma_{3} \wedge \textrm{Vol}_{\Sigma_2} - 2 A_1 \wedge d \sigma_{3} \big) - 3 L^4 \sin\chi \ A_1 \wedge \sigma_{3} \wedge d\chi   \Big) \ .
 \end{aligned}
\end{equation}
We close this section by observing that, if the coordinate $\rho$
takes values in a finite interval, then the radius of the 
M-theory circle, $e^{\frac{4}{3} \widehat{\Phi}}$
never blows up, because the function $\Delta$ that appears in the expression 
of the dilaton is positive definite. We have checked that the equations of motion of the 11-dimensional Supergravity are solved by this background.
\subsection{Uplift of the NATD of the Donos-Gauntlett solution}
\label{upliftnatdDG}
Here, the NSNS fields of the 10-dimensional theory have 
been written in detail in eq. \eqref{s5}. 
In order to complete the description of the M-theory background 
we need also to consider the potentials 
$\widehat{C}_1$ and $\widehat{C}_3$ that are encoded in the RR fields of 
eq. \eqref{s11}. Hence from the RR 2-form potential 
we can easily read $\widehat{C}_1$,
\begin{equation}
  \widehat{C}_1 = \frac{L^2 \l \, \mathcal{B}_{-}}{6 \a'^{\frac{1}{2}}} \, d\beta - \frac{L^4}{27 \a'^{\frac{3}{2}}} \, \s_3 \ ,
\end{equation}
where the function $\mathcal{B}_{-}$ has been defined in 
eq. \eqref{s1}. Also, from the RR 4-form we can obtain 
the potential $\widehat{C}_3$ which in this case is,
\begin{equation}
 \begin{aligned}
  \widehat{C}_3 & = \frac{e^{-2B-V} L^4 \l}{36 \a'^{\frac{3}{2}}} \Big( 2 e^{2B + 2V} L^2 \alpha - 3 \l \a' \r \cos\chi \Big) \textrm{Vol}_{AdS_3} 
\\[5pt]
      & - \frac{\a'^{\frac{1}{2}} L^2 \r \sin\chi}{18 \Delta} \Big( 18 \a' \l \mathcal{B}_{-} \big( \s_3 + d\xi \big) \wedge d\beta + 4 L^2 \s_{3} \wedge d\xi \Big) \wedge \Sigma_1
\\[5pt]
      & + \frac{L^2 \l \a'^{\frac{1}{2}}}{12 }  \Big(\mathcal{B}_{-}^2 + \mathcal{B}^2 + \r^2 \sin^2\chi \Big) d\beta \wedge d\s_{3} \ .
 \end{aligned}
\end{equation}
Here for brevity we have defined the 1-form $\Sigma_1$ in the following way:
\begin{equation}
\Sigma_1 = 6 \a' e^{2V} \mathcal{B} \ \big(  \sin\chi d\r +\r \cos\chi d\chi  \big) - 9 \a' e^{2U} \r \sin\chi \ \big(  \cos\chi d\r -\r \sin\chi d\chi  \big)\ .
\end{equation}
Finally, we observe that the radius of the 
M-theory circle is finite, for reasons similar to those discussed in
the  previous example. We have checked that the equations of motion of the 11-dimensional Supergravity are solved by this background.

This completes our presentation of this set of new and exact solutions. 
 The expressions for these backgrounds
in cartesian and cylindrical coordinates are written in Appendix \ref{cylcart} .
 A summary of all the solutions can be found in Table \ref{tabla1}.

We will now move on to the second part of this paper. We will study 
aspects of the field theories that our new and smooth backgrounds 
are defining.

\part{Field Theory Observables}

\label{qftxaxa}

\section{General comments on the Quantum Field Theory}\label{comentarios}
Let us start our study of the correspondence between our new metrics 
with their respective field theory  dual. We will state some general points that these field theories will fulfill.

In the case of the backgrounds corresponding to the compactifications described in Section \ref{section211xx}, our field theories are obtained by a twisted KK-compactification on a two dimensional manifold---that can be $H_2, \  S^2$ or $T^2$. The original field theory is, as we mentioned, the Klebanov-Witten quiver, that controls the  high energy dynamics of our system. The bosonic part of the global symmetries for this  QFT in the UV are 
\beq
SO(1,3)\times SU(2)\times SU(2)\times U(1)_R\times U(1)_B ,
\eeq
where, as we know the $SO(1,3)$ is enhanced to $SO(2,4)$.
The theory contains two vector multiplets ${\cal W}^i=(\lambda^i, A_\mu^i)$, for $i=1,2$, together with four chiral multiplets 
${\cal A}_\alpha=(A_\alpha,\psi_{\alpha})$ for $\alpha=1,2$ and ${\cal B}_{\dot{\alpha}} =(B_{\dot{\alpha} } 
,\chi_{\dot{\alpha} }   )$ with $\dot{\alpha}=1,2$.

These fields transform as vector, spinors and scalars under $SO(1,3)$---that is $A_\alpha, B_{\dot{\alpha}}$ are singlets, the fermions transform in the $\bf(\frac{1}{2}, 0) \oplus (0,\frac{1}{2})$ and the 
vectors in the $\bf (\frac{1}{2},\frac{1}{2})$. The transformation  under the 'flavor' quantum numbers 
$SU(2)\times SU(2)\times U(1)_R\times U(1)_B $ is,
\bea
& & A_\alpha=(2,1,\frac{1}{2}, 1),\;\;\;  B_{\dot{\alpha}}=(1,2,\frac{1}{2},-1),\nonumber\\
& & \psi_\alpha=(2,1,-\frac{1}{2},1),\;\;\;\chi_{\dot{\alpha}}=(1,2,-\frac{1}{2},-1),\\
& & \lambda^i=(1,1,1,0),\;\;\; A_\mu^i=(1,1,0,0).
\nonumber
\label{transflawsxx}
\eea

The backgrounds in Section \ref{section211xx},
are describing the strong coupling regime of the field theory above, in the case in which we compactify the  D3 branes
on $\Sigma_2$ twisting the theory. This means, mixing the R-symmetry $U(1)_R$ with the $SO(2)$ isometry of $\Sigma_2$.
This twisting is reflected in the metric fibration between 
the $\eta$-direction (the Reeb vector) and the $\Sigma_2$.
The fibration is implemented by a vector field $A_1$ 
in eq. \eqref{NN02}. The twisting mixes the R-symmetry of the QFT, 
represented by $A_1$ in the dual description,  
with (part of) the Lorentz group.
In purely field theoretical terms, we are modifying the covariant derivative of different fields that under the  combined action of 
the spin connection and the R-symmetry (on the curved part of the space) will read $D_\mu\sim \partial_\mu+\omega_\mu+ A_\mu$.

In performing this twisting, the fields decompose under 
$SO(1,3)\to SO(1,1)\times SO(2)$. 
The decomposition is straightforward for the bosonic fields. 
For the fermions, we have that $\bf(\frac{1}{2}, 0) $ 
decomposes as ${\bf (+,\pm)}$ and similarly 
for the $(0,\frac{1}{2})$ spinors. 

The twisting itself is the 'mixing' between the $\pm$ charges of the spinor
and its R-symmetry charge. There is an analog operation for the vector and 
scalar fields. Some fields are scalars under the diagonal group in
$U(1)_R \times SO(2)_{\Sigma_2}$. Some are spinors and some are vectors.
Only the scalars under the diagonal group are massless. These determine the 
SUSY content of the QFT.
This particular example 
amounts to preserving two supercharges. 
There are two massless vector multiplets and two massless matter multiplets. The rest of the fields get a mass whose set by the inverse size of the compact manifold. In other words, the field theory at low energies is a two dimensional  CFT (as indicated by the $AdS_3$ factor), preserving $(0,2)$ SUSY and  obtained by a twisted compactification
of the Klebanov-Witten CFT. The QFT is deformed in the UV by a relevant operator of dimension two, as we can read from eq. \eqref{O20x}. 

An alternative way to think about this QFT is as the one describing
the low energy excitations of a stack of $N_c$ D3 branes wrapping  
a calibrated space $\Sigma_2$
inside a Calabi-Yau 4-fold.

The situation with the metrics in Section \ref{sectionDG} is 
more subtle. In that case there is also a flow from the Klebanov-Witten quiver to a two-dimensional CFT preserving (0,2) SUSY. The difference is that this second QFT is not apparently obtained via a twisting procedure. As emphasized by the authors of \cite{Donos:2014eua}, 
the partial breaking of SUSY is due (from a five-dimensional 
supergravity perspective) to 'axion' fields depending on the torus directions. These axion fields are proportional to a  
deformation parameter---that we called $\lambda$ in eq. \eqref{fluxes-bef}. 
The deformation in the UV QFT is driven 
by an operator of dimension four that was identified to be $Tr(W_1^2-W_2^2)$ and a  dimension six operator that acquires a VEV, as discussed in \cite{Donos:2014eua}. 

To understand the dual field theory to the  IIA backgrounds obtained after non-Abelian T-duality and presented in Section \ref{s2h2natd} involves more intricacy. Indeed, it is at present unclear what is the analog field theoretical operation of non-Abelian T-duality. There are, nevertheless, important hints. 
Indeed, the foundational paper of Sfetsos and Thompson \cite{Sfetsos:2010uq}, that sparked the interest of the uses of non-Abelian T-duality in quantum field theory duals, showed that if one starts with a background of the form $AdS_5\times S^5/Z_2$, a particular solution of the Gaiotto-Maldacena system (after lifting to M-theory) is generated \cite{Gaiotto:2009gz}.
This is hardly surprising, as the backgrounds of eleven dimensional supergravity with an $AdS_5$ factor and preserving $\mathcal{N}=2$ SUSY in four dimensions, have been classified. What is interesting is that the solution generated by Sfetsos and Thompson
appears as a zoom-in on the particular class of solutions in \cite{Maldacena:2000mw}. This was extended in 
\cite{Itsios:2013wd} that computed the action of non-Abelian T-duality on the end-point of  the flow from the $\mathcal{N}=2$ conformal quiver with adjoints to the Klebanov-Witten CFT. Again, not surprisingly, the  backgrounds obtained correspond to the $\mathcal{N}=1$ version of the Gaiotto $T_N$ theories---
these were called {\it Sicilian} field theories by Benini, Tachikawa and Wecht
in  \cite{Benini:2009mz}, see Fig. \ref{RoadMap}.
It is noticeable, that while the Sicilian theories can be 
obtained by a twisted compactification of M5 branes on $H_2, S^2, T^2$, 
the case obtained using non-Abelian T-duality corresponds only 
to M5 branes compactified on $S^2$ and preserving minimal SUSY in four dimensions. What we propose in this paper is that the twisted compactification on $\Sigma_2$ of a Sicilian gauge theory
 can be studied by using the backgrounds we discussed in Section \ref{s2h2natd} and their M-theory counterparts.
We will elaborate more about the 2-d CFTs and 
their flows in the coming sections.
 
 In the following, we will calculate different observables of  these QFT's by using the backgrounds as a 'definition' of the 2d SCFT. The backgrounds are smooth and thus the observables have trustable results. 
Hence, we are defining a QFT by its observables, 
calculated in a consistent way by the dual solutions. 
The hope is that these calculations together with 
other efforts can help map the space of 
these new families of CFTs. To the study of these observables we turn now.

\section{Quantized charges}\label{chargessectionxx}
In this section, we will study the quantized charges on the string side.
This analysis appears in the field theory part of the paper because these charges will, as in the
canonical case of $AdS_5\times S^5$, translate into the ranks of the gauge theory local symmetry groups.

The NATD produced local solutions to the 10-dim SUGRA equations of motion. Nevertheless, it is still not known how to obtain the global properties of these new geometries. 
Some quantities associated to a particular solution, like the Page charges below, are only well-defined when the global properties of the background are known. Since we have only a local description of our solution, we will propose very reasonable global results for the Page charges, mostly based on physical intuition.

Let us start by analyzing a quantity that is proposed to be periodic in the string theory.
We follow the ideas introduced in \cite{Lozano:2014ata} and further elaborated in \cite{Macpherson:2014eza}. 
To begin with, we focus on the NATD version of the twisted solutions; described in Section \ref{s2h2natd}. Let us define the quantity,
 \begin{equation}
b_0= \frac{1}{4 \pi^2 \alpha'} \int_{\Pi_2} B_2  ~~ \in  \ [0,1] \ ,
\label{page9b}
\end{equation}
where the cycle $\Pi_2$ is defined as,
 \begin{equation}
\Pi_2=S^2   ~~~  \Big\{  \chi , \xi  , \alpha=0, \rho=\textrm{const} , d\beta=-\frac{1}{3z} d\xi    \Big\} \ .
\label{page9ceeett}
\end{equation}
As the topology of the NATD theory is not known, we propose that this cycle is present in the geometry. This cycle will have a globally defined volume form, which in a local description can be written as  $\textrm{Vol}_{\Pi_2}=\sin \chi \ d \xi \wedge d\chi$. 
We then find,
 \begin{equation}
b_0=\frac{1}{4 \pi^2 \alpha'} \int_{\Pi_2} \alpha' \rho \sin \chi d\xi \wedge d \chi= \frac{\rho}{\pi}  ~~ \in  \ [0,1].
\label{page9deee}
\end{equation}
Again, we emphasize that this is a {\it proposal} made in \cite{Lozano:2014ata} and used in \cite{Macpherson:2014eza}. Moving further than $\pi$ along the variable $\rho$ can be 'compensated' by performing a large gauge transformation on the $B_2$-field,
\begin{equation}
B_2 \rightarrow B_2'=B_2 - \alpha '  n \pi \sin \chi d \xi \wedge d\chi.
\label{page10scsc}
\end{equation}
We will make extensive use of this below. 


 Let us now focus on the conserved magnetic charges defined for our backgrounds. We will start the analysis for the case of the solutions in Section \ref{section211xx}.

\subsection{Page charges for the twisted solutions}
We will perform this study for the solutions before and 
after the NATD, and we will obtain how the Page charges transform under the NATD process.
Page charges  (in contrast to Maxwell charges) have
 proven fundamental in understanding aspects of the dynamics of 
field theories---see for example \cite{Benini:2007gx}.

In the following, we use as definition of Page charge,
\begin{equation}
\begin{aligned}
&  N_{Dp} \big{|}_{\Pi_{8-p}}=\frac{1}{2\kappa_{10}^2 T_{Dp}} \int_{\Pi_{8-p}}  \left( \sum\limits_{i} F_i  \right) \wedge  e^{-B_2},      ~~~~~~         \left. N_{NS5} \right|_{\Pi_3} =  \frac{1}{2\kappa_{10}^2 T_{NS5}} \int_{\Pi_3} H_3 \ ,
\\[5pt]
& 2\kappa_{10}^2 =(2\pi)^{7} \alpha'^{4},      ~~~~~~   T_{Dp} =\frac{1}{(2\pi)^{p} \alpha'^{\frac{p+1}{2}}  },  ~~~~~~ T_{NS5} = T_{D5} \ .
\end{aligned}
\label{Page5g}
\end{equation}
In particular, for D3 branes we have,
\begin{equation}
N_{D3} \big{|}_{\Pi_5}=\frac{1}{2\kappa_{10}^2 T_{D3}} \int_{\Pi_5} \left( F_5 - B_2 \wedge F_3 + \frac{1}{2}B_2 \wedge B_2 \wedge F_1 \right).\;\;
\nonumber
\label{page201tgrtg}
\end{equation}

The topology of the internal space is 
$\Sigma_2 \times S^2 \times S^3$.  
First, we consider the cycle $S^2 \times S^3$. 
Second, we can consider some cycles given by the product of $\Sigma_2$ with a generator of $H_3 \left( S^2 \times S^3 , \mathbb{Z}  \right)$. Notice that the $S^2 \times S^3$ is realized as a $U(1)$ fibration over an $S_1^2 \times S_2^2$ base. A smooth 3-manifold, $S_1^3$,  that can be used to generate $H_3 \left( S^2 \times S^3 , \mathbb{Z}  \right)$ is provided by the circle bundle restricted to the $S_1^2$ factor. We can also choose $S_2^3$, defined to be the circle bundle restricted to the $S_2^2$ factor of the base space. In summary, the relevant cycles are,
\begin{equation}
\begin{aligned}
\Pi_5^{(1)}=S^2 \times S^3  ~\Big\{\theta_1,\phi_1,\theta_2,\phi_2, \psi \Big\}, \;\;\;\Pi_5^{(2)}&=\Sigma_2 \times S_1^3                ~  \Big\{\alpha,\beta, \theta_1,\phi_1, \psi \Big\}, \;\;\;
\Pi_5^{(3)}=\Sigma_2 \times S_2^3   ~ \Big\{ \alpha,\beta ,\theta_2,\phi_2, \psi \Big\} \ .
\label{page6tt}
\end{aligned}
\end{equation}
The background fields $B_2$, $F_1$ and $F_3$ are vanishing, and only $F_5$ contributes. 
The specific components of $F_5$ in eq. \eqref{NN01} that have non vanishing pullback on these cycles are,
\begin{equation}
F_5= \frac{L^4}{9} \ \textrm{Vol}_1 \wedge \textrm{Vol}_2 \wedge \eta + \frac{L^4}{6} z \textrm{Vol}_{\Sigma_2} \wedge \left( \textrm{Vol}_1 + \textrm{Vol}_2 \right) \wedge \eta + ...
\label{page6a}
\end{equation}
We explicitly see that it is a globally defined form, as all the involved quantities ($\eta$, $\textrm{Vol}_{\Sigma_2}$, $\textrm{Vol}_1$, $\textrm{Vol}_2$) are well defined globally.
The associated Page charges of D3 branes for the background  around 
eq. \eqref{NN01} are,
\begin{equation}
N_{D3} \big{|}_{\Pi_5^{(1)}}=\frac{4 L^4}{27 \alpha'^2 \pi} \ ,~~~~~~~~~~~~~~ \widehat{N}_{D3} \big{|}_{\Pi_5^{(2)}}=\widetilde{N}_{D3} \big{|}_{\Pi_5^{(3)}}=\frac{L^4}{\alpha'^2} \frac{z \  vol \left( \Sigma_2 \right)}{18 \pi^2} \ ,
\label{page9dvdfv}
\end{equation}
where $vol\left( \Sigma_2 \right)$ is the total volume of the two-manifold $\Sigma_2$.\footnote{Notice that we use Vol for volume elements (differential forms) and $vol$ for the actual volumes of the manifolds (real numbers).}
As usual, the first relation quantizes the size of the space,
\begin{equation}
L^4=\frac{27 \pi}{4} \alpha'^2 N_{D3}.
\label{page10xdf}
\end{equation}
We have then defined three D3-charges. The one associated with $N_{D3}$ is the usual one appearing
also in the $AdS_5\times S^5$ case. The other two can be thought as charges  'induced'  by the wrapping of the D3 branes on the Riemann surface. The reader may wonder  whether these charges are present in the backgrounds found after NATD. We turn to this now.

The particular  expressions for Page charges in Type IIA are,
\begin{equation}
\begin{aligned}
N_{D6} \big{|}_{\Pi_2}&=\frac{1}{2\kappa_{10}^2 T_{D6}} \int_{\Pi_2} \left( \widehat{F}_2-\widehat{F}_0 \ \widehat{B}_2\right), 
\\[5pt]
N_{D4} \big{|}_{\Pi_4}&=\frac{1}{2\kappa_{10}^2 T_{D4}} \int_{\Pi_4} \left( \widehat{F}_4- \widehat{B}_2 \wedge \widehat{F}_2 + \frac{1}{2} \widehat{F}_0 \  \widehat{B}_2 \wedge \widehat{B}_2 \right), 
\\[5pt]
N_{D2} \big{|}_{\Pi_6}&=\frac{1}{2\kappa_{10}^2 T_{D2}} \int_{\Pi_6} \left( \widehat{F}_6- \widehat{B}_2 \wedge \widehat{F}_4+\frac{1}{2} \widehat{B}_2 \wedge \widehat{B}_2 \wedge \widehat{F}_2- \frac{1}{6} \widehat{F}_0 \ \widehat{B}_2 \wedge \widehat{B}_2 \wedge \widehat{B}_2  \right).
\label{page201uu}
\end{aligned}
\end{equation}
We label the radius of the space of the geometry in eq. \eqref{NN11} by $\widehat{L}$, to distinguish it from $L$, the quantized radius before the NATD.

In order to properly define the cycles to be considered, we should know the topology of this NATD solution. However, we have obtained only a local expression for this solution, and we do not know the global properties. As we explained above, we will  present a proposal to define the Page charges that would explain the transmutation of branes through the NATD. We propose the relevant cycles to be, \footnote{Intuitively, we can think that the branes transform under NATD as 3 consecutive T-dualities. For example, in the first 5-cycle of eq. (\ref{page6tt}), the NATD is performed along 3 of the coordinates, ($\theta_2, \phi_2, \psi$), in such a way that we end up with a 2-cycle, the first cycle in eq.(\ref{page7r}), associated with D6 branes. In the second 5-cycle of eq.(\ref{page6tt}) the NATD only affects the $\psi$-direction, so it disappears, and two more directions are added in order to complete the 3 T-dualities, ending up with a 6-cycle in eq.(\ref{page7r}), associated with D2-branes.}
\begin{equation}
\Pi_2^{(1)}  \ \  \Big\{\theta_1,\phi_1 \Big\} ,  ~~~~ 
      \Pi_6^{(2)}= \Big\{\alpha,\beta,\theta_1, \phi_1, \rho, \xi,\chi=\frac{\pi}{2} \Big\} ,  ~~~~  \Pi_2^{(3)} \  \  \Big\{\alpha,\beta \Big\} .
\label{page7r}
\end{equation}
The associated charges are,
\begin{equation}
\label{page8zr}
 \begin{aligned}
  & N_{D6} \big{|}_{\Pi_2^{(1)}}=\frac{2 \widehat{L}^4}{27 \alpha'^2 } \ ,    \qquad \quad  \widetilde{N}_{D6}\big{|}_{\Pi_2^{(3)}}=\frac{\widehat{L}^4}{\alpha'^2} \frac{z \ vol \left( \Sigma_2 \right)}{36 \pi} \ ,
  \\[10pt]
  & \widehat{N}_{D2} \big{|}_{\Pi_6^{(2)}}=\frac{\widehat{L}^4}{\alpha'^2} \frac{z \  vol \left( \Sigma_2 \right)  vol \left(\rho, \xi \right)}{ 144 \pi^4}=\frac{n^2}{4}\frac{\widehat{L}^4}{\alpha'^2} \frac{z \  vol \left( \Sigma_2 \right)}{36 \pi}.
 \end{aligned}
\end{equation}
%
%
In the last expression, we performed the integral over the $\rho$-coordinate in the interval $[0, n\pi]$.
These three charges are in correspondence with the ones before the NATD in eq. \eqref{page9dvdfv}.
Indeed, we can compute the quotients,
\bea
& & \frac{\widehat{N}_{D3} }{N_{D3}}=\frac{\widetilde{N}_{D3}}{N_{D3}}=\frac{3}{8\pi}z \  vol \left( \Sigma_2 \right), \qquad \frac{4 \widehat{N}_{D2} }{n^2 N_{D6}}=\frac{\widetilde{N}_{D6}}{N_{D6} }=\frac{3}{8\pi}z \  vol \left( \Sigma_2 \right).
\label{manaxx}
\eea
These quotients indicate a nice correspondence between charges before
and after the duality.

Using the first relation in eq. \eqref{page8zr}  we quantize the size $\widehat{L}$ of the space after NATD to be
$\widehat{L}^4=\frac{27}{2} \alpha'^2 N_{D6}$.

A small puzzle is presented by the possible existence of charge for D4 branes, as there would be no quantized number before the NATD to make them correspond to. To solve this puzzle, we propose that there should be a globally defined closed non-exact form that allows us to perform a large gauge transformation for the $\widehat{B}_2$, in such a way that all the D4 Page charges vanish. In local coordinates, we have a gauge transformation,
\begin{equation}
\widehat{B}_2 \rightarrow \widehat{B}'_2 = \widehat{B}_2 + \alpha' d\left[ \rho \cos \chi \ \tilde{\sigma}_3 \right],
\label{page9ss9}
\end{equation}
written in such a way that the integrand has at least one leg along a non-compact
coordinate,
\begin{equation}
\widehat{F}_4-\widehat{B}'_2 \wedge \widehat{F}_2=\frac{z \widehat{L}^4}{6 \sqrt{\alpha'}} e^{-2B} \left(\cos \chi d\rho - \rho \sin \chi d \chi \right) \wedge \textrm{Vol}_{AdS_3}.
\label{page9adtrtg}
\end{equation}
Hence, any Page charge for D4 branes is vanishing. To be precise, the  D2 charge $\widehat{N}_{D2}$ must be computed after choosing this gauge, as it depends on $\widehat{B}_2$, but it turns out to be the same as calculated in eq. \eqref{page8zr}.

The motion in the $\rho$-coordinate, as we discussed above, can be related to large gauge transformations of the $\widehat{B}_2$-field.
The  large transformation that 'compensates' for motions in $\rho$, namely
$
 \widehat{B}_2 \rightarrow \widehat{B}'_2=\widehat{B}_2 - \alpha '  n \pi \sin \chi d \xi \wedge d\chi,
$
has the effect of changing the Page charges associated with D4 branes, that were initially vanishing. Indeed, if we calculate  for the following four cycles,
\begin{equation}
\Pi_4^{(1)}  ~~  \Big\{\theta_1,\phi_1, \chi, \xi \Big\} \ , ~~~~~~ ~~
\Pi_4^{(2)} ~~  \Big\{\alpha,\beta , \chi, \xi \Big\} \ , ~~~~
\label{page12y}
\end{equation}
the Page charge of D4 branes varies according to,
\begin{equation}
\begin{aligned}
\Delta N_{D4} \big{|}_{\Pi_4^{(1)}}&=\frac{1}{2\kappa_{10}^2 T_{D4}} \int_{\Pi_4^{(1)}} \left(- \Delta  \widehat{B}_2 \wedge \widehat{F}_2  \right) =- n \frac{L'^4}{\alpha'^2}\frac{2}{27} =-n N_{D6},   
\\[10pt]
\Delta N_{D4} \big{|}_{\Pi_4^{(2)}}&=\frac{1}{2\kappa_{10}^2 T_{D4}} \int_{\Pi_4^{(2)}} \left(- \Delta  \widehat{B}_2 \wedge \widehat{F}_2  \right) = -n \frac{L'^4}{\alpha'^2}\frac{z}{36 \pi}  vol \left( \Sigma_2 \right)=-n \widetilde{N}_{D6}  .
\label{page11n}
\end{aligned}
\end{equation}
We can interpret these findings in the following way. Our QFT  (after the NATD) can be thought as living on the world-volume of a superposition of D2 and D6 branes. Motions in the $\rho$-coordinate induce charge of D4 branes, which can be interpreted as new gauge groups appearing. This suggest that we are working with a linear quiver, with many gauge groups. Moving $n \pi$-units in $\rho$ generates or 'un-higgses'  new gauge groups of rank $ n N_{D6}$ and  $n \widetilde{N}_{D6}$. Computing volumes or other observables that involve integration on the $\rho$-coordinate amounts to working with a  QFT with different gauge group, depending on the range in $\rho$ we decide to integrate over. Notice that $\rho$ is not a holographic coordinate. Motions in $\rho$ are not changing the energy in the dual QFT. For the $AdS$-fixed points the theory is conformal and movements in $\rho$ do not change that.

In the paper \cite{Macpherson:2014eza} , the motion in the $\rho$-coordinate was argued to be related to a form of 'duality' (a Seiberg-type of duality was argued to take place, in analogy with the mechanism of the Klebanov-Strassler duality cascade, but in a CFT context). That can not be the whole story as other observables, like for example the number of degrees of freedom in the QFT, change according to the range of the $\rho$-integration. Hence the motion in $\rho$ can not be just a duality. 
 We are proposing here that moving in the $\rho$-coordinate amounts to changing the quiver, adding gauge groups, represented by the increasing D4 charge.

We will now present the same analysis we have performed above, but for the case of the background in Section \ref{sectionDG}.

\subsection{Page charges for the Donos-Gauntlett solution}
Part of the analysis that follows was carefully done  in the original work of \cite{Donos:2014eua} and here we will extend the study for the solution after the NATD that we presented in Section \ref{natdDG}. We will give an outline of the results as the general structure is similar to the one displayed by the twisted solutions of the previous section.

We focus on the original Donos-Gauntlett background first. We denote by $d_{\alpha}$, $d_{\beta}$ the two radii of the torus (the cycles of the torus are then of size $2\pi d_\alpha$ and $2\pi d_\beta$ respectively) and consider
five different cycles in the geometry,
%
%
\begin{equation}
\label{page20ab}
 \begin{aligned}
& \Pi_5^{(1)}=S^2 \times S^3  ~~  \Big\{\theta_1,\phi_1,\theta_2,\phi_2, \psi \Big\} ,  ~~
\Pi_5^{(2)}=T^2 \times S_1^3                ~~   \Big\{\alpha,\beta, \theta_1,\phi_1, \psi \Big\},     ~~
\Pi_5^{(3)}=T^2 \times S_2^3   ~~ \Big\{ \alpha,\beta ,\theta_2,\phi_2, \psi \Big\} ,
\\[10pt]
& \Pi_3^{(4)}=S_1^1 \times s(S)  ~~ \Big\{ \alpha, \theta_1= \theta_2, \phi_1= - \phi_2 , \psi=\textrm{const} \Big\},   ~~
\Pi_3^{(5)}=S_2^1 \times s(S)  ~~ \Big\{ \beta, \theta_1= \theta_2, \phi_1= - \phi_2 , \psi=\textrm{const} \Big\} .
 \end{aligned}
\end{equation}
%
%
The  Page charges associated with $D$3, $D5$ and $NS5$ branes are,
%
%
\begin{equation}
\label{page21a}
 \begin{aligned}
  & N_{D3} \big{|}_{\Pi_5^{(1)}}=\frac{4 L^4}{27 \alpha'^2 \pi} \ ,  \qquad \widehat{N}_{D3} \big{|}_{\Pi_5^{(2)}}= - \widetilde{N}_{D3} \big{|}_{\Pi_5^{(3)}}=  2 \ \frac{L^4}{\alpha'^2} \frac{\lambda^2 \ d_{\alpha} d_{\beta}}{9} \ ,
  \\[10pt]
 & N_{NS5} \big{|}_{\Pi_3^{(4)}}=- 2 \ \frac{L^2 \lambda \ d_{\alpha}}{3 \alpha'} \ , \qquad 
     N'_{D5} \big{|}_{\Pi_3^{(5)}}=2 \ \frac{L^2 \lambda \ d_{\beta}}{3 \alpha'} \ .
 \end{aligned}
\end{equation}
After the NATD, we focus on the background around eqs. (\ref{s5}) (\ref{s11}). 
We consider the  following cycles in the geometry,
%
\begin{equation}
\label{page273}
 \begin{aligned}
  & \Pi_2^{(1)}  ~~   \Big\{ \theta_1,\phi_1 \Big\}, ~~~~
\Pi_6^{(2)}         ~~   \Big\{ \alpha, \beta, \theta_1,\phi_1, \rho, \xi , \chi=\frac{\pi}{2}  \Big\}, ~~~~
\Pi_2^{(3)}    ~~ \Big\{ \alpha,\beta \Big\}, ~~~~
\\[10pt]
  & \Pi_6^{(4)} ~~ \Big\{ \beta, \theta_1, \phi_1, \rho, \chi, \xi \Big\}, ~~~~
\Pi_2^{(5)}  ~~ \Big\{ \beta, \chi, \rho=\rho_0 \Big \}.   ~~~~~~~~~~~~~~~~~~~~~~~~~~~~~~~
 \end{aligned}
\end{equation}
The  correspondent Page charges defined on them (the $\rho$-coordinate is taken in the $[0, n\pi]$ interval),
%
%
\begin{equation}
\label{page8khg}
 \begin{aligned}
   & N_{D6} \big{|}_{\Pi_2^{(1)}}=\frac{2 \widehat{L}^4}{27 \alpha'^2 } \ ,    \qquad     \widetilde{N}_{D6}\big{|}_{\Pi_2^{(3)}}=  - \frac{\widehat{L}^4}{\alpha'^2} \frac{ \lambda^2  \pi}{18} d_{\alpha} d_{\beta} \ , \qquad  N'_{D6} \big{|}_{\Pi_2^{(5)}}=\frac{ L'^2 }{ \alpha' } \frac{\lambda}{3} \rho_0 \ d_{\beta} \ ,
   \\[10pt]
   & \widehat{N}_{D2} \big{|}_{\Pi_6^{(2)}}=-\frac{\widehat{L}^4}{\alpha'^2} \frac{  \lambda^2 \   vol \left( T^2 \right)  vol  \left(\rho , \xi \right)}{288 \pi^4}  =- \frac{\widehat{L}^4}{\alpha'^2} \lambda^2 \frac{n^2 \pi}{72} d_{\alpha} d_{\beta} ,
   \\[10pt]
   &  N_{D2} \big{|}_{\Pi_6^{(4)}}= \frac{\widehat{L}^2}{\alpha'} \frac{\lambda}{24 \pi^3} d_{\beta} \  vol \left( \rho, \chi, \xi \right)= \frac{\widehat{L}^2}{\alpha'} \lambda \frac{n^3 \pi}{18} d_{\beta}.  ~~~~~~~~~~~
 \end{aligned}
\end{equation}
From the first relation we obtain,
$
\widehat{L}^4=\frac{27}{2} \alpha'^2 N_{D6}$.  Like in the case of the twisted solutions, we can choose a gauge for the $\widehat{B}_2$ field
%
%
\begin{equation}
\label{page9joijo}
 \begin{aligned}
  & \widehat{B}_2 \rightarrow \widehat{B}'_2=\widehat{B}_2 + \delta \widehat{B}_2 \ ,
\\[5pt]
  & \delta \widehat{B}_2=\frac{9 \alpha'^2 \lambda}{2 L^2} \ d \beta \wedge \Big( \rho \sin \chi \ d \big( \rho \sin \chi \big)+ \mathcal{B} \ d \mathcal{B} \Big)+ L^2 \frac{ \lambda}{6} \big( d \alpha \wedge \sigma_{3} - \alpha \ \sigma_{1}\wedge \sigma_{2} \big), 
 \end{aligned}
\end{equation}
such that the Page charge of D4 branes, when computed on  every possible compact  4-cycle, is vanishing. 
Indeed, after the gauge transformation, we have
\begin{equation}
\widehat{F}_4-\widehat{B}'_2 \wedge \widehat{F}_2  =\frac{L^4 \lambda}{36 \sqrt{\alpha'}} e^{V}\left( \frac{2 L^2}{\alpha'}d\alpha -3 e^{-2B-2V} \lambda \ d(\rho \cos \chi) \right) \wedge \textrm{Vol}_{AdS_3}.
\label{page9hu}
\end{equation}
Any integral over compact manifolds is vanishing. Like in the case of the twisted solutions, $\widehat{N}_{D2}$ and $N_{D2}$ should be recalculated after choosing this gauge;  but their value turns out to be unchanged.

We can apply the same  string theory considerations on the quantity $b_0$ that now is defined as an integral over the cycle,
\begin{equation}
\Pi_2=S^2 \ ,  \qquad \{  \chi , \xi , \alpha = \textrm{const},\rho= \textrm{const}    \}.
\label{page9c}
\end{equation}
We then calculate,
 \begin{equation}
b_0=\frac{1}{4 \pi^2 \alpha'} \int_{\Pi_2} \alpha' \rho \sin \chi d\xi \wedge d \chi= \frac{\rho}{\pi}  ~ \in  \ [0,1].
\label{page9d0}
\end{equation}
If we move further than $\pi$ along the variable $\rho$, we can compensate this by performing the large gauge transformation
$
\widehat{B}_2 \rightarrow \widehat{B}'_2=\widehat{B}_2 - \alpha '  n \pi \sin \chi d \xi \wedge d\chi .
$
We now consider the correspondent variation of Page charges for D4 branes, that can be calculated using the following cycles,
\begin{equation}
\Pi_4^{(1)} ~~ \{ \theta_1,\phi_1, \chi, \xi \} \ , ~~~~~~~~ 
\Pi_4^{(2)} ~~ \{\alpha,\beta , \chi, \xi \}, ~~~~
\label{page1200}
\end{equation}
to be,
\begin{equation}
\begin{aligned}
\Delta Q^P_{D4} \big{|}_{\Pi_4^{(1)}}&=\frac{1}{2\kappa_{10}^2 T_{D4}} \int_{\Pi_4^{(1)}} \left(- \Delta  \widehat{B}_2 \wedge \widehat{F}_2  \right) =- n \frac{L^4}{\alpha'^2}\frac{2}{27}   =- n N_{D6} \ ,
\\[5pt]
\Delta Q^P_{D4} \big{|}_{\Pi_4^{(2)}}&=\frac{1}{2\kappa_{10}^2 T_{D4}} \int_{\Pi_4^{(2)}} \left(- \Delta  \widehat{B}_2 \wedge \widehat{F}_2  \right) =-  n \frac{L^4}{\alpha'^2}\frac{\pi \lambda^2}{18 }  d_{\alpha} d_{\beta}=n \widetilde{N}_{D6} \ .
\label{page1100}
\end{aligned}
\end{equation}
The variation of the Page charges of D2 branes vanishes under this large gauge transformation. For the Donos-Gauntlett solution we observe a structure very similar to the one discussed for the twisted solutions. 
Again, here we would propose that the NATD background 'un-higgses'
gauge groups of rank $n N_{D6}$ as we move in units of $n\pi$ in the $\rho$-coordinate.

We move now to the study of another important observable of 
our dual 2-d and 4-d CFTs.

\section{Central charges and c-theorem}
\label{centralchargessectionxx}
In this section, we will study the central charge, an important observable of the different strongly coupled QFTs that our backgrounds in the Part I of the paper are defining.

Let us start with a brief summary of the ideas behind this observable. 
The RG-flow can be understood as the motion of the different couplings of the QFT $\lambda_i$  in terms of a parameter $t=-\ln \mu$, such that for a given operator $\widehat{O}$,
\beq
\frac{d\widehat{O}}{dt}=-\beta_i(\lambda) \frac{\partial \widehat{O}}{\partial\lambda_i}.
\eeq
Hence, the beta functions $\beta_i(\lambda)$ are the 'velocities' of the motion towards the IR.
It is interesting to define a 'c-function' $c(t)$ with the property that it decreases when flowing to low energies,
\beq
\frac{dc(t)}{dt} \leq 0.
\eeq
Such that at stationary points $\frac{dc(t)}{dt}=0$ implies that $\beta_i(\lambda)=0$ and vice versa.
The intuition behind this quantity is that massless degrees of freedom are lifted by relevant deformations, the flow to low energies then coarse-grains away these lifted modes. This intuition is realized in different situation: the two dimensional case, with Zamolodchikov's definition of $c(g)$ 
\cite{Zamolodchikov:1986gt} or Cardy's conjecture for the 'a-theorem' 
\cite{Cardy:1988cwa},  proven by Komargodski and Schwimmer in
\cite{Komargodski:2011vj}. There are 
different versions of the c-theorem, varying in 'strength' and generality. See the paper 
\cite{Barnes:2004jj} for a summary.

As we discussed, the  c-function is properly defined
only at the conformal points of a QFT. 
Hence, we can define it properly in
all of our backgrounds only at the $AdS_5$ and $AdS_3$ fixed points. In those cases the central charge is basically proportional to the volume of the 'internal manifold' $M_d$ (the complement-space of $AdS_5$ or $AdS_3$). 
 Indeed, there exists a well-established formalism to calculate central charges
 that uses the relation between this quantity and the Weyl anomaly in the QFT when placed on a generic curved space. This was first discovered in \cite{Brown:1996ata}  (before the Maldacena conjecture was formulated!) and a  complete understanding was  developed in \cite{Skenderis:1998ata}.
 Indeed, for conformal field theories in two and four dimensions, associated with $AdS_3$
and $AdS_5$ geometries respectively, we have \footnote{The central charges $a$ and $c$ are equal
at the leading order in an  $N_c$-expansion. This is the result captured by the Supergravity approximation used in this paper. Also notice that the L's entering in this formulas are relative to an $AdS$ space expressed in the canonical form. }
\begin{equation}
\label{skenderishenningson}
\hskip -4.2cm \textrm{2 dimensions:}\ , \qquad \mathcal{A}=-\frac{L \bar{R}}{16 \pi G_N^{(3)}}=-\frac{c}{24 \pi}\bar{R} , ~~~~~ ~~~~~~~~~~~~~~~ c=\frac{3}{2} \frac{L}{G_N^{(3)}} \ ,
\end{equation}
and
\begin{equation}
\begin{aligned}
& \textrm{4 dimensions:} \ , \qquad   \mathcal{A}=-\frac{L^3 }{8 \pi G_N^{(5)}}  \left( - \frac{1}{8}\bar{R}^{ij}\bar{R}_{ij} +\frac{1}{24} \bar{R}^2 \right)  =\frac{1 }{16 \pi^2 }\Bigg[c \Big( \bar{R}^{ijkl}\bar{R}_{ijkl} -2 \bar{R}^{ij}\bar{R}_{ij} +\frac{1}{3}\bar{R}^2 \Big) 
\\[5pt]
& \qquad\qquad\qquad\qquad\quad \; - a \Big( \bar{R}^{ijkl}\bar{R}_{ijkl} -4 \bar{R}^{ij}\bar{R}_{ij} + \bar{R}^2 \Big)  \Bigg] \ ,   ~~~~~~~    c=a=\frac{\pi}{8}\frac{L^3}{G_N^{(5)}} \ , 
\end{aligned}
\end{equation}
where  $\bar{R}_{ijkl}$, $\bar{R}_{ij}$ and $\bar{R}$ 
are the Riemann and Ricci tensors and scalar of the boundary metric. 
The Newton constant in different dimensions is calculated according to
(we take $g_s=1$ as in the rest of this paper),
 \begin{equation}
G_N^{(10)}=8 \pi^6  \alpha'^4,  ~~~~~ G_N^{(10-d)}=\frac{G_N^{(10)}}{vol (M_d)} . 
\label{Central1a}
\end{equation}
For solutions presenting a flow between these fixed points (or generically, an RG flow), a quantity that 
 gives an idea of the number of degrees of freedom can also be defined. This quantity measures an 'effective volume' of the internal space, that is a volume that is weighted together with the dilaton
and  other factors in the metric. To define such a quantity one goes back to a proposal by Freedman, Gubser, Pilch and Warner \cite{Freedman:1999gp}---see also the paper \cite{Alvarez:1998wr} for earlier attempts. Indeed, for any background (that
should be considered to be a solution of a D-dimensional supergravity, possibly connected with string theory) of the form,
\beq
ds_{D}^2= e^{2A(r)}dx_{1,D-2}^2 + dr^2 \ ,
\label{manaza}
\eeq
and assuming that the matter fields satisfy certain Energy conditions
\cite{Freedman:1999gp}, it was proven  using the Einstein equations that the quantity,
\beq
c\sim \frac{1}{(A')^{D-2}},
\label{zaza}
\eeq 
is monotonically increasing towards the UV \cite{Freedman:1999gp}. This proposal was extended by 
Klebanov, Kutasov and Murugan in the paper \cite{Klebanov:2007ws}, to account for an RG-flow in a $d+1$ dimensional QFT, dual to a generic metric and dilaton of the form,
\beq
ds^2= \alpha_0(r)\Big[ dx_{1,d}^2+ \beta_0(r) dr^2  \Big]
+ g_{ij}(r,\vec{\theta}) d\theta^id\theta^j,\;\;\;\; \Phi(r).
\label{ppeerroo}
\eeq
In these cases and in cases where the functions $\alpha_0, \Phi$ depend also on the internal coordinates $\alpha_0(r,\vec{\theta}), \Phi(r,\vec{\theta})$, the formulas of \cite{Klebanov:2007ws} were extended in \cite{Macpherson:2014eza} \footnote{It would be interesting to find a generalization to the case in which also the function $\beta(r,\vec{\theta})$.} to be,
\beq
c= d^d \frac{ \beta_0(r)^\frac{d}{2}  \widehat{H}^{\frac{2d+1}{2}}}{ \pi G_N^{(10)} (\widehat{H}')^d}, \;\;\;\; \widehat{H}= \Big( \int d\vec{\theta} \sqrt{e^{-4\Phi} \det[g_{int} ]\alpha_0^d} \Big)^2.
\label{kkmcentral}
\eeq
At conformal points, i.e. when calculated in $AdS$ backgrounds, this gives a constant result, in agreement with eq. \eqref{skenderishenningson}. For backgrounds
with a flow, the quantity in eq. \eqref{kkmcentral} gives an idea of the number of degrees of freedom that participate in the dynamics of the QFT at a given energy.

In the following sections, 
we will quote the results for central charges 
according to eq. \eqref{skenderishenningson} 
for the conformal field theories in two and four dimensions. Following that,
we will write the result that eq. \eqref{kkmcentral} gives for the flows between theories.

\subsection{Central charge at conformal points}

As anticipated, we will quote here the results for eq. \eqref{skenderishenningson} for the different $AdS_3$ and $AdS_5$ fixed points. We start with the twisted solutions of Section \ref{section211xx}. We will use that the volume of the $T^{1,1}$ space is $ vol(T^{1,1})=16\pi^3/27$.

\noindent{\underline{Twisted geometries}}

 For the IR $AdS_3$ fixed point, the volume of the internal compact manifold is $
vol (M_7)= \frac{1}{3}   L^7  vol (\Sigma_2) vol (T^{1,1})$, and the central charge is,
\begin{equation}
 c= 9 \big{|} N_{D3} \widehat{N}_{D3} \big{|}  =  \frac{L^8}{\alpha'^4} \frac{ vol (\Sigma_2) vol (T^{1,1})}{24 \pi^6}  .
\label{Central42}
\end{equation}
At the UV $AdS_5$ fixed point, the volume of the internal compact manifold is $vol (M_5)= L^5  vol (T^{1,1})$, and the result for the central charge is,
\begin{equation}
c=\frac{27}{64} N_{D3}^2=\frac{L^8}{\alpha'^4} \frac{ vol (T^{1,1})}{64 \ \pi^5} \ .
\label{Central3aoee}
\end{equation}
After the NATD, we must  consider the new radius of the space $\widehat{L}$ and the volume of the new 5-dim compact space $4 \pi^2 vol (\rho, \chi, \xi)$. The computations turn out to be similar as the previous ones, and we obtain that the central charges before and after NATD, for both the two and four dimensional CFTs, are related by,
\begin{equation}
\frac{\hat{c}}{c}=\frac{\widehat{L}^8}{L^8} \frac{4 \pi^2  vol (\rho, \chi, \xi)}{ vol (T^{1,1})}.
\label{Central5aa}
\end{equation}
Let us comment on the quantity $vol (\rho, \chi, \xi)$, that  appears 
in the calculation of the Page charges in Section \ref{chargessectionxx}---see for example, eq. \eqref{page8khg} and also in the computation of the 
entanglement entropy of Section \ref{sectionEEWilson}.
Indeed, if we calculate,  \footnote{We identify the integral with the volume of the manifold spanned by the new coordinates. This becomes more apparent if we use the expressions in the appendix, in different coordinate systems. }
\begin{equation}
\begin{aligned}
& vol (\rho, \chi, \xi)=\int_{0}^{n\pi} \rho^2 d\rho \int_{0}^{2\pi} d\xi \int_{0}^{\pi}\sin\chi d\chi= \frac{4\pi^4}{3}n^3 \ ,
\\[10pt]
& \frac{\hat{c}}{c}=\frac{\widehat{L}^8}{L^8} \frac{4 \pi^2  vol (\rho, \chi, \xi)}{ vol (T^{1,1})}= \frac{36 \pi N_{D6}^2}{N_{D3}^2} n^3 \ .
\end{aligned}
\label{xxyz}
\end{equation}
We have performed the $\rho$-integral in the interval $[0,n\pi]$. 
The logic behind this choice was spelt out 
in Section \ref{chargessectionxx}, see below eq. \eqref{page11n}. 
The proposal is that moving in units of $\pi$ in the $\rho$-coordinate implies 'un-higgsing' a gauge group, hence we would have a linear quiver gauge theory. The central charge captures this un-higgsing procedure, increasing according to how many groups we 'create'. What is interesting is the $n^3$
behavior in eq. \eqref{xxyz}. Indeed, if $n$ were associated with the rank 
of a gauge group, this scaling would be precisely the one obtained 
in Gaiotto-like CFTs (also valid for the $\mathcal{N}=1$ 'Sicilian' theories 
of \cite{Benini:2009mz}). Indeed,   
the NATD procedure when applied to the $AdS_5\times T^{1,1}$ 
background creates  metric and fluxes  similar to those characterizing 
the Sicilian CFTs. The backgrounds in Part I of the paper
are dual to a compactification of the Klebanov Witten CFT  
and (using the NATD background) the Sicilian CFT 
on a two space $\Sigma_2$. The two-dimensional 
IR fixed point of these flows is described by our 'twisted $AdS_3$' and its NATD. The central charge  of the Sicilian CFT and its compactified version is presenting a behavior that goes like
a certain rank to a third power $c\sim n^3$. 
This suggest that crossing $\rho=\pi$ amounts to adding D4 branes and 
Neveu-Schwarz five branes and $n$ is the number of  branes 
that were added ---see eqs. (\ref{page11n}) and (\ref{page1100})---  or crossed\footnote{Thanks to Daniel Thompson for a 
discussion about this.}. Had we integrated on the interval $[n\pi, (n+1)\pi]$, we would have obtained a scaling like
$c\sim N_{D_4}^2$ at leading order in $n$.

\noindent {\underline{Donos-Gauntlett geometry}}

We study here the central charge for the Donos-Gauntlett background in Section \ref{sectionDG}.
For the $AdS_3$ fixed point, the volume of the internal compact manifold $vol (M_7)= L^7 \left(4/3 \right)^{5/4} 4 \pi^2 d_{\alpha} d_{\beta} \ vol (T^{1,1})$, and the central charge is,
\begin{equation}
c= 3 \ \big{|} N_{D3} \widehat{N}_{D3} \big{|}=\frac{2}{3} \frac{L^8}{\alpha'^4} \frac{d_{\alpha} d_{\beta} vol (T^{1,1})}{\pi^4} \ . 
\label{Central3ao1}
\end{equation}
For the $AdS_5$ fixed point, the volume of the internal compact manifold $vol (M_5) =  L^5  vol (T^{1,1})$, and the central charge results in,
\begin{equation}
c=\frac{27}{64} N_{D3}^2=\frac{L^8}{\alpha'^4} \frac{ vol (T^{1,1})}{64 \ \pi^5} \ .
\label{Central3ao}
\end{equation}

The quotient of central charges before and after the NATD for the Donos-Gauntlett QFT, are given by a similar expression to that in eq. \eqref{Central5aa}.

We move now to study a quantity that gives an idea 
of the degrees of freedom along a flow.

\subsection{Central charge for flows across dimensions}
\label{CCdifdims}
In the previous section, we calculated the central charge for two and four dimensional CFTs dual to the $AdS_3$ and $AdS_5$ fixed points of the flow. In this section, we will use the developments in  \cite{Freedman:1999gp} and \cite{Klebanov:2007ws}, to write a c-function along the flows between these fixed points. We will find various subtleties,
\begin{itemize}
\item{When considered as a low energy two-dimensional CFT, the definition of the c-function evaluated on the flows will not detect the presence of the four dimensional CFT in the far UV.}
\item{We attempt to generalize the formula of \cite{Klebanov:2007ws}
for anisotropic cases (that is for field theories that undergo a spontaneous compactification on $\Sigma_2$). This new definition will detect both the two dimensional and four dimensional conformal points, but will not necessarily be decreasing towards the IR. This is not in contradiction with 'c-theorems' that assume Lorentz invariance.}
\end{itemize}
We move into discussing these different points in our particular examples.
To start, we emphasize that the formulas in eqs.(\ref{manaza})-(\ref{kkmcentral}), contain the same information. Indeed, the authors of 
\cite{Klebanov:2007ws} present a 'spontaneous compactification' of a higher dimensional Supergravity (or String theory) to $d+2$ dimensions, see eq. \eqref{ppeerroo}. Moving the reduced system to Einstein frame and observing that the $T_{\mu\nu}$ of the  matter in the lower dimension satisfies  certain positivity conditions imposed in \cite{Freedman:1999gp}, use of  eq. \eqref{zaza} implies eq. \eqref{kkmcentral}. Hence, we will apply eqs. \eqref{ppeerroo}-\eqref{kkmcentral} to our different compactifications in Part I.

{\underline{Twisted and Donos-Gauntlett solutions.}}
For the purpose of the flows both twisted and Donos-Gauntlett solutions present
a similar qualitative behavior.
We start by considering the family of backgrounds in eq. \eqref{NN02} as dual to field theories in $1+1$ dimensions.
In this case the quantities relevant for the calculation of the central charge are,
\begin{equation}
\begin{aligned}
& d=1,\;\; \alpha_0 =L^2 e^{2A},\;\;\; \beta_0=e^{-2A},\;\;\; e^{\Phi}=1,
\\[5pt]
& \frac{ds_{int}^2}{L^2}=e^{2 B} ds^2_{\Sigma_{{2}}}+e^{2U}ds_{KE} ^2+e^{2V} \left( \eta + z A_1 \right)^2.
\end{aligned}
\end{equation}
We calculate the quantity
\beq
\widehat{H}= {\cal N}^2 e^{2(B+4U+V+A)},     \qquad      {\cal N}=\frac{(4\pi)^3 vol (\Sigma_2) L^8}{108}.
\label{ccc}
\eeq
Then, we obtain
\beq
c=\frac{{\cal N} e^{2B+4U+V}}{2 \pi G_N^{(10)}(2B'+4U'+V'+A')}.
\eeq
Using the BPS equations describing these flows in eq. \eqref{O17} we can get an expression without derivatives. Specializing for the solution with an $H_2$ in Section \ref{section211xx}, we find
\beq
c=\frac{{\cal N}}{9 \pi G_N^{(10)}}\frac{(1+e^{2r})^2}{1+2e^{2r}}.
\label{vava}
\eeq
We can calculate this for the background we obtained in 
Section \ref{s2h2natd}, by application of NATD. 
The result and procedure will be straightforward, 
but we will pick a factor of the volume of the space 
parametrized by the new coordinates,
$
vol (\rho, \chi, \xi)$.

For the purposes of the RG-flow, the quotient of the central charges will 
be the same as the quotient in eq. \eqref{xxyz}. This was indeed observed 
in the past \cite{Itsios:2013wd}, \cite{Macpherson:2014eza}
and is just a consequence of the invariance of the quantity 
$\big(e^{-2\Phi}\sqrt{\det[g]}\big)$ under NATD.

Coming back to eq. \eqref{vava}, we find that in the far IR, represented by $r\to -\infty$, 
the central charge is constant. But in the far UV ($r\to \infty$), 
we obtain a result that is not characteristic of a CFT. 
Hence, this suggest that the definition is only capturing the behavior 
of a 2-dim QFT. In other words, the four dimensional QFT may 
be thought as a two dimensional QFT, but  with an infinite number of fields. 

The absence of the four dimensional fixed point in our eq. \eqref{vava} can be accounted if we generalize the prescription to calculate central charges for an anisotropic 4-dim QFT. Holographically
this implies working with a background of the form,\footnote{A natural generalization of eq. \eqref{olaqase} is $
ds^2 =  -\alpha_0 dt^2 +\alpha_1 dy_1^2+....+\alpha_d dy_d^2
 +\Pi_{i=1}^{d}\alpha_i^{\frac{1}{d}}
\beta dr^2 + g_{ij}d\theta^i d\theta^j.$}
\beq
ds_{10}^2 =  -\alpha_0 dy_0^2 +\alpha_1 dy_1^2+\alpha_2 ds_{\Sigma_2}^2
 +\left( \alpha_1 \alpha_2^2 \right)^{\frac{1}{3}}
\beta_0 dr^2 + g_{ij}d\theta^i d\theta^j.
\label{olaqase}
\eeq
In this case we define,
\begin{equation}
\begin{aligned}
 & G_{ij}d\xi_i d\xi_j=\alpha_1 dy_1^2 + \alpha_2 ds_{\Sigma_2}^2
 + g_{ij}d\theta^i d\theta^j,  
\\[5pt]
 & \widehat{H}= \left( \int d\theta^i \sqrt{e^{-4\Phi} \det[G_{ij}] }  \right)^2, 
\\[5pt]
 & c=d^d\frac{\beta_0^{\frac{d}{2}} \widehat{H}^{\frac{2d+1}{2}}}{\pi G_N^{(10)}(\widehat{H}')^d}.
\end{aligned}
\label{centralanisotropic}
\end{equation}

We can apply this generalized definition to the flow for the twisted 
$H_2$ background of Section \ref{section211xx}, and Donos-Gauntlett background of Section \ref{sectionDG} (for more examples the reader is referred to appendix \ref{AppCentralC} ).
In this case, we consider them as dual to a field theory in $3+1$  
anisotropic dimensions (two of the dimensions are compactified on a $\Sigma_2$).
The quantities relevant for the calculation of the central charge are,
\begin{equation}
\begin{aligned}
& & d=3,\;\; \alpha_1 =L^2 e^{2A},\;\;\; \alpha_2=L^2 e^{2B} , \;\;\; \beta_0=e^{\frac{-2A-4B}{3}},\;\;\; e^{\Phi}=1,
\\[5pt]
& &G_{ij}d\xi_i d\xi_j=L^2 \left( e^{2A} dy_1^2 +e^{2B} ds^2_{\Sigma_2}+e^{2U}ds_{KE} ^2+e^{2V} \left( \eta + z A_1 \right)^2\right).
\end{aligned}
\end{equation}
We calculate 
\beq
\widehat{H}= {\cal N}^2 e^{2(2B+4U+V+A)},\;\;\; 
{\cal N}=\frac{(4\pi)^3  L^8}{108}.
\label{cccc}
\eeq
Then,  we obtain
\beq
c=\frac{ 27 {\cal N} e^{4U+V}}{8 \pi G_N^{(10)} (2B'+4U'+V'+A')^3}.
\label{c000}
\eeq
Focusing on the $H_2$ case, if we use the solution that 
describe this flow---see eq. \eqref{O19} we get an analytical expression,
\beq
c=\frac{\mathcal{N}}{\pi \ G_N^{(10)}} \left( \frac{1+e^{2r}}{1+2e^{2r}} \right)^3 \ .
\eeq
Notice that, by definition, this quantity gives the correct central charge 
in the UV (a constant, characterizing the 4-d fixed point). 
In the IR, the quantity turns out to be constant too, 
so it is capturing the presence of a 2-d fixed point. 
Nevertheless, it is probably not an appropriate candidate for 
a 'c-function between dimensions' 
as it is not necessarily increasing towards the UV. 
This is not in contradiction with the logic of 'a-theorems' and proofs like the
ones 
in \cite{Freedman:1999gp} or \cite{Komargodski:2011vj}, 
as the metric does not respect Lorentz invariance. 
Hence, it is not satisfying the assumptions of the theorems.
For the Donos-Gauntlett case analogous things happen. 
It would be very nice to try to apply the recent ideas of 
\cite{Gukov:2015qea}
to this flow. Notice that this feature of a 'wrong monotonicity' for the central
charge was also observed---for theories breaking Lorentz invariance in Higher Spin theories---see the papers \cite{Gutperle:2011kf}.

Let us move now to study other observables defining the 2-d and 4-d
QFTs.

\section{Entanglement entropy and Wilson loops}\label{sectionEEWilson}
In this section, we will complement the work done above,
by studying a couple of fundamental observables in the QFTs defined by 
the backgrounds in Part I of the paper.

Whilst at the conformal points the functional dependence of
results is determined by the symmetries, the interest will be 
in the coefficients accompanying the dependence. Both observables
interpolate smoothly between the fixed points.

\subsection{Entanglement entropy}
\label{EE}

\def\reg{\mathrm{reg}}

The aim of this section is to compute the 
entanglement entropy on a strip, which extends along the 
direction $y_1 \in [-\frac{d}{2},\frac{d}{2}]$, and study 
how this observable transforms under NATD. 
The input backgrounds for our calculations will be the Donos-Gauntlett and the $H_2$ flow as well as their non-Abelian T-duals. Since the procedure is the same in all of the cases and due to the similarity of the geometries we will present the results in a uniform way.

For the computation of the entanglement entropy one has to apply the 
Ryu-Takayanagi formula \cite{Ryu:2006ef} for non-conformal metrics \cite{Nishioka:2006gr}. This prescription states that the holographic entanglement entropy between the strip and its complement is given by the minimal-area static surface that hangs inside the bulk, and whose boundary coincides with the boundary of the strip. The general form of the entanglement entropy for the non-conformal case is,

\begin{equation}
 \label{prescription} S = \frac{1}{4G_N^{(10)} } \int d^8 \sigma e^{-2
\Phi} \sqrt{G^{(8)}_{\rm ind}}~.
\end{equation}

For the strip, we chose the embedding functions to be $y_0 = \textrm{const}$ and $r=r(y_1)$ and then using the conservation of the 
Hamiltonian  we arrive at an expression for $r(y_1)$ that makes the area minimal under that embedding. With that we compute,

\begin{equation}
\label{Ryu-Takayanagi}
 S= \frac{\tilde{L}^8}{54}\frac{\pi}{G^{(10)}_N} vol (\Sigma_2)V_3 \int\limits_{r_*}^{\infty} dr \ e^{-A} \frac{G^2}{\sqrt{G^2 - G^2_*}} \ ,
\end{equation}
where   the form of the function $G$ depends on the geometry of the background 
that we consider, 
\begin{equation}
 G = \left\{ \begin{array}{ll}
                   e^{A+2B} & \textrm{for the twisted solutions}
                   \\[5pt]
                   e^{A+2B+4U+V} & \textrm{for the DG solution}
                  \end{array}
        \right. \ .
\end{equation}
Above, $r_*$ is the radial position of the hanging surface tip and we define $G_*=G(r_*)$. Also, with $vol(\Sigma_2)$ we denote the volume of 
the Riemann surface $\Sigma_2$. Notice that the form of the 
function $G$ is the same before and after NATD. 
Moreover we consider,
\begin{equation}
\label{Radii}
 \tilde{L} = \left\{ \begin{array}{ll}
                     L                     & \textrm{before NATD}
                   \\[5pt]
                   \widehat{L}& \textrm{after NATD}
                  \end{array}
        \right. \ ,
\end{equation}
%
The quantity $V_3$ is defined as,
\begin{equation}
\label{3dimVol}
 V_3 = \left\{ \begin{array}{ll}
                       16 \ \pi^2 & \textrm{before NATD}
                       \\[5pt]
                       \int d \chi \ d \xi \ d \r \ \r^2 \sin\chi & \textrm{after NATD}
                      \end{array}
            \right. \ .
\end{equation}
Before the NATD transformation $V_3$ comes from the 3-dimensional 
submanifold that is spanned by the coordinates 
$(\th_2,\phi_2,\psi)$, while after the NATD it comes 
from the submanifold that is spanned by the dual 
coordinates $(\r,\chi,\xi)$. This 
implies that the entropies before and after NATD are proportional all along the flow, 
for any strip length. 
A discussion on possible values of the  quantity 
$\int d \chi \ d \xi \ d \r \ \r^2 \sin\chi$ 
can be found in the quantized charges 
Section \ref{chargessectionxx} and in the discussion 
on central charges in Section \ref{centralchargessectionxx}.

When computed by the Ryu-Takayanagi formula eq. \eqref{Ryu-Takayanagi} the entanglement entropy 
is UV divergent. In order to solve this we compute the 
regularized entanglement entropy ($S^{\reg}$) by subtracting 
the divergent part of the integrand of  eq. \eqref{Ryu-Takayanagi}. 
The regularized entanglement entropy is given by,
\begin{equation}
\label{RegEE}
 S^{\reg} = \frac{\tilde{L}^8}{54}\frac{\pi}{G^{(10)}_N} vol (\Sigma_2)V_3 \Bigg\{ \int\limits_{r_*}^{\infty} dr \   \Big( \frac{e^{-A} G^2}{\sqrt{G^2 - G^2_*}} - F_{UV}  \Big)    - \int\limits^{r_*} dr \ F_{UV}  \Bigg\} \ ,
\end{equation}
where the last integral is an indefinite integral with the result being evaluated at $r=r_*$ and 
\begin{equation}
F_{UV}= \left\{ \begin{array}{ll}
                  \frac{1+e^{2r}}{3} & \textrm{for the $H_2$ twisted solution}
                   \\[5pt]
                   e^{2r}+\frac{\lambda^2}{3}& \textrm{for the Donos-Gauntlett solution}
                  \end{array}
        \right. \ .
\end{equation}
From the formulas \eqref{Radii}, \eqref{3dimVol} and \eqref{RegEE} it is obvious that the regularized entanglement entropies before and after the NATD transformation differ by the factor
\begin{equation}
\label{EEquotient}
 \frac{\widehat{S}^{\reg}}{S^{\reg}} = \frac{\widehat{L}^8}{L^8}\frac{\int d \chi \ d \xi \ d \r \ \r^2 \sin\chi}{16 \ \pi^2} \ .
\end{equation}
In the formula above we denoted by $\widehat{S}^{\reg}$ the value of the
entanglement entropy after the NATD transformation. 
 As discussed below eq. (\ref{vava}), the quantity $\big(e^{-2\Phi}\sqrt{\det[g]}\big)$ is invariant under NATD, and this explains why the ratio (\ref{EEquotient}) is constant along the flow.

At this point let us normalize the regularized 
entanglement entropy by defining the quantity,
\begin{equation}
\label{SPrimeDef}
 S' = \frac{54}{\tilde{L}^8} \frac{G^{(10)}_N}{\pi \ vol(\Sigma_2) V_3} \ S^{\reg} \ .
\end{equation}
In what follows we present the behavior of $S'$ in the UV and the IR for the geometries of interest. We express the results in terms of the width of the strip $d$,
%
\begin{equation}
 d = 2 G_* \int\limits_{r_*}^{\infty} dr \ \frac{e^{-A}}{\sqrt{G^2 - G_{*}^2}} \ .
\end{equation}
The UV/IR behavior written in 
eqs \eqref{TwistedUVEE}, \eqref{DGUVEE}, \eqref{TwistedIREE} 
and \eqref{DGIREE} below, 
are just consequences of the fact that in far UV and far 
IR the dual QFT is conformal. 
The functional forms are universal, so our main interest 
is the constant appearing in them, 
and also as a cross-check of numerical results.

\subsubsection*{Behavior in the UV}

\underline{Twisted geometries}

In the case of the twisted $H_2$ geometry  we find that the width of the strip is,
\begin{equation}
 d = e^{-r_*} \int\limits_{1}^{\infty} \frac{du}{u^2} \frac{2}{\sqrt{u^6 -1}} = \frac{2 \sqrt{\pi} \Gamma\big(  \frac{2}{3} \big)}{\Gamma\big(  \frac{1}{6} \big)} \ e^{-r_*} \ .
\end{equation}
Here in the integration we changed the variable $r$ by $u = \frac{e^r}{e^{r_*}}$. From the calculation of the normalized entropy $S'$ we observe that in the UV this behaves like $\frac{1}{d^2}$, namely
\begin{equation}
\label{TwistedUVEE}
 S' = - \frac{\pi^{3/2}}{6} \Bigg(  \frac{\Gamma\big(  \frac{2}{3} \big)}{\Gamma\big(  \frac{1}{6} \big)}  \Bigg)^3 \frac{1}{d^2} + \frac{1}{3} \ln d + \frac{1}{3} \ln \Bigg(  \frac{\Gamma\big(  \frac{1}{6}  \big)}{2 \sqrt{\pi}} \Gamma\Big(  \frac{2}{3}  \Big)  \Bigg) \ ,
\end{equation}
where we also included subleading and next-to subleading terms.

\noindent \underline{Donos-Gauntlett geometry}

Similarly in the case of the Donos-Gauntlett geometry we find that the width of the strip in terms of $r_*$ (considering also a subheading term) is,
\begin{equation}
 d = \frac{2\sqrt{\pi} \Gamma\big(  \frac{2}{3} \big)}{\Gamma\big(  \frac{1}{6}  \big)} e^{-r_*} + \l^2 \big(  \frac{5}{72} + \frac{11}{24} I_1  \big) e^{-3 r_*} \ ,
\end{equation}
where
\begin{equation}
 I_1 = \int\limits_{1}^{\infty} dz \ \frac{z^2}{(z^4 + z^2 + 1)\sqrt{z^6-1}} = 0.1896 \ldots
\end{equation}
Here as well we end up with a $\frac{1}{d^2}$ behavior, a logarithmic subleading contribution and a constant $c_1$ for the regularized entropy,
\begin{equation}
\label{DGUVEE}
 S' = - 2 \pi^{3/2} \Bigg(  \frac{\Gamma\big(  \frac{2}{3} \big)}{\Gamma\big(  \frac{1}{6} \big)}  \Bigg)^3 \frac{1}{d^2} + \frac{\l^2}{3} \ln d + c_1 \ ,
\end{equation}
where $c_1$ ,
\begin{equation}
 c_1 = \lambda^{2}\left(-\frac{1}{3}\ln\left(\frac{2\sqrt{\pi}\Gamma\left(\frac{2}{3}\right)}{\Gamma\left(\frac{1}{6}\right)}\right)+\left(\frac{\ln2}{9}+\frac{11}{48}I_{1}\right)+\frac{\Gamma\left(-\frac{1}{3}\right)}{6\Gamma\left(\frac{2}{3}\right)}\left(\frac{5}{72}+\frac{11}{24}I_{1}\right)\right) \ .
\end{equation}

\subsubsection*{Behavior in the IR}

\underline{Twisted geometries}

The calculation for the IR limit is more tricky. The origin of the subtlety is that the integrals we have to evaluate now run all along the flow and we do not know the analytical properties of the integrands. In order to address this issue we split the integration into the intervals $[r_*,a]$ and $[a,+\infty)$ choosing $a$ to be in the deep IR but always greater than $r_*$. 
See \cite{Bea:2013jxa} for details of this procedure.

Following this prescription in the calculation, for the 
width of the strip we find,
\begin{equation}
 d = 
 \frac{4}{3} \ e^{-\frac{3 r_{*}}{2}}.
\end{equation}
%
If we do the same analysis when we calculate the normalized entropy we find,
\begin{equation}
\label{TwistedIREE}
 S' = \frac{2}{9} \ln d + \frac{2}{9} \ln \frac{3}{2} \ .
\end{equation}
%
%
The logarithmic dependence  of the leading term on $d$ is the expected for a $1+1$ theory.

\noindent \underline{Donos-Gauntlett geometry}

We close this section presenting the corresponding results for the Donos-Gauntlett geometry. As in the case of the twisted geometries we split the integrations in the same way. Then for the width of the strip we find,
    \begin{equation}
 d = e^{-a_0} \frac{2\sqrt{2}}{3^\frac{3}{4}} e^{-\frac{3^\frac{3}{4}}{\sqrt{2}}r_*} \ .
    \end{equation}
The normalized entropy displays again a logarithmic behavior in terms of the width of the strip,
\begin{equation}
\label{DGIREE}
 S' = \frac{8}{9} \ln d + \frac{8}{9} \ln \Big(  e^{a_0} \frac{3^{\frac{3}{4}}}{\sqrt{2}}  \Big) + c_2 \ ,
\end{equation}
where the constant $c_2$ has the value,
\begin{equation}
 c_2= \int_{0}^{\infty} \left( e^{-A} G-  e^{2r}-\frac{4}{3} \right)dr + \int_{-\infty}^{0} \left( e^{-A} G- e^{2r}-\left(\frac{4}{3}\right)^{\frac{5}{4}} \right)dr = -0.0312 \ldots
\end{equation}
We will now perform a similar analysis for Wilson loops.

\begin{figure}
\centering
\label{fig: S'} 
\begin{subfigure}[b]{0.49\textwidth}
\centering
\includegraphics[width=\textwidth]{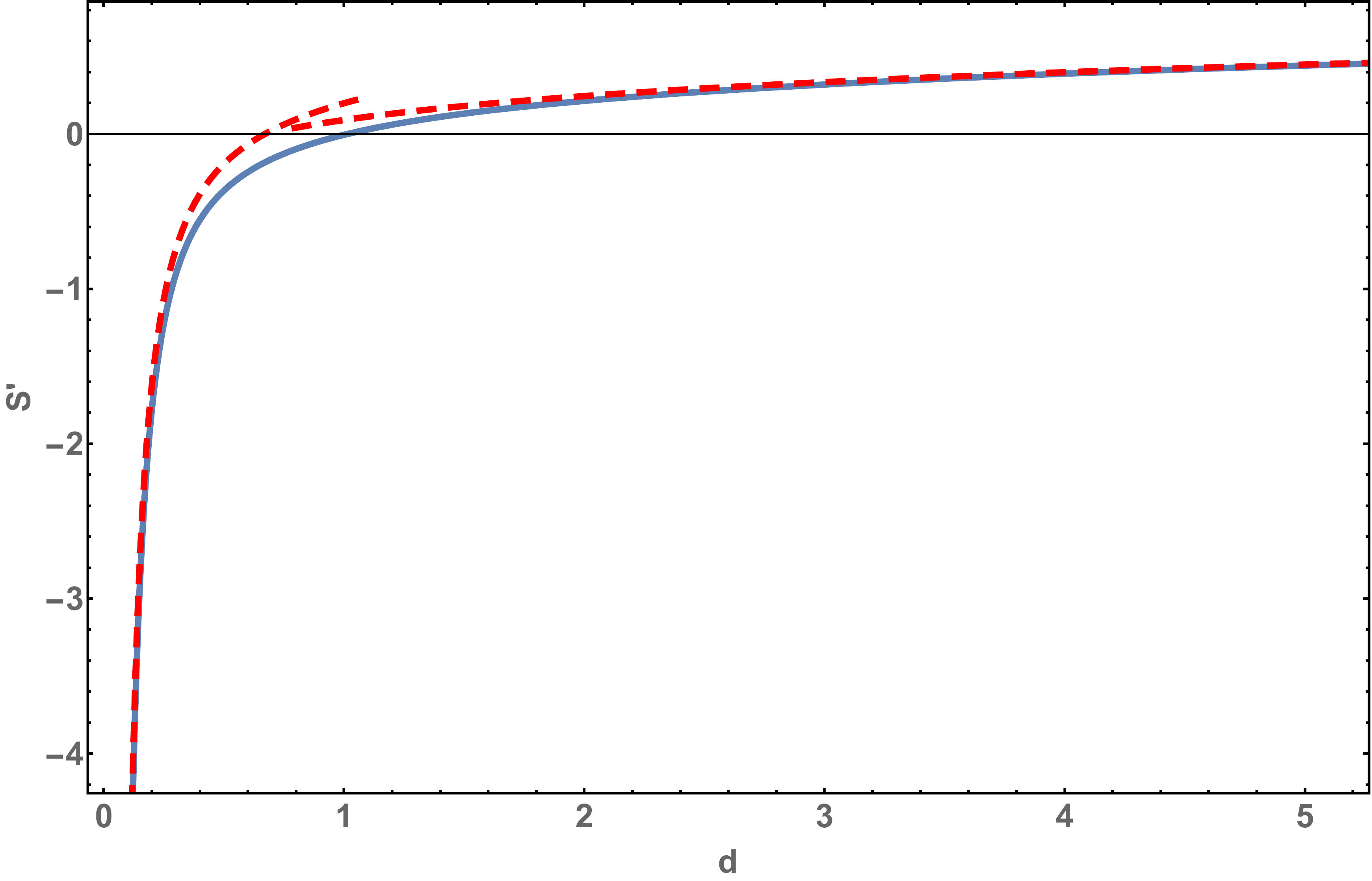}
\end{subfigure}
~
\begin{subfigure}[b]{0.49\textwidth}
\centering
\includegraphics[width=\textwidth]{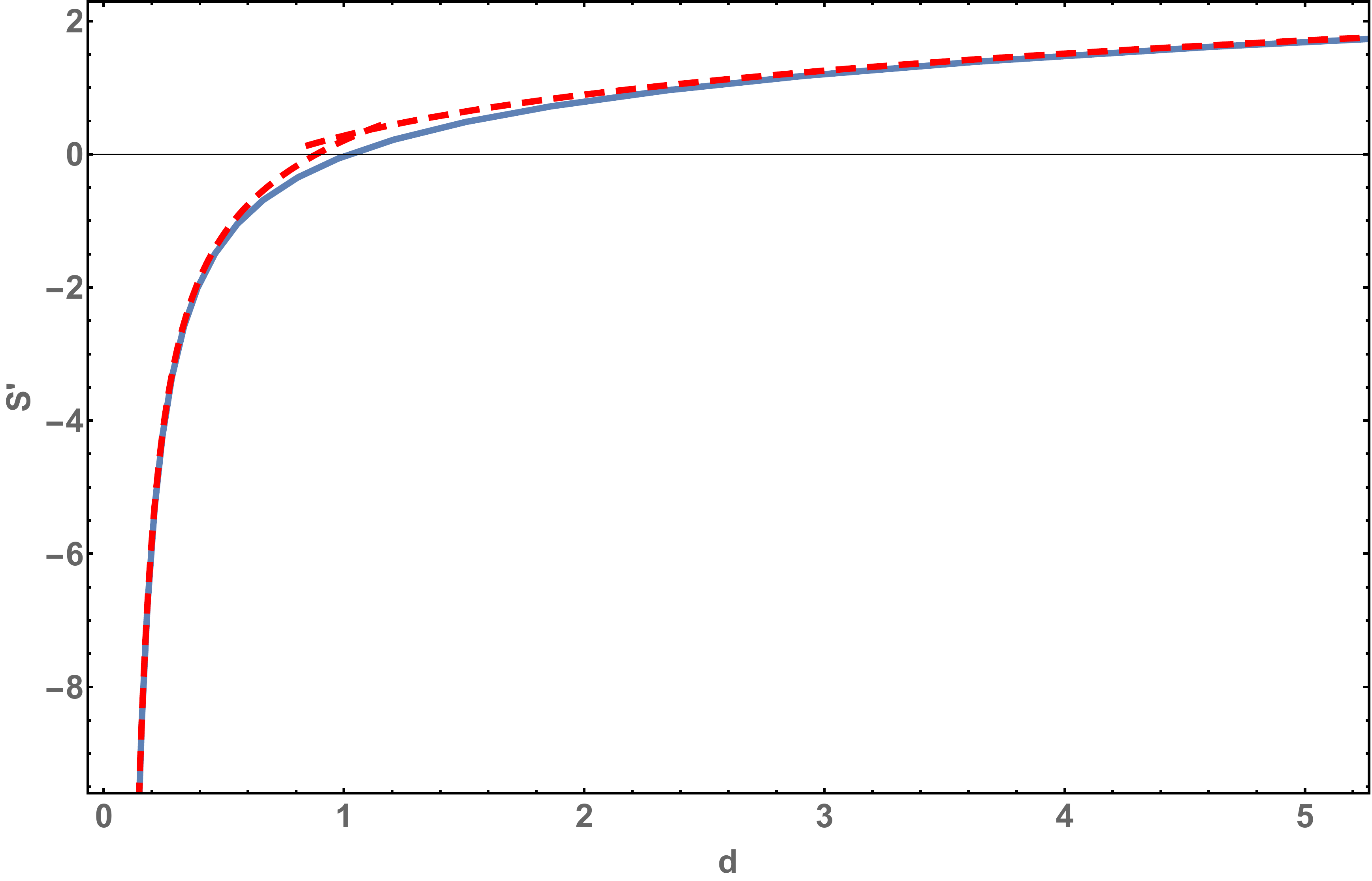}
\end{subfigure}
\caption{$S'$ as a function of $d$ for twisted $H_2$ (left) and Donos-Gauntlett solution (right).
The continuous curves correspond to the numerical value, while the dashed red ones to the UV and IR limits.
}
\end{figure}

\subsection{Wilson loop}

In this section we calculate the potential energy as a function 
of the separation in the $y_1$ direction (denoted as $d$)
for two non-dynamical sources added to the QFT \cite{Maldacena:1998im}. 
In holography this observable can be represented 
by a hanging string whose ends are separated by a 
distance $d$ along the $y_1$ direction. 
In our calculations we consider 
an embedding of the form $r = r(y_1)$ for the string. Such an embedding gives rise to the following induced metric on the string,
\begin{equation}
\label{InducedStringMetric}
 ds_{\textrm{st}}^2 = \tilde{L}^2 \Big(  e^{-2A} \ dy_0^2 + \big(  e^{2A} + r'^2  \big) \ dy_1^2  \Big) \ ,
\end{equation}
where $\tilde{L}$ is defined in \eqref{Radii}. It is obvious that the above induced metric is the same for 
all of the geometries that we have discussed in this paper so far (even for the duals). For this reason we believe that it is not necessary to make any distinction with respect to these geometries for the moment. Moreover, this means that the observable is 'uncharged' under NATD and thus it has the same functional form when computed in the initial and dualized geometries. The interest will be 
in the numerical coefficients our calculation will give.

The Nambu-Goto Lagrangian density for the string takes the form,
\begin{equation}
 \mathcal{L} = \frac{1}{2\pi\alpha'} \sqrt{-\det(g_{\textrm{ind}})} = \frac{\tilde{L}^2}{2\pi \alpha'} \ e^{A} \ \sqrt{e^{2A} + r'^2} \ ,
\end{equation}
where $g_{\textrm{ind}}$ stands for the induced metric \eqref{InducedStringMetric}. The conservation of the Hamiltonian implies that,
\begin{equation}
 \frac{e^{3A}}{\sqrt{e^{2A} + r'^2}} = e^{2 A_*} \ ,
\end{equation}
where $A_*$ is the value of the function $A(r)$ at the tip of the hanging string $r=r_*$. We can solve the last equation for $r'$ and use the result to calculate the distance between the endpoints of the string. If we do this we can express $d$ in terms of $r_*$
\begin{equation}
\label{QuarkDistance}
 d = e^{2 A_*} \int\limits_{r_*}^{\infty} dr \ \frac{e^{-A}}{\sqrt{e^{4A} - e^{4A_*}}} \ .
\end{equation}

The Nambu-Goto action now reads
\begin{equation}
\label{NambuGoto}
 S = \frac{T \tilde{L}^2}{\pi \alpha'} \int\limits_{r_*}^{\infty} dr \ \frac{e^{3A}}{\sqrt{e^{4A} - e^{4A_*}}} \ ,
\end{equation}
where $T = \int dt$. The integral in eq. \eqref{NambuGoto} is divergent since we are considering quarks of infinite mass sitting at the endpoints of the string. We can regularize  this integral by subtracting the mass of the two quarks and dividing by $T$ as it is shown below
\begin{equation}
\label{QuarkAntiquarkEnergy}
 \frac{E}{\tilde{L}^2} \alpha' = \frac{1}{\pi} \int\limits_{r_*}^{\infty} dr e^A \ \Big(  \frac{e^{2A}}{\sqrt{e^{4A} - e^{4A_*}}} - 1  \Big) - \frac{1}{\pi} \int\limits_{-\infty}^{r_*} dr \ e^{A} \ .
\end{equation}
This formula gives us the quark-antiquark energy. In order to calculate the same observable starting
with the NATD geometries one must take into account that the $\textrm{AdS}$ radius $L$ is different from that of the original geometries. In fact both results are related in the following way
\begin{equation}
\frac{\widehat{E}}{E} = \frac{\widehat{L}^2}{L^2} \ .
\end{equation}
In the last expression the hats refer to the dual quantities.

At this point we will explore the UV and IR limits of the quark-antiquark energy both for the twisted and the  Donos-Gauntlett geometries.

\subsubsection*{Behavior in the UV}

\underline{Twisted and Donos Gauntlett geometry}

First we focus on the twisted solution where the Riemann surface is the hyperbolic space, i.e. $\Sigma_2 = H_2$. In Section \ref{section211xx} we saw that in this case the function $A(r)$ behaves like $A(r) \sim r$. Taking this into account we can compute the distance between the quarks from the formula \eqref{QuarkDistance}. The result of this is
\begin{equation}
 d = \frac{2 \sqrt{2} \ \pi^\frac{3}{2}}{\Gamma \Big(  \frac{1}{4} \Big)^2} e^{-r_*} \ .
\end{equation}
Solving this equation for $r_*$ we can substitute into the result coming from the formula \eqref{QuarkAntiquarkEnergy}. This will give the quark-antiquark energy in terms of $d$ which in our case is
\begin{equation}
\label{QQb1}
 E = - \frac{\tilde{L}^2}{\alpha'} \ \frac{4 \ \pi^2}{ \Gamma  \Big(  \frac{1}{4} \Big)^4} \frac{1}{d} \ ,
\end{equation}
as expected for a CFT. The main point of interest in the previous formula is in the numerical coefficient.


Similar considerations for the case of the Donos-Gauntlett geometry give the same results as in the twisted case above. This is because the asymptotic behavior of the function $A(r)$ in the UV is the same in both cases.

\subsubsection*{Behavior in the IR}

\underline{Twisted geometry}

Again in the IR region we address again the same difficulty that we found in the computation of the entanglement entropy. We use the same trick to overcome it, that is we split the integrations into the intervals $[r_*,a]$ and $[a, +\infty)$ where $a$ has value in the deep IR but always greater than $r_*$. 

In Section \ref{section211xx} we saw that in the case where $\Sigma_2 = H_2$, the IR behavior of the function $A(r)$ is $A(r) \sim \frac{3}{2} \ r$. Applying this into the formula \eqref{QuarkDistance} we obtain the following result,
\begin{equation}
 d = \frac{4 \sqrt{2} \ \pi^\frac{3}{2}}{3 \ \Gamma \Big(  \frac{1}{4} \Big)^2} e^{-\frac{3}{2} r_*} \ .
\end{equation}
As before we solve the previous result for $r_*$ and we substitute it into the expression that we find from the calculation of the quark-antiquark potential. This way we express the energy as a function of the distance between the quarks,
\begin{equation}
\label{QQb2}
 E = - \frac{\tilde{L}^2}{\alpha'} \ \frac{16 \ \pi^2}{ 9 \ \Gamma \Big(\frac{1}{4} \Big)^4} \frac{1}{d} \ .
\end{equation}

\noindent \underline{Donos-Gauntlett geometry}

Repeating the same steps for the case of the Donos-Gauntlett geometry we find that the distance between the quarks is,
\begin{equation}
 d = \frac{4 \pi^\frac{3}{2}}{ 3^\frac{3}{2} \Gamma \Big( \frac{1}{4} \Big)^2} e^{-a_0 - \frac{3^{\frac{3}{4}}}{\sqrt{2}} r_*} \ .
\end{equation}
Then, expressing the energy in terms of the distance $d$ we find again a dependence proportional to $\frac{1}{d}$,
\begin{equation}
\label{QQb3}
 E = - \frac{\tilde{L}^2}{\alpha'} \frac{8 \pi^2}{3^{\frac{3}{2}} \Gamma \Big(  \frac{1}{4}  \Big)^4} \frac{1}{d} \ .
\end{equation}

Let us point out that the behavior in eqs \eqref{QQb1}, \eqref{QQb2} and \eqref{QQb3} are just consequences of the fact that far in the UV and far in the IR the QFT is conformal.

\begin{figure}[h]
\label{fig: WL joint}
\begin{subfigure}[h]{0.49\textwidth}
\centering
\includegraphics[width=\textwidth]{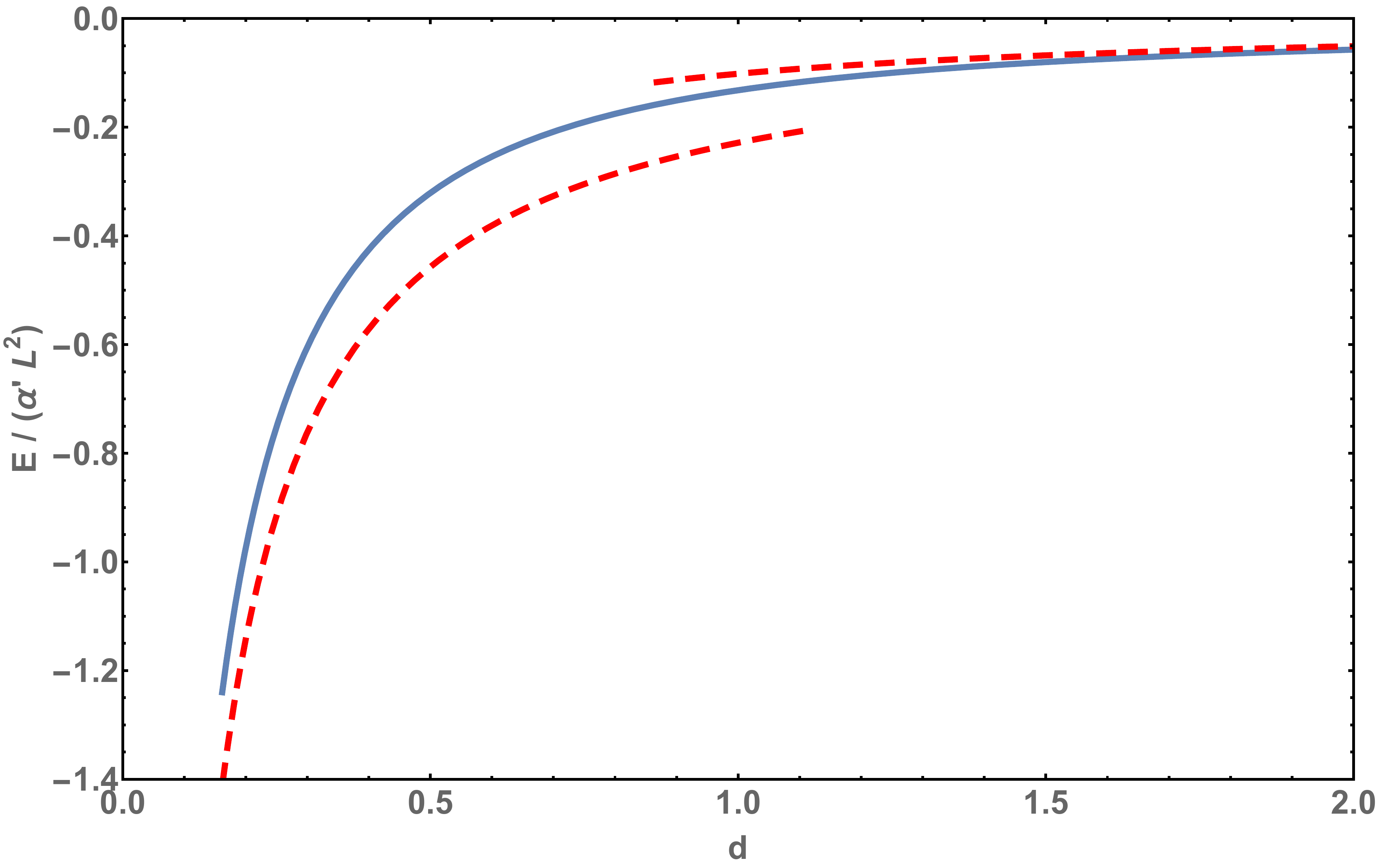}
\end{subfigure}
\begin{subfigure}[h]{0.49\textwidth}
\centering
\includegraphics[width=\textwidth]{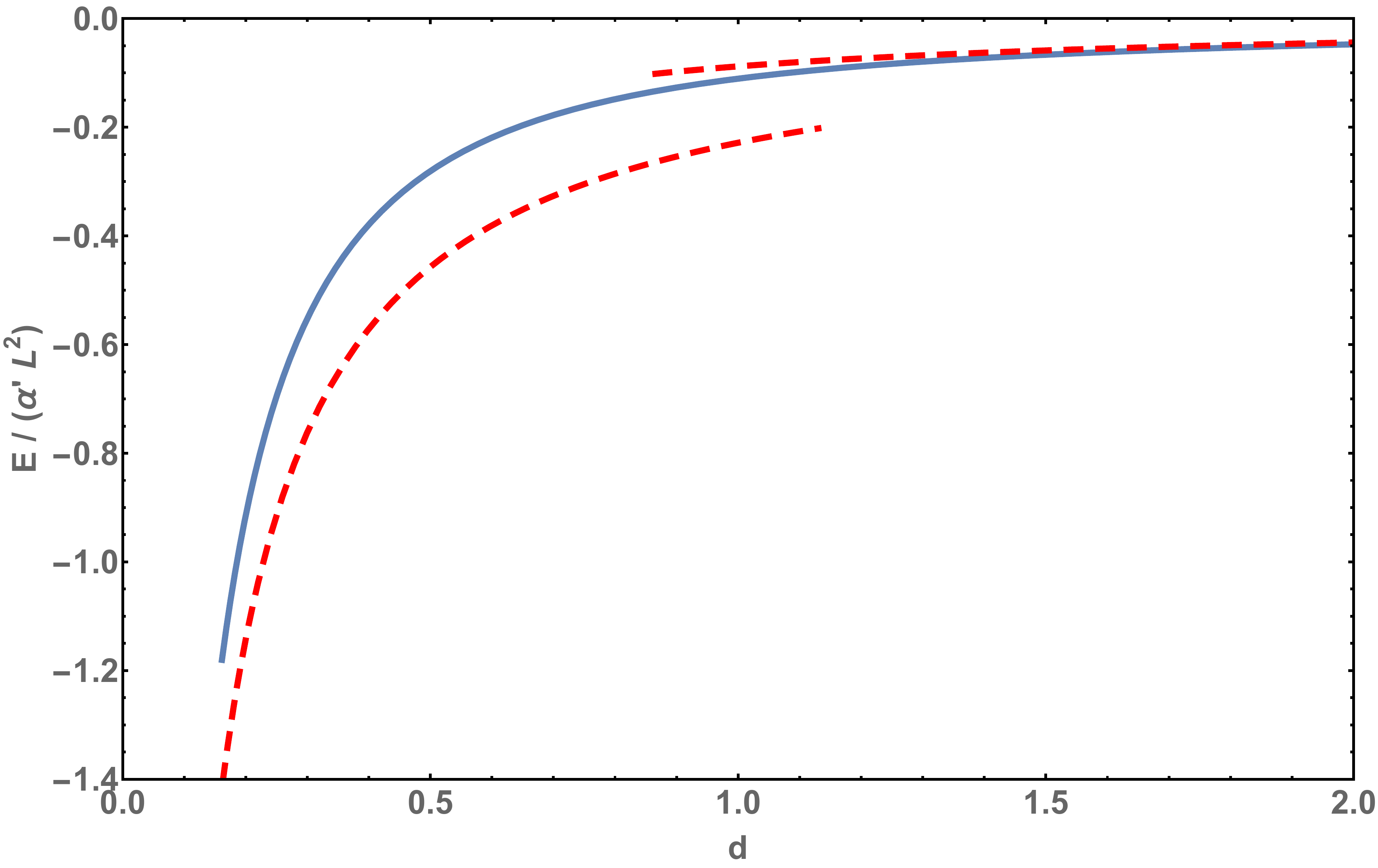}
\end{subfigure}

\caption{The quark-antiquark potential $\frac{E}{\alpha' L^{2}}$ as a function of the distance $d$ in the cases of the twisted $H_2$ (left) and Donos-Gauntlett (right) solutions. The continuous curves correspond to the numerical results and the dashed ones to the UV and IR limits.}
\end{figure}
%
%
%
%
%
%
%
%

\section{Conclusions and future directions}\label{conclusiones}

Let us start with a  brief summary. We studied backgrounds dual to 
two-dimensional SUSY CFTs. The 2-d CFTs were obtained
by compactification of the four-dimensional Klebanov-Witten CFT
on a torus or on a compact hyperbolic plane. The 2-d CFT preserves (0,2)
SUSY.

On those Type IIB backgrounds we performed a NATD transformation,
using an $SU(2)$-isometry of the 'internal space' (or conversely, 
a global symmetry of the dual CFT). As a result, we constructed 
{\it new, smooth} and SUSY preserving backgrounds in Type IIA and M-theory
with an $AdS_3$ fixed point. A further T-duality was used to construct new, smooth and SUSY
Type IIB background whose IR is of the form $AdS_3\times M_7$ and all fluxes are
active. Moreover, the uplift to 11 dimensions of the new Type IIA solutions is presented.

We analyzed the dual QFT by computing its observables, using
the smooth backgrounds mentioned above. By studying the Page charges, we observed that there is a correspondence between the branes of the starting Type IIB solution and those of the Type IIA solution after NATD.

The behavior of the central charge in the original CFT 
(the compactification of the Klebanov-Witten theory to 2-d) 
is $c\sim N_{D3}^2$, while
after the NATD goes like $c\sim N_{D6}^2 n^3$. This new (cubic)
dependence suggest a relation with long linear quivers, 
which would imply that $n$ is measuring the number of  D4 and $NS_{5}$ branes.
The picture that emerges is that of a 2-d CFT living on the intersection
of D2 , D6 and $NS_5$ branes, with induced D4 charge every time 
an NS-brane is crossed. Quantized charges support this interpretation. 

Entanglement entropy and Wilson loops
had the expected universal dependence on $d$ (the quark-antiquark separation or
the length of  strip separating the two regions) at the fixed points. The interest of the expressions
is on the coefficients, not determined by conformal invariance. Interestingly, along the flow the observables smoothly interpolate between the IR and the UV behaviors, which are fixed points of different dimensionality. Both for the entanglement entropy and the Wilson loops we found that the quotient of their values before and after NATD is constant along the flow, as it is expected.

In the future it would be interesting to:
\begin{itemize}
\item{Connect our study with previous calculations done for 
$AdS_3$, either at the sigma model or the supergravity level.
We are presenting new backgrounds, hence new 2-d CFTs on which 
studies done in the past could be interesting to revisit.}
\item{Make more precise the QFT dual to our backgrounds. Apply 
the findings in this paper to other examples, perhaps more symmetric.}
\item{Extend the classification of $AdS_3$ backgrounds 
in Type II String Theory
or M-theory. Our backgrounds suggest more general metrics 
for the internal space. Check whether our solutions fall within 
the existent classifications of, for example \cite{Gauntlett:2006ux},  \cite{Beck:2015hpa}.}
\item{Find new observables that select (or explore) the values of $\rho=n\pi$
argued in this work to be special values of the $\rho$-coordinate. 
Explore further the prescription based on the periodicity of $b_0$.}
\item{Explore the definition for a c-function in the anisotropic case.
Is our 
definition of any use? Can a definition for anisotropic backgrounds or flows between dimensions be engineered by using the ideas in \cite{Myers:2012ed}?}
\item{It would be interesting to apply the c-extremization formalism of \cite{Benini:2013cda} to our examples with less SUSY. In the same vein,
the formalism of \cite{Karndumri:2013iqa} seems suited for our
examples. Indeed, a reduction of our IIB and IIA backgrounds in
the sense of \cite{Karndumri:2013dca} seems possible. This 3d-supergravity may be
the suitable arena \cite{Colgain:2015ela} to study c-extremization in our cases. One may even speculate with the same idea applied to our non-SUSY examples.}
\item{Find other observables in the CFTs that can be 
studied in our new $AdS_3$ backgrounds. 
Study the fate of the known observables under  the NATD. 
Are those operators charged or uncharged under the 
global symmetries used to dualize?}
\item{ In the previous literature \cite{Bobev:2014jva}, flows  to $(0,2)$ SCFTs have been found, starting  from the  Leigh-Strassler fixed point ---something quite similar in spirit to our flows. Also, flows from $N=4$ SYM and the $(0,2)$ SCFT in six dimensions to 2d SCFTs with (0,2) have been discussed in \cite{Benini:2013cda}. Finally, the paper \cite{Kutasov:2013ffl} discusses flows from other $N=1$ QFTs to $(0,2)$ theories triggered by the presence of a magnetic field on a torus (similar to the Donos-Gauntlett flow and its NATD). It would be very interesting to relate these  different flows. Our NATD geometries bring a new element, incorporating linear quivers. }
\item{I would be of interest to break conformality, starting from our $AdS_3$ backgrounds and flowing down to a smooth space. This would mimic a (0,2) confining theory. Many exact methods to compute observables exist---see for example  \cite{Gadde:2014ppa}. It would be instructive to compare them with holographic computations. }
\end{itemize}
These, and various others, appear as directions worth
exploring at the moment of writing. We hope that readers appreciate the
interest of this line of research and contribute with their ideas.

\section*{Acknowledgments}
We would like to mention discussions with various colleagues that helped improving the presentation of this work. We thank
Thiago Araujo, Justin David, Jerome Gauntlett, Gaston Giribet, Amihay Hanany, Prem Kumar, Yolanda Lozano, Niall T. Macpherson, Horatiu Nastase, Alfonso V. Ramallo, Diego Rodriguez-Gomez, Alexandre Serantes, Konstantinos Sfetsos, Aninda Sinha and Daniel Thompson for these conversations, ideas and discussions. Y. B., G. I., J. E. and J. A. S. G. are funded by the Spanish grant FPA2011-22594, by the Consolider-Ingenio 2010 Programme CPAN (CSD2007-00042), by Xunta de Galicia (Conselleria de Educaci\'on and grant INCITE09-206-121-PR ), and by FEDER. Y. B. is also supported by the foundation Pedro Barri\'e de la Maza and the Spanish FPU fellowship FPU12/00481. J. A. S. G. acknowledges also support from the Spanish FPI fellowship from FEDER grant FPA2011-22594. C. N. is Wolfson Fellow of the Royal Society. K. K. is being funded with an STFC studentship. The research of D. S. is implemented under the ARISTEIA action of the operational programme education and lifelong learning and is co-funded by the European Social Fund (ESF) and National Resources. Y. B. and J. A. S. G. want to thank the University of Swansea for hospitality. Y. B. also wants to thank the University of S\~ao Paulo for hospitality at the final stages of this work.

\appendix

\section{SUSY analysis}\label{appendixsusy}

\subsection*{SUSY preserved by the twisted solutions}\label{susyvar}
In this appendix we write explicitly the variations of the dilatino and gravitino for the ansatz (2.1-2.4), for the 3 cases $H_2$, $S^2$ and $T^2$.  The SUSY transformations for the dilatino $\lambda$ and the gravitino $\psi_m$ for Type IIB SUGRA in string frame are \cite{3},
%
\bea
 & & \delta_{\epsilon}\lambda = \left[ \frac{1}{2}\Gamma^m \partial_m \Phi + \frac{1}{4\cdot3!}H_{mnp}\Gamma^{mnp} \tau_3 - \frac{e^{\Phi} }{2}  F_m \Gamma^m(i\tau_2) -\frac{e^{\Phi}}{4\cdot3!} F_{mnp}\Gamma^{mnp} \tau_1 \right] \epsilon \ , 
\\[5pt]
 & & \delta_{\epsilon}\psi_m =\left[  \nabla_m+ \frac{1}{4\cdot2!}H_{mnp}\Gamma^{np} \tau_3 + \frac{e^{\Phi}}{8} \left( F_n \Gamma^n (i\tau_2)+\frac{1}{3!}F_{npq}\Gamma^{npq} \tau_1 + \frac{1}{2\cdot5!}F_{npqrt}\Gamma^{npqrt} (i\tau_2)  \right) \Gamma_m \right] \epsilon \ ,
 \nonumber
 \label{susy00}
\eea
%
where $\tau_i \ , \; i = 1,2,3$, are the Pauli matrices. Let us consider the $H_2$ case in detail (the $S^2$ case is obtained analogously) . Recall that the vielbein is written in (\ref{vielbein00}).

The dilatino variation vanishes identically, as the fields involved are vanishing. The $m=0$ component of the gravitino reads,
\begin{equation}
\delta_{\epsilon} \psi_0=\left[  \frac{A'}{2L} \Gamma_{04} - \frac{e^{-4U-V}}{2L} \Gamma_{04} \Gamma_{0123} i \tau_2 + \frac{e^{-2B-2U-V}}{16L} z \big(\Gamma_{014} - \Gamma_{239} \big) \big(   \Gamma_{78} - \Gamma_{56} \big) \Gamma_0 i\tau_2 \right] \epsilon \ .
\label{gravitino0}
\end{equation}
First, we use the chiral projection of Type IIB,
\begin{equation}
\Gamma_{11}\epsilon=\epsilon \ ,
\label{gravitino00}
\end{equation}
where we define $\Gamma_{11}=\Gamma_{0123456789}$. We also impose the following projections (K\"ahler projections),
\begin{equation}
\Gamma_{56}\epsilon= - \Gamma_{78}\epsilon= - \Gamma_{49}\epsilon \ .
\label{gravitino01}
\end{equation}
Then, expression (\ref{gravitino0}) simplifies to,
\begin{equation}
\delta_{\epsilon} \psi_0= \Gamma_{04}\left[ \frac{A'}{2L}  - \frac{e^{-V-4U}}{2L}+\frac{e^{-2B-2U-V}}{4L}z \Gamma_{0178} i \tau_2 \right]\epsilon \ .
\label{gravitino02}
\end{equation}
We now impose the usual projection for the D3-brane,
\begin{equation}
\Gamma_{0123} \ i \tau_2 \epsilon= \epsilon \ ,
\label{gravitino03}
\end{equation}
and also a further projection related to the presence of the twisting,
\begin{equation}
\Gamma_{23}\epsilon= \Gamma_{78}\epsilon \ .
\label{gravitino04}
\end{equation}
Then, imposing that expression (\ref{gravitino02}) vanishes we obtain,
\begin{equation}
A'-e^{-V-4U}+\frac{z}{2} e^{-2B-2U-V}=0 \ .
\label{gravitino05}
\end{equation}
For the component $m=1$ of the gravitino equation, we obtain that it is zero when we impose the projections  and equation (\ref{gravitino05}). For the component  $m=2$ we have,
\begin{equation}
\delta_{\epsilon} \psi_2=\left[  \frac{B'}{2L} \Gamma_{24} +\frac{e^{-2B+V}}{4L} z \Gamma_{39} - \frac{e^{-V-4U}}{2L} \Gamma_{24}\Gamma_{0123} i \tau_2 - \frac{ e^{-2B-2U-V}}{4L} z \Gamma_{24} \Gamma_{0178} i\tau_2 \right] \epsilon \ .
\label{gravitino06}
\end{equation}
 Combining projections (\ref{gravitino01}) and (\ref{gravitino04}) we get  $\Gamma_{39} \epsilon=-\Gamma_{24} \epsilon$. Then, (\ref{gravitino06}) gives the condition,
 \begin{equation}
B'-e^{-V-4U}- \frac{z}{2}  e^{-2B-2U-V} -\frac{z}{2} e^{-2B+V}=0 \ .
\label{gravitino07}
\end{equation}
For $m=3$, after imposing the projections and equation \eqref{gravitino07} we arrive at,
\begin{equation}
\delta_{\epsilon} \psi_3= - \frac{e^{-B}}{2L} \cot \alpha \ (1+3z) \Gamma_{23} \epsilon \ .
\label{gravitino08}
\end{equation}
There are two contributions to this term, one coming from the curvature of the $H_2$ (through the spin connection) and another coming from the twisting $A_1$. That is, here we explicitly see that the twisting is introduced to compensate the presence of the curvature, in such  a way that some SUSY can be still preserved. Then, we impose,
\begin{equation}
z=-\frac{1}{3} \ .
\label{gravitino09}
\end{equation}
For $m=4$, the variation is,
\begin{equation}
\delta_{\epsilon} \psi_4= \frac{1}{L}\partial_r \epsilon - \frac{1}{2L} \left[ e^{-4U} + 2e^{-2B -2U-V} \right]  \epsilon \ .
\end{equation}
From the condition $\delta_{\epsilon} \psi_4=0$, we obtain a differential equation for $\epsilon$. Solving for it we arrive at the following form for the Killing spinor,
\begin{equation}
\epsilon= e^{1/2 \int (e^{-4U} + 2e^{-2B -2U-V} )  dr} \ \epsilon_0 \ ,
\end{equation}
where $\epsilon_0$ is spinor which is independent of the coordinate $r$. For $m=5,6,7,8$ the variations vanish as long as ,
\begin{equation}
U'+e^{-V-4U}-e^{V-2U}=0 \ .
\label{gravitino011}
\end{equation}
 Finally, for m=9 the graviton variation vanishes if,
\begin{equation}
V'-3e^{-V}+2e^{V-2U}+e^{-V-4U}-\frac{z}{2} e^{-2B-2U-V} +\frac{z}{2}e^{-2B+V}=0 \ .
\label{gravitino012}
\end{equation}
Summarizing, the variations of the dilatino and gravitino vanish if we impose the next projections on the Killing spinor,
\begin{equation}
\Gamma_{11} \epsilon = \epsilon \ , ~~~~~~ \Gamma_{56} \epsilon = - \Gamma_{78} \epsilon= - \Gamma_{49} \epsilon \ ,  ~~~~~~ \Gamma_{0123} \ i \tau_2 \epsilon = \epsilon \ , ~~~~~~  \Gamma_{23} \epsilon =\Gamma_{49} \epsilon \ ,
\label{gravitino013}
\end{equation}
and the BPS equations (\ref{O17}), together with the condition $z=-1/3$.
%
%
For the case of the 2-torus, if we focus on the $m=3$ component,
\begin{equation}
\delta_{\epsilon} \psi_3= \left[   -\frac{z e^{-2B+V}}{4L}\Gamma_{29}  -\frac{3 z e^{-B}}{2L} \ \alpha \ \Gamma_{78} + \frac{B'}{2L} \Gamma_{34}  - \frac{e^{-V-4U}}{2L} \Gamma_{34} - \frac{z e^{-2B-2U-V}}{4L} \Gamma_{34} \right] \epsilon \ ,
\label{gravitino014}
\end{equation}
we see that there is one term depending on $\alpha$, due to the twisting. Contrary to the $H_2$ and $S^2$ cases, here there is no curvature term that could cancel it. This will force $z=0$, obtaining $A'=B'$, which does not permit an $AdS_3$ solution. 

Finally, after all this analysis we deduce that the Killing spinor does not depend explicitly on the coordinates $(\theta_2,\phi_2,\psi)$ on which we perform the NATD transformation. In fact it only has a dependence on the coordinate $r$.

\subsection*{SUSY preserved by the NATD solutions}

In the above subsection we calculated the amount of SUSY that is preserved by the Type IIB supergravity solutions of the Section \ref{section211xx} by examining the dilatino and the gravitino variations. Here we compute the portion of SUSY that is preserved by a supergravity solution after a NATD transformation following the argument of \cite{Sfetsos:2010uq}, which has been proven in \cite{Kelekci:2014ima}. According to this one just has to check the vanishing of the Lie-Lorentz (or Kosmann) derivative \cite{Kosmann} of the Killing spinor along the Killing vector that generates the isometry of the NATD transformation. More concretely, suppose that we want to transform a supergravity solution by performing a NATD transformation with respect to some isometry of the background that is generated by the Killing vector $k^\m$. Then there is a simple criterion which states that if the Lie-Lorentz derivative of the Killing spinor along $k^\m$ vanishes, then the transformed solution preserves the same amount of SUSY as the original solution. In the opposite scenario one has to impose more projection conditions on the Killing spinor in order to make the Lie-Lorentz derivative vanish. Thus in that case the dual background preserves less supersymmetry than the original one.

We recall that given a Killing vector $k^\m$ the Lie-Lorentz derivative on a spinor $\epsilon$ along $k^\m$ maps the spinor $\epsilon$ to an other spinor and is defined as,
\begin{equation}
\label{KosmannDer}
 \mathcal{L}_{k} \epsilon =  k^\m D_\m \epsilon + \frac{1}{4} \big( \nabla_\m k_\n \big) \Gamma^{\m\n} \epsilon =        k^\m D_\m \epsilon + \frac{1}{8} (dk)_{\m\n} \Gamma^{\m\n} \epsilon \ ,
\end{equation}
where $D_\m \epsilon = \partial_\m \epsilon + \frac{1}{4} \omega_{\m\r\s} \Gamma^{\r\s} \epsilon$. For further details about the Lie-Lorentz derivative we urge the interested reader to consult \cite{Ortin:2002qb}.  

In this paper we constructed new Type IIA supergravity solutions by applying a NATD transformation with respect to the $SU(2)$ isometry of the original backgrounds that corresponds to the directions $(\th_2,\phi_2,\psi)$. The non-vanishing components of the associated Killing vectors are,
\begin{equation}
\label{KillingV}
 \begin{array}{lll}
   k_{(1)}^{\th_2} = \sin\phi_2 \ , & k_{(1)}^{\phi_2} = \cot\th_2 \cos\phi_2 \ , & k_{(1)}^{\psi} = - \frac{\cos\phi_2}{\sin\th_2} \ ,
   \\[10pt]
   k_{(2)}^{\th_2} = \cos\phi_2 \ , & k_{(2)}^{\phi_2} = -\cot\th_2 \sin\phi_2 \ , & k_{(2)}^{\psi} = \frac{\sin\phi_2}{\sin\th_2} \ ,
   \\[10pt]
   k_{(3)}^{\phi_2} = 1 \ .  & \textrm{}   & \textrm{}
 \end{array}
\end{equation}
In what follows we will compute the Lie-Lorentz derivative along the three Killing vectors $(k_{(1)},k_{(2)},k_{(3)})$ using the geometries of the Sections \ref{section211xx} and \ref{sectionDG}. It turns out that in all cases the Lie-Lorentz derivative vanishes without the requirement of imposing further projections on the Killing spinor. This means that the new solutions that we found using the technique of NATD preserve the same SUSY as the original solutions.

\subsubsection*{The NATD of the twisted solutions}

Let us now compute the Lie-Lorentz derivative along the Killing vector \eqref{KillingV} for the twisted geometries that are described by the formulas \eqref{NN02}-\eqref{vielbein00}. In the previous section, which deals with the supersymmetry of the starting solutions, we mentioned that the Killing spinor does not depend on the isometry coordinates $(\th_2,\phi_2,\psi)$. This means that the first term in \eqref{KosmannDer} reduces to,
\begin{equation}
 k_{(i)}^\m D_\m \epsilon = \frac{1}{4} \ \omega_{\m\r\s} \ k_{(i)}^\m \ \Gamma^{\r\s} \epsilon \ , \quad i=1,2,3 \ .
\end{equation}
Hence for each of the three Killing vectors we find,
\begin{equation}
 \label{KosmannTwisted1}
 \begin{aligned}
  &  k_{(1)}^\m D_\m \epsilon = \frac{z}{12} \ e^{2V-2B} \ \cos\phi_2 \sin\theta_2 \Gamma^{23} \epsilon - \frac{e^{2V-2U}}{6} \ \cos\phi_2 \sin\theta_2 \big(\Gamma^{56} - \Gamma'^{78} \big)  \epsilon
  \\[5pt] 
  & \qquad \quad - \frac{1}{2} \ \cos\phi_2 \sin\theta_2 \Gamma'^{78} \epsilon + \frac{e^{V-U}}{2 \sqrt{6}} \ \big(\cos\th_2 \cos\phi_2 \Gamma'^8 - \sin\phi_2 \Gamma'^7 \big) \Gamma^9 \epsilon 
  \\[5pt]
  & \qquad \quad + \frac{e^U \ U'}{2 \sqrt{6}} \big( \cos\theta_2 \cos\phi_2 \Gamma'^7 + \sin\phi_2 \Gamma'^8 \big) \Gamma^4 \epsilon + \frac{e^V \ V'}{6} \sin\theta_2 \cos\phi_2 \Gamma^{49} \epsilon \ ,
  \\[5pt]
 &  k_{(2)}^\m D_\m \epsilon = -\frac{z}{12} \ e^{2V-2B} \ \sin\phi_2 \sin\theta_2 \Gamma^{23} \epsilon + \frac{e^{2V-2U}}{6} \ \sin\phi_2 \sin\theta_2 \big( \Gamma^{56} - \Gamma'^{78} \big) \epsilon
  \\[5pt] 
  & \qquad \quad + \frac{1}{2} \ \sin\phi_2 \sin\theta_2 \Gamma'^{78} \epsilon - \frac{e^{V-U}}{2 \sqrt{6}} \ \big( \cos\th_2 \sin\phi_2 \Gamma'^8 + \cos\phi_2 \Gamma'^7 \big) \Gamma^9 \epsilon 
  \\[5pt]
  & \qquad \quad - \frac{e^U \ U'}{2 \sqrt{6}} \big( \cos\theta_2 \sin\phi_2 \Gamma'^7 - \cos\phi_2 \Gamma'^8 \big) \Gamma^4 \epsilon - \frac{e^V \ V'}{6} \sin\theta_2 \sin\phi_2 \Gamma^{49} \epsilon \ ,
  \\[5pt]
 &  k_{(3)}^\m D_\m \epsilon = -\frac{z}{12} \ e^{2V-2B} \ \cos\theta_2 \Gamma^{23} \epsilon + \frac{e^{2V-2U}}{6} \ \cos\theta_2 \big( \Gamma^{56} - \Gamma'^{78} \big) \epsilon + \frac{1}{2} \ \cos\theta_2 \Gamma'^{78} \epsilon
  \\[5pt] 
  & \qquad \quad + \frac{e^{V-U}}{2 \sqrt{6}} \ \sin\th_2 \Gamma'^8 \Gamma^9 \epsilon - \frac{e^U \ U'}{2 \sqrt{6}} \sin\theta_2 \Gamma^4 \Gamma'^7 \epsilon  - \frac{e^V \ V'}{6} \cos\theta_2 \Gamma^{49} \epsilon \ .
 \end{aligned} 
\end{equation}
For convenience we have defined the rotated $\Gamma$-matrices,
\begin{equation}
\label{GammaRot}
 \Gamma'^7 = \cos\psi \ \Gamma^7 + \sin\psi \ \Gamma^8 \ , \quad \Gamma'^8 = -\sin\psi \ \Gamma^7 + \cos\psi \ \Gamma^8 \ .
\end{equation}
Let us now compute the 1-forms that are dual to the Killing vectors. What one has to do is to lower the index of the Killing vectors \eqref{KillingV} using the metric \eqref{NN02} which gives the following result,
\begin{equation}
\label{KosmannTwisted2}
 \begin{aligned}
  & k_{(1)} = - \frac{L^2}{3} e^{2V} \sin\theta_2 \cos\phi_2 \ \big(\eta + z A_1\big) + \frac{L^2}{6} e^{2U} \big(  \sin\phi_2 d\theta_2 + \sin\theta_2 \cos\theta_2 \cos\phi_2 d\phi_2  \big) \ ,
  \\[5pt]
  & k_{(2)} = \frac{L^2}{3} e^{2V} \sin\theta_2 \sin\phi_2 \ \big(\eta + z A_1\big) + \frac{L^2}{6} e^{2U} \big(  \cos\phi_2 d\theta_2 - \sin\theta_2 \cos\theta_2 \sin\phi_2 d\phi_2  \big) \ ,
  \\[5pt]
  & k_{(3)} = \frac{L^2}{3} e^{2V} \cos\theta_2 \ \big(\eta + z A_1\big) + \frac{L^2}{6} e^{2U} \sin^2\theta_2 d\phi_2 \ .
 \end{aligned}
\end{equation}
The second term of \eqref{KosmannDer} can be computed by acting with the exterior derivative on the above 1-forms and contracting the result with $\Gamma$-matrices. Notice that in order to compare with \eqref{KosmannTwisted1} one has to express the components of $dk_{(i)}, \,\, i=1,2,3$ using the flat frame \eqref{vielbein00}. Finally for the second term of \eqref{KosmannDer} we find,
\begin{equation}
\label{KosmannCheck}
  \frac{1}{8} (dk_{(i)})_{\m\n} \Gamma^{\m\n} \epsilon = - k_{(i)}^\m D_\m \epsilon, \,\, i=1,2,3,
\end{equation}
which means that the Lie-Lorentz derivative along the Killing vectors $k_{(i)}, \,\, i=1,2,3$ vanishes.

\subsubsection*{The NATD the Donos-Gauntlett solution}

In the case of the Donos-Gauntlett geometry \eqref{metric-bef} we notice that all the necessary expressions are quite similar to those computed in the previous subsection. This is because the only significant difference between the line element of the twisted geometries and that of the Donos-Gauntlett geometry is just a fiber term. As in the example of the previous subsection, the Killing spinor does not depend on the isometry coordinates $(\theta_2,\phi_2,\psi)$. This implies that the derivative term $k^\m \partial_\m \epsilon$ in eq. \eqref{KosmannDer} has no contribution to the result. Then the first term of eq. \eqref{KosmannDer}, for each of the three Killing vectors, can be easily obtained from eq. \eqref{KosmannTwisted1} by setting $z=0$. Similarly, if we set $z=0$ into eq. \eqref{KosmannTwisted2} we find the 1-forms $k_{(1)}, k_{(2)}, k_{(3)}$ for the Donos-Gauntlett case. Once we know these 1-forms we can follow the same prescription as in the previous subsection and compute the second term of \eqref{KosmannDer} for each Killing vector. It happens again that this term, when computed for every Killing vector, is related to the first term by a minus sign and thus the Lie-Lorentz derivative vanishes without imposing further projections on the Killing spinor.

\section{Details on the Donos-Gauntlett solution}
\label{appendixDG}
Here, we describe the numerical flow of Donos-Gauntlett. We also present an analytic approximation to the numerical solution. This complements the study in Section \ref{sectionDG}.

We start by writing the BPS equations,
\begin{equation}
\begin{aligned}
& A'=	\frac{1}{4} \lambda ^2 e^{-2 B-2 U-V}+e^{-4 U-V},\;\;\;
B' =		e^{-4 U-V}-\frac{1}{4} \lambda ^2 e^{-2 B-2 U-V},	 
\\[5pt]
& U'=	e^{V-2 U}-e^{-4 U-V},	 \;\; V' = 	-\frac{1}{4} \lambda ^2 e^{-2 B-2 U-V}-e^{-4 U-V}-2 e^{V-2 U}+3 e^{-V}.
\label{BPSx}
\end{aligned}
\end{equation}
The above functions satisfy the following equations of motion,
\begin{equation}
\label{EOM}
\begin{aligned}
& A'' =	\frac{1}{2} \lambda ^2 e^{-2 B-4 U}+\frac{1}{8} \lambda ^4 e^{-4 B-4 U-2 V}+4 e^{-8 U-2 V}-A' \left( 2A'+2B'+4U'+V' \right),	 
 \\[7pt]
& B''=	-\frac{1}{2} \lambda ^2 e^{-2 (B+2 U)}-\frac{1}{8} \lambda ^4 e^{-2 (2 (B+U)+V)}+4 e^{-2 (4 U+V)}-B' \left(2A'+2B'+4U'+V' \right),	 
 \\[7pt]
& U''=		-\frac{1}{2} \lambda ^2 e^{-2 (B+2 U)}-4 e^{-2 (4 U+V)}-2 e^{2 V-4 U}+6 e^{-2 U}-U' \left(2A'+2B'+4U'+V' \right),	 
 \\[7pt]
& V''= 	\frac{1}{2} \lambda ^2 e^{-2 (B+2 U)}-\frac{1}{8} \lambda ^4 e^{-2 (2 B+2 U+V)}-4 e^{-2 (4 U+V)}+4 e^{2 V-4 U} 
-V' \left(2A'+2B'+4U'+V' \right),	 
\end{aligned}
\end{equation}
and the constraint
\begin{equation}
\begin{aligned}
 & 8 e^{2 (B+2 U+V)} \Bigg[e^{2 B+4 U} \Big(2 A' \left(2 B'+4 U'+V'\right)+A'^2+2 V' \big(B'+2 U' \big)+8 B' U'
\\[5pt]
 &\qquad\quad + B'^2+6 U'^2 \Big) -12 e^{2 (B+U)}+\lambda ^2 \Bigg]+16 e^{4 (B+U+V)}+32 e^{4 B}+\lambda ^4 e^{4 U} =0.
\end{aligned}
\end{equation}

\subsection*{The SUSY Flow}\label{subsec:SUSYflow}

\begin{figure}[h]
    \centering
     \includegraphics[width=0.7\textwidth]{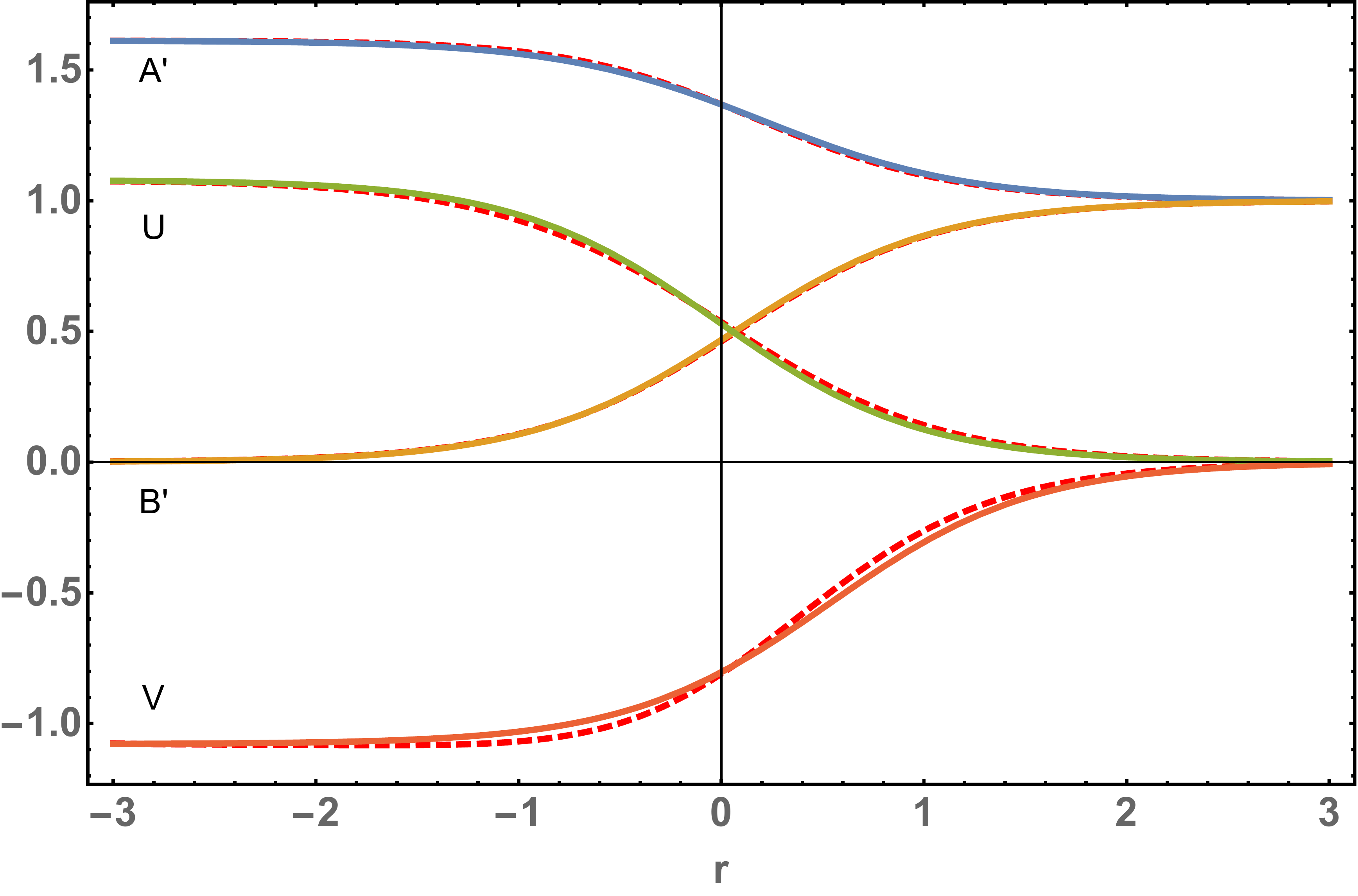}
         \caption{From top to bottom: the numerical and analytical approximation
values of the functions $A',U,B',V$. The solid lines correspond to the numerical solutions and the dashed lines to the analytical approximations. Notice that the plots for the functions $U$ and $V$ are rescaled by a factor of $15$.}
     \label{Fig:Flow}
\end{figure}

Let us now discuss the numerical flow 
between the two asymptotic solutions.  The appropriate asymptotics 
in the IR ($r\rightarrow-\infty$) are given by (here we can set $\lambda=2$ without loss of generality)
\begin{equation}
\label{IRasym}
\begin{aligned}
& A=		a_0 + \frac{r}{R_{DG}}	+	\frac{3}{2} s_1 e^{2r/R_{DG}}+ \dots + \frac{1}{4} \left(-3 + \sqrt{5}\right) s_2 e^{(-1 + \sqrt{5}) r/R_{DG}} + \dots \, 	,
 \\[7pt]
& B=		\frac{1}{4}\ln\left(\frac{4}{3}\right) +		s_1 e^{2r/R_{DG}} + \dots + s_2 e^{(-1 + \sqrt{5})r/R_{DG}} + \dots \,		 ,
 \\[7pt]
& U=		\frac{1}{4}\ln\left(\frac{4}{3}\right) -		s_1 e^{2r/R_{DG}} + \dots + \frac{1}{4} \left(2 - \sqrt{5}\right) s_2 e^{(-1 + \sqrt{5})r/R_{DG}} + \dots \,		,
 \\[7pt]
& V=		-\frac{1}{4}\ln\left(\frac{4}{3}\right) -		s_1 e^{2r/R_{DG}} + \dots + \frac{1}{4} \left(-9 + 4\sqrt{5}\right) s_2 e^{(-1 + \sqrt{5})r/R_{DG}} + \dots \, ,
\end{aligned}
\end{equation}
where $R_{DG}=\frac{\sqrt{2}}{3^{3/4}}$, and $s_1$, $s_2$ and $a_0$ are integration constants to be fixed.  In the UV ($r\rightarrow\infty$) are given by (leaving $\lambda$-dependence intact)
\begin{equation}
\label{UVasym}
\begin{aligned}
A&=	r - \frac{5}{48} \lambda^2 e^{-2r} 	+		\frac{287}{18432} \lambda^4 e^{-4r} 	- 	\frac{5953}{2211840} \lambda^6 e^{-6r}	+\dots\,	,
\\[7pt]
B&=	r + \frac{7}{48} \lambda^2 e^{-2r} 	-		\frac{385}{18432} \lambda^4 e^{-4r} 	+ 	\frac{8267}{2211840} \lambda^6 e^{-6r}	+\dots\,	,
\\[7pt]
U&=	\frac{1}{48} \lambda^2 e^{-2r} 	-		\frac{13}{1536}  \lambda^4 e^{-4r} 	+\left(	\frac{c_u}{\lambda^6} + \frac{3}{1280}r	\right)  \lambda^6	 e^{-6r}		+\dots\, ,
\\[7pt]
V&=	-\frac{1}{24} \lambda^2 e^{-2r} 	+		\frac{37}{1536}  \lambda^4 e^{-4r} 	- 	\left( \frac{4 c_u}{ \lambda^6}	- \frac{9023}{3317760}				+ \frac{3}{320}r	\right)  \lambda^6 e^{-6r}		+\dots\, .
\end{aligned}
\end{equation}
where there is only one integration constant given by $c_u$.  In Fig. \ref{Fig:Flow}, we plot the behavior of the background functions for this flow, using the BPS equations and numerically solving for $c_u\approx 0.104892 $.
%
%
%

\subsection*{A possible deformation of the SUSY flow }
\label{subsec:nonSUSYflow}

Here we present the UV expansion when we instead solve the full EOM given in eq. \eqref{EOM}, as opposed to the BPS equations given in eq. \eqref{BPS}.  So, in the UV ($r\rightarrow\infty$), using a similar expansion as in the SUSY case we find (leaving the $\lambda$-dependence intact again)
\begin{equation}
\label{ns-UVasym}
\begin{aligned}
A&=	r - \frac{5}{48} \lambda^2 e^{-2r} 	-	\left(\frac{c_b}{\lambda^4}  +	 \frac{49}{9216} \right) \lambda^4 e^{-4r} 	- 	 \left( \frac{c_b}{40 \lambda^4} + \frac{1777}{552960} \right) \lambda^6 e^{-6r}	+\dots\,	,
 \\[7pt]
B&=	r + \frac{7}{48} \lambda^2 e^{-2r} 	+		c_b e^{-4r} 	- 	 \left( \frac{c_b}{40 \lambda^4} - \frac{889}{276480} \right) \lambda^6 e^{-6r}	+\dots\,	,
 \\[7pt]
U&=	\frac{1}{48} \lambda^2 e^{-2r} 	-		\frac{13}{1536}  \lambda^4 e^{-4r} 	+\left(	\frac{c_u}{\lambda^6} - \frac{1}{4} \left(\frac{c_b}{5 \lambda^4} - \frac{479}{92160} \right) r	\right)  \lambda^6	 e^{-6r}		+\dots\, ,
 \\[7pt]
V&=	-\frac{1}{24} \lambda^2 e^{-2r} 	+		\frac{37}{1536}  \lambda^4 e^{-4r} 	- 	\left( \frac{4 c_u}{ \lambda^6} + \frac{3 c_b }{20\lambda^4} + \frac{9023}{3317760}		-	\left(	\frac{c_b}{5 \lambda^4} - \frac{343}{829440} \right) r	\right)  \lambda^6 e^{-6r}		
 \\[7pt]
&  + \left(		\frac{c_v}{\lambda^8}	- \left(\frac{7 c_b}{240 \lambda^4}	-	\frac{3353}{4423680}\right) r	 \right)   \lambda^8 e^{-8r} +\dots\, .
\end{aligned}
\end{equation}
Now we have three integration constants $c_b$, $c_u$ and $c_v$.  We can recover the original UV expansions presented in eq. \eqref{UVasym} by setting
\begin{equation}
c_b \rightarrow -\frac{385 }{18432} \lambda^4 , 	\qquad 	c_v \rightarrow \frac{7 c_u }{12} \lambda^2 - \frac{72617 }{63700992} \lambda^8 ,
\label{SUSYrecover}
\end{equation}
leaving one integration constant $c_u$ as before.  It would be interesting to see if a family of flows can be found using these expansions, and see if any new IR fixed points appear, as limits. 

\subsection*{Analytic approximation for the Donos-Gauntlett flow
}
\label{subsec:Analytic approximation}

Below, we discuss a simple, analytic approximation for the functions $A,B,U,V$ of the Donos-Gauntlett solution. 
After this, we explain how an analogy of those functions with some thermodynamic quantities of a two-level Maxwell-Boltzmann system appears.

\subsubsection*{Analytic expressions}
\label{AnalyticAppSection}

In our ansatz for the analytic approximation eq. \eqref{AnalyticApproxAnsatz} we consider the following values for the various constants that appear,
\begin{equation}
 \mu_{A}=\mu_B=\mu_U=\mu_V=\frac{1}{2} \ , \qquad r_{A}=\frac{a_{0}}{1-\frac{1}{R}}=0.212 \ , \qquad r_{B}=\frac{1}{2}\ln\frac{2}{\sqrt{3}} \ .
\end{equation}
These values come from the zeroth order of the fixed point expansions. \footnote{ 
For the expansions around the UV and the IR we find the following values for the parameters $\m_A, \ \m_B, \ \m_U, \ \m_V$:
\[
\begin{aligned}
 & \mu_{A}=\mu_B=\mu_U=\mu_V=\frac{1}{2} \ , \qquad \textrm{(UV)} \ ,
 \\[5pt]
 & \mu_{A}=\mu_B=\mu_U=\mu_V=  \frac{\sqrt{2}}{\big(\sqrt{5} - 1\big) \ 3^{\frac{3}{4}} } = 0.502 \ldots  \ , \qquad \textrm{(IR)} \ .
\end{aligned}
\]
In what follows we chose the values that correspond to the UV.
} 
The parameters $r_U$ and $r_V$ are not fixed and we chose $r_U = -0.019$, $r_V = 0.537$. The comparison with the numerics is depicted in the Fig. \ref{Fig:Flow}.
%
%



\subsubsection*{Two-level system thermodynamic analogy}

By thermodynamic analogy we mean the association of some thermodynamic quantities of the two-level Maxwell-Boltzmann system with the functional forms of the warping factors of the geometry. The important quantities in our scheme are the free energy ($F$) and the occupation probability ($P$),
\begin{equation}
\begin{aligned}
 & F_A= -\frac{1}{2} \ \ln \Big[1+e^{-\frac{r-r_{A}}{\m_A}}\Big] \ ,
\qquad
F_B= -\frac{1}{2} \ \ln \Big[ 1+e^{- \frac{r-r_{B}}{\m_B}}\Big] \ ,
\\[10pt]
& P_U= \frac{1}{1+e^{ \frac{r-r_{U}}{\m_U}}} \ ,
\qquad\qquad\qquad \;\;
P_V= \frac{1}{1+e^{ \frac{r-r_{V}}{\m_V}}} \ .
\end{aligned}
\end{equation}
 The warping factors of the geometry can be written in terms of the above quantities in the following way,
%
%
\begin{equation}
e^{2A} =e^{2(r+\frac{1-R_{DG}}{R_{DG}}F_{A})} \ , \qquad
e^{2B} = e^{2(r-F_{B})} \ , \qquad
e^{2U} =\left[\frac{2}{\sqrt{3}}\right]^{P_{U}} \ , \qquad
e^{2V} =\left[\frac{\sqrt{3}}{2}\right]^{P_{V}} \ .
\end{equation}
The interpretation of this rewriting  is as follows: for
each function $A,B,U,V$ we have a two-level system with energies
$E_1=0$ and $E_2 = r + \textrm{const}$. The factors
$e^{\frac{2(1-R_{DG})}{R_{DG}}F_{A}},\, e^{-2F_{B}}$ are reminiscent of two-state
transition theory. Also, the temperature of the systems is $k T = \m = 1/2$ and $k$ is the Boltzmann's constant.

It would be interesting to exploit this formal analogy further. In
particular, if having thermal equilibrium populations is 'extremal' in
some sense, if the variation of some relevant QFT quantity extremized
along this 'isothermal' flow, or  some result of two state transition
theory in statistical mechanics may be applied. \\
\subsubsection*{The 
two-level system analogy in the exact twisted $H_2$ flow}

This case is an exact and simplified version of the approximated solution
and the two-level system analogy we found for the Donos-Gauntlett solution. Let us consider the solution of eq. \eqref{O19} with $a_{0}=0$.

%
The free energy for the two-level system with temperature $2 k T=1$ and energy levels $E_1=0, \ E_2= r$ is,
\begin{equation}
  F =-\frac{1}{2}\ln\left[1+e^{-2r}\right] \ .
\end{equation}
The warp factors of the geometry now take the following form,
\begin{equation}
  e^{2 A} = e^{2r+F} \ ,  \qquad
  e^{2 B} = \frac{1}{3} \ e^{2(r-F)} \ .
\end{equation}

We saw how we can approximate the solutions of the BPS systems of eq. \eqref{O25} and eq. \eqref{BPS} using the ansatz of eq. \eqref{AnalyticApproxAnsatz}.
%
%
%
%
%
%

%

\section{Expressions after NATD in other coordinates}\label{cylcart}
In this appendix, we quote the expressions of the backgrounds obtained after NATD, in cartesian and cylindrical coordinates. 
Recall that in all of our examples, during the NATD transformation, we considered a gauge fixing of the form $\theta_2 = \phi_2 = \psi = 0$. As a result the Lagrange multipliers $x_1, x_2$ and $x_3$ play the r\^ole of the dual coordinates in the new background generated by the NATD transformation.
In this appendix we work out the details of the NATD backgrounds that we found expressing $\big(   x_1, x_2,x_3 \big)$ both in the Cartesian and in the cylindrical coordinate systems.
%
\vskip 10pt
\noindent\underline{NATD of the twisted solutions: eqs \eqref{NN11}-\eqref{NN12}}
\vskip 10pt
\noindent\emph{a) Cartesian coordinates}
\vskip 10pt

In this coordinate system the NSNS sector of the solution becomes.
%
\bea
 \label{NN2}
 &  & e^{-2 \widehat{\Phi}}=\frac{L^2}{324 \alpha'^3} \Delta, \qquad \Delta =L^4 + 18 \alpha'^2 \Big(3 x_1^2+ 3 x_2^2 + 2 x_3^2 \Big) \ ,
\nonumber\\[5pt]
 &  & \frac{d\hat{s}^{2}}{L^2}=e^{2 A} \left(- dy_0^2+d y_1^2\right)+e^{2 B} ds^2_{\Sigma_2}+ dr^2+\frac{1}{6} \left(\sigma_{1}^2+\sigma_{2}^2\right) +\frac{\alpha'^2}{\Delta} \Bigg[ 6 \Bigg( \Big(x_1^2+x_2^2 \Big)\tilde{\sigma}_{3}^2
\nonumber\\[5pt]
&  & \qquad -2 \tilde{\sigma}_{3} \Big(x_2 dx_1-x_1 dx_2 \Big)+dx_1^2 + dx_2^2 \Bigg) + 9 d x_3^2 + \frac{324 \alpha'^2}{L^4} \Big(x_1 d x_1+ x_2 d x_2 + x_3 d x_3 \Big)^2 \Bigg] \ , 
\\[5pt]
&  &  \frac{\Delta \widehat{B}_2}{\alpha'^3}=36 x_3 \Bigg(\tilde{\sigma}_{3}\wedge \Big(x_1 dx_1+x_2 dx_2 + x_3 dx_3 \Big) - d x_1 \wedge d x_2 \Bigg) + \left( \frac{L^4}{\alpha'^2} \tilde{\sigma}_{3} + 54 (x_2 dx_1-x_1 dx_2) \right)\wedge dx_3 .
\nonumber
\eea
On the other hand, the RR sector is,
\begin{equation}
 \label{NN3}
\begin{aligned}
 & \widehat{F}_0=0, \qquad \widehat{F}_2=\frac{L^4}{54\alpha'^{\frac{3}{2}}} \left( 2 {\sigma_{1}\wedge \sigma_{2}} + 3 z \textrm{Vol}_{\Sigma_2} \right),
\\[5pt]
 & \widehat{F}_4=\frac{L^4}{18 \sqrt{\alpha'}} \Bigg[  3 z e^{-2B} dx_3  \wedge \textrm{Vol}_{AdS_3} + z \ x_3 \ \textrm{Vol}_{\Sigma_2} \wedge {\sigma_{1} \wedge \sigma_{2}}  \\[5pt]
 & \quad \; -\frac{18 \alpha'^2 }{\Delta} \Bigg( z \textrm{Vol}_{\Sigma_2}  + \frac{2}{3} {\sigma_{1} \wedge \sigma_{2}} \Bigg) \wedge \Bigg( 2 x_3 \Big(dx_1 \wedge dx_2+\big( x_1 dx_1 + x_2 dx_2 \big)\wedge \tilde{\sigma}_{3} \Big) 
\\[5pt]
 & \quad \; +3 \Big(x_1 dx_2-x_2 dx_1 + \big( x_1^2+x_2^2 \big)\tilde{\sigma}_{3} \Big) \wedge d x_3 \Bigg) \Bigg] \ .
\end{aligned}
\end{equation}
To explore the singularity structure of this metric we compute the Ricci scalar (we only present it in Cartesian coordinates),
 \begin{equation}
 \begin{aligned}
 & \mathcal{R}=\frac{2}{L^2 \Delta^2} \Bigg( - \frac{27}{2}z^2 e^{-4B} \alpha'^2 \left( x_1^2+x_2^2 \right) \Delta + e^{-2B}\Delta^2 \kappa + 71 L^8 
  \\[5pt]
 &   \quad + 90 L^4 \alpha'^2 \Big( 9 \left( x_1^2+x_2^2 \right)+ 28 x_3^2 \Big) 
 +2916 \alpha'^4 \left( 18 x_3^2 \left(x_1^2+x_2^2\right)+7 \left(x_1^2+x_2^2\right)^2+12 x_3^4 \right)
 \\[5pt]
 & \quad -\Delta^2\left( 2 \left(A''+B''\right)+4 A' B'+3A'^2+3 B'^2\right) \Bigg) \ ,
\end{aligned}
\end{equation}
where $\kappa = -1, 0, 1$ for each of the three cases $\Sigma_2 = H_2, T^2, S^2$ respectively.

\vskip 10 pt 

\noindent\emph{b) Cylindrical coordinates}

\vskip 10pt

Let us now see the form of the fields in the cylindrical coordinates,
\begin{equation}
\label{NN5}
    x_1= \rho \cos \xi \ , \qquad x_2= \rho \sin \xi \ , \qquad x_3=x_3 \ .
\end{equation}
%
The  NSNS sector in this coordinate system reads,
\begin{equation}
\label{NN7}
\begin{aligned}
& e^{-2 \widehat{\Phi}}=\frac{L^2}{324 \alpha'^3} \Delta \ , \qquad  \Delta =L^4 + 18 \alpha'^2 \left(3 \rho^2 + 2 x_3^2\right) \ , 
\\[5pt]
& \frac{d\hat{s}^{2}}{L^2}=e^{2 A} \left(- dy_0^2+d y_1^2\right)+e^{2 B} ds^2_{\Sigma_2} + dr^2+\frac{1}{6} \left(\sigma_{1}^2+\sigma_{2}^2\right)
\\[5pt]
& \qquad +\frac{\alpha'^2}{\Delta} \Bigg(6  \Big(\rho^2 \big( d \xi +\tilde{\sigma}_{3} \big)^2+ d \rho ^2 \Big)+9  d x_3^2 + \frac{324 \alpha'^2}{L^4} \big(\rho  d \rho +x_3 d x_3 \big)^2\Bigg)  \ ,
\\[5pt]
& \widehat{B}_2=\frac{\alpha'^3}{\Delta} \Bigg(36 x_3 \Big(\tilde{\sigma}_{3}\wedge \big(\rho d \rho+ x_3 dx_3 \big)+\rho d\xi\wedge d\rho \Big)+\Big(\frac{L^4}{\alpha'^2} \tilde{\sigma}_{3} -54 \rho^2 d\xi \Big)\wedge dx_3 \Bigg).
\end{aligned}
\end{equation}
Also the RR fields take the form,
\begin{equation}
\label{NN8}
\begin{aligned}
& \widehat{F}_0=0, \qquad
\widehat{F}_2=\frac{L^4}{54\alpha'^{\frac{3}{2}}} \Big( 2 {\sigma_{1}\wedge \sigma_{2}} + 3 z \textrm{Vol}_{\Sigma_2} \Big) \ , 
\\[5pt]
& \widehat{F}_4=\frac{L^4}{18 \sqrt{\alpha'}} \Bigg[  3 z e^{-2B} dx_3  \wedge \textrm{Vol}_{AdS_3} + z x_3 \textrm{Vol}_{\Sigma_2} \wedge {\sigma_{1} \wedge \sigma_{2}}
\\[5pt]
& \quad \; -\frac{18 \alpha'^2 \rho}{\Delta} \Big( z \textrm{Vol}_{\Sigma_2}  +\frac{2}{3} {\sigma_{1} \wedge \sigma_{2}} \Big) \wedge \Big( 2 x_3 d \rho -3 \rho d x_3 \Big) \wedge \Big( d \xi + \tilde{\sigma}_{3} \Big)  \Bigg] \ . 
\end{aligned}
\end{equation}

\vskip 10pt
\noindent\underline{NATD of the Donos-Gauntlett: eqs \eqref{s5}-\eqref{s11}}
\vskip 10pt

\noindent\emph{a) Cartesian coordinates}

\vskip 5pt
The NSNS sector of this solution in the Cartesian coordinate system takes the form,
\begin{equation}
  \begin{aligned}
 &  e^{-2 \widehat{\Phi}} =\frac{L^2}{324 \alpha'^3} \Delta \ , \qquad \Delta =L^4 e^{4 U+2 V}+54 \alpha'^2 e^{2 U} \big(x_1^2+x_2^2 \big)+36 \alpha'^2 \mathcal{B}^2  \ ,  
\\[5pt]
 & \frac{d\hat{s}^{2}}{L^2} =e^{2 A} \left(- dy_0^2+d y_1^2\right)+e^{2 B} \left( d \alpha^2+ d \beta^2\right)+ dr^2+\frac{e^{2 U}}{6} \left(\sigma_{1}^2+\sigma_{2}^2\right) + \frac{\alpha'^2}{\Delta} \Bigg[ 6 e^{2 U+2 V} \Big( \big(x_1^2+x_2^2 \big) \sigma_{3}^2
\\[5pt]
 & \quad \; -2 \sigma_{3} \big( x_2 dx_1-x_1 dx_2 \big)+dx_1^2 + dx_2^2 \Big)+9 e^{4 U} d x_3^2 + \frac{324 \alpha'^2}{L^4} \Big(x_1 d x_1+ x_2 d x_2 +\mathcal{B} d x_3 \Big)^2 \Bigg]  \ , 
\\[5pt]
 & \widehat{B}_2 =L^2\frac{\lambda}{6} \ \alpha \ {\sigma_{1}\wedge \sigma_{2}}+\frac{\alpha'^3}{\Delta} \Bigg[ 36 \mathcal{B} e^{2V} \Big(\sigma_{3}\wedge \big(x_1 dx_1+x_2 dx_2 + \mathcal{B} dx_3 \big) - d x_1 \wedge d x_2 \Big) 
\\[5pt]
 &  \quad \; + e^{2U} \Big( e^{2V+2U} \frac{L^4}{\alpha'^2} \sigma_{3} + 54 \big(x_2 dx_1-x_1 dx_2 \big) \Big)\wedge dx_3 \Bigg] \ .
 \end{aligned}
\label{c5}
\end{equation}
The fields of the RR in this coordinates become,
%
\begin{align}
\label{c11}
& \widehat{F}_0 =0 \ , \qquad \widehat{F}_2 =\frac{L^2}{6\alpha'^{\frac{3}{2}}} \Bigg( \lambda \alpha' d\mathcal{B}_{-}\wedge d\beta + \frac{2}{9} L^2 {\sigma_{1}\wedge \sigma_{2}} \Bigg) \ ,
\nonumber\\[5pt]
& \widehat{F}_4 =\frac{L^4 \lambda}{36 \sqrt{\alpha'}} \Bigg[ e^{V} \Bigg( \frac{2 L^2}{\alpha'}d\alpha -3 e^{-2B-2V} \lambda dx_3 \Bigg) \wedge \textrm{Vol}_{AdS_3}- \frac{6 \alpha'}{L^2} \Big( x_1 dx_1+x_2 dx_2 + \mathcal{B} d \mathcal{B} \Big)  \wedge d \beta \wedge {\sigma_{1} \wedge \sigma_{2}}
\nonumber\\[5pt]
& \quad \;+\frac{36 \alpha'^2 }{\Delta} \Bigg( \frac{3 \alpha'}{L^2} d \beta \wedge d\mathcal{B}_{-}  -\frac{2}{3 \lambda} {\sigma_{1} \wedge \sigma_{2}} \Bigg) \wedge \Bigg( 2 e^{2V}\mathcal{B}\Big(dx_1 \wedge dx_2 + \big(x_1 dx_1 + x_2 dx_2 \big) \wedge \sigma_{3} \Big)
\\[5pt]
& \quad \; +3 e^{2U} \Big(x_1 dx_2-x_2 dx_1 + \big( x_1^2+x_2^2 \big) \sigma_{3} \Big) \wedge d x_3 \Bigg) \Bigg] \ .
\nonumber
\end{align}
%
Also now the functions $\mathcal{B}_\pm$ become,
\begin{equation}
\label{CalBpm}
 \mathcal{B}_{\pm}=x_3\pm\frac{L^2 \lambda}{6 \alpha'} \ \alpha \ , \qquad \mathcal{B}=\mathcal{B}_{+} \ .
\end{equation}

\vskip 10 pt \noindent\emph{b) Cylindrical coordinates}

\vskip 10pt

The NSNS sector of this solution in cylindrical coordinates, eq. \eqref{NN5}, is,
\begin{equation}
\label{5}
\begin{aligned}
 & e^{-2 \widehat{\Phi}} =\frac{L^2}{324 \alpha'^3} \Delta \ , \qquad \Delta =L^4 e^{4 U+2 V}+6 \alpha'^2 \left(9 \rho ^2 e^{2 U}+6 \ \mathcal{B}^2 e^{2 V}\right) \ ,
\\[5pt]
 & \frac{d\hat{s}^{2}}{L^2} =e^{2 A} \left(- dy_0^2+d y_1^2\right)+e^{2 B} \left( d \alpha^2+ d \beta^2\right)+ dr^2+\frac{e^{2 U}}{6} \left(\sigma_{1}^2+\sigma_{2}^2\right)
\\[5pt]
 & \qquad +\frac{\alpha'^2}{\Delta} \Bigg[6 e^{2 U+2 V} \Big(\rho^2 \big( d \xi +\sigma_{3} \big)^2+ d \rho ^2 \Big)+9 e^{4 U} d x_3^2 + \frac{324 \alpha'^2}{L^4} \Big(\rho  d \rho +\mathcal{B} \ d x_3 \Big)^2 \Bigg] \ ,
\\[5pt]
 & \widehat{B}_2 =L^2\frac{\lambda}{6} \alpha \ {\sigma_{1}\wedge\sigma_{2}}+\frac{\alpha'^3}{\Delta} \Bigg[36\mathcal{B} \ e^{2V} \Big(\sigma_{3}\wedge \big(\rho d \rho+\mathcal{B}dx_3 \big)+\rho d\xi\wedge d\rho \Big)
 \\[5pt]
 & \quad \; +e^{2U} \Bigg(e^{2V+2U}\frac{L^4}{\alpha'^2}\sigma_{3} -54 \rho^2 d\xi \Bigg)\wedge dx_3 \Bigg] \ .
\end{aligned}
\end{equation}
The RR sector has the form,
\begin{equation}
\label{11}
\begin{aligned}
 & \widehat{F}_0 =0 \ , \qquad \widehat{F}_2 =\frac{L^2}{6\alpha'^{\frac{3}{2}}} \Bigg( \lambda \alpha' d\mathcal{B}_{-} \wedge d\beta + \frac{2}{9} L^2 {\sigma_{1}\wedge \sigma_{2}} \Bigg) \ ,
\\[5pt]
 & \widehat{F}_4 =\frac{L^4 \lambda}{36 \sqrt{\alpha'}} \Bigg[ e^{V} \Bigg( \frac{2 L^2}{\alpha'}d\alpha -3 e^{-2B-2V} \lambda dx_3 \Bigg) \wedge \textrm{Vol}_{AdS_3}-  \frac{6 \alpha'}{L^2} \Big( \rho d \rho + \mathcal{B} d \mathcal{B} \Big)  \wedge d \beta \wedge {\sigma_{1} \wedge \sigma_{2}}
\\[5pt]
 & \quad \; +\frac{36 \alpha'^2 \rho}{\Delta} \Bigg( \frac{3 \alpha'}{L^2} d \beta \wedge d\mathcal{B}_{-}  -\frac{2}{3 \lambda} {\sigma_{1} \wedge \sigma_{2}} \Bigg) \wedge \Big( 2 e^{2V}\mathcal{B}d \rho -3 e^{2U} \rho d x_3 \Big) \wedge \Big( d \xi + \sigma_{3} \Big) \Bigg] \ .
\end{aligned}
\end{equation}
Also, in this coordinates the functions $\mathcal{B}_\pm$ have the same form as in the Cartesian coordinates. Hence their expressions are given in eq. \eqref{CalBpm}. 

\vskip 10pt
\noindent\underline{The NATD-T solution: eqs \eqref{NN2TT}-\eqref{NN3TT}}
\vskip 10pt

\noindent\emph{a) Cartesian coordinates}

\vskip 10pt

In Cartesian coordinates the 1-form $A_3$ defined in eq. \eqref{NN1TT} and the function $\Delta_T$ defined in eq. \eqref{moduluskillingvector} take the following form,
\begin{equation}
\begin{aligned}
& A_3 =3 z \sinh a \ dx_3 - db \ ,  
\\[5pt]
& \Delta_T = 54 \alpha'^2 z^2 (x_1^2+x_2^2)  \sinh^2 a + e^{2B} \cosh^2 a \ \Delta \ ,
\end{aligned}
\end{equation}
and $\Delta$ is given in eq. \eqref{NN2}.

The NSNS sector of the NATD-T solution in Cartesian coordinates takes the form,
%
\begin{align}
     e^{-2 \widetilde{\Phi}} & = \frac{L^4}{324 \alpha'^4} \Delta_T  \ ,
\nonumber\\[5pt]
    \frac{d\tilde{s}^{2}}{L^2} & = e^{2 A} \left(- dy_0^2+d y_1^2\right)+e^{2 B} da^2+ dr^2+\frac{1}{6} \left(\sigma_{1}^2+\sigma_{2}^2\right) 
\nonumber\\[5pt]
    & + \frac{\alpha'^2}{\Delta_T} \Bigg[  \frac{\Delta}{L^4} d b^2 + 6 e^{2B} \cosh^2 a  \Big( \big(x_1^2+x_2^2 \big)\sigma_{3}^2-2 \sigma_{3} \big(x_2 dx_1-x_1 dx_2 \big)+dx_1^2 + dx_2^2 \Big)
\nonumber\\[5pt]
    & +  9\Big( z^2 \sinh^2 a+e^{2B} \cosh^2 a  \Big) \Big(d x_3^2 + \frac{36 \alpha'^2}{L^4} \big( x_1 d x_1+ x_2 d x_2 + x_3 d x_3 \big)^2 \Big) 
\\[5pt]
    & -  6 z \sinh a d b  \Big(d x_3 + \frac{36 \alpha'^2}{L^4} x_3 \big(x_1 d x_1+ x_2 d x_2 + x_3 d x_3 \big) \Big)\Bigg] \ ,  
\nonumber\\[5pt]
    \widetilde{B}_2 & =  \frac{\alpha'^3}{\Delta_T}     \Bigg[  e^{2B} \cosh^2 a   \Bigg( 36 x_3 \Big( \sigma_{3}\wedge \big(x_1 dx_1+x_2 dx_2 + x_3 dx_3 \big) - d x_1 \wedge d x_2 \Big)   
\nonumber\\[5pt]
    & +  \Big( \frac{L^4}{\alpha'^2} \sigma_{3} + 54 \big(x_2 dx_1-x_1 dx_2 \big) \Big)\wedge dx_3 \Bigg) - 18 z \sinh a \Big(  \big(x_1 dx_2 - x_2 d x_1 \big) \wedge A_3 + \big(x_1^2 + x_2^2 \big) d b \wedge \sigma_{3}   \Big) \Bigg] \ .
\nonumber
\end{align}
%
The RR sector reads
%
\begin{align}
\label{NN3TTbx}
\widetilde{F}_1&=\frac{z L^4}{18 \alpha'^2} \cosh a \ d a \ ,
\nonumber\\[5pt]
\widetilde{F}_3&=\frac{L^4}{54\alpha'} \left(  2 A_3 + 3 z x_3 \cosh a d a  \right)    \wedge         \sigma_{1}\wedge \sigma_{2}
\nonumber\\[5pt]
&+\frac{ z L^4 \alpha'  \cosh a }{\Delta_T} \Bigg[    e^{2B} \cosh^2 a   \Bigg( 2 x_3 \Big( \sigma_{3}\wedge \big(x_1 dx_1+x_2 dx_2 \big) - d x_1 \wedge d x_2 \Big) 
\nonumber\\[5pt]
& - 3 \Big( \big(x_1^2+x_2^2 \big) \sigma_{3} + x_1 dx_2-x_2 dx_1 \Big)\wedge dx_3 \Bigg) \wedge d a
  - z \sinh a \Big( \big(x_1^2+x_2^2 \big) \sigma_{3} + x_1 dx_2-x_2 dx_1 \Big)\wedge A_3\wedge da \Bigg] \ , 
\nonumber\\[5pt]
\widetilde{F}_5&= \frac{L^4}{6} e^{-2B}  \left( 24 e^{4B} \cosh a d a \wedge (x_1 dx_1+x_2 dx_2+x_3 dx_3) + z dx_3 \wedge d b \right) \wedge \textrm{Vol}_{AdS_3} 
\\[5pt]
 & + \frac{L^4 \alpha'^2 \cosh a}{18 \Delta_T} \Bigg[   e^{2B} \cosh a \Bigg(  dx_1 \wedge dx_2  \wedge \Big[ z \Big( \frac{L^4}{\alpha'^2} + 54 \big(x_1^2 + x_2^2 \big) \Big) \cosh a \ d a  - 24 \ x_3 A_3 \Big] 
\nonumber\\[5pt]
 & + 18  \Big( x_1 dx_2 - x_2 d x_1 \Big) \wedge \Big( 3 z x_3 \cosh a d a -2 d b \Big) \wedge dx_3 \Bigg) 
\nonumber\\[5pt]
 & + 18 z^2 \sinh a \Big( 3 z \sinh a \big(x_1^2+x_2^2 \big) dx_1 \wedge dx_2 + x_3 \big(  x_2 d x_1 - x_1 d x_2 \big) \wedge A_3 \Big) \wedge d \a  \Bigg] \wedge  \sigma_{1}\wedge \sigma_{2} \ .
\nonumber
\nonumber
\end{align}
    
\noindent\emph{b) Cylindrical coordinates}

\vskip 5pt
 
 In cylindrical coordinates the 1-form $A_3$ in eq. \eqref{NN1TT} and the function $\Delta_T$  in eq. \eqref{moduluskillingvector} are
\begin{equation} 
\label{NN2TTbx33}
A_3 = z \sinh a \ dx_3 - db \ , \qquad  \Delta_T = 54 \alpha'^2 z^2 \rho^2  \sinh^2 a + e^{2B} \cosh^2 a \ \Delta \ ,
\end{equation}
and the function $\Delta$ is given in eq. \eqref{NN7}.

The NSNS sector of the solution in this coordinate system is written below,
%
\begin{align}
    & e^{-2 \widetilde{\Phi}} =\frac{L^4}{324 \alpha'^4} \Delta_T \ ,  
\nonumber\\[5pt]
    & \frac{d\tilde{s}^{2}}{L^2} =e^{2 A} \left(- dy_0^2+d y_1^2\right)+e^{2 B} d a^2+ dr^2+\frac{1}{6} \left(\sigma_{1}^2+\sigma_{2}^2\right)
\nonumber\\[5pt]
    & \quad \;\; +\frac{\alpha'^2}{\Delta_T} \Bigg[  \frac{\Delta}{L^4} db^2 + 6 e^{2B} \cosh^2 a  \Big(\rho^2 \big(\sigma_{3}+d\xi \big)^2+ d\rho^2 \Big)
\\[5pt]
    & \quad \;\; +9\Big( z^2 \sinh^2 a+e^{2B} \cosh^2 a  \Big) \Bigg(d x_3^2 + \frac{36 \alpha'^2}{L^4} \Big(\rho d\rho + x_3 d x_3 \Big)^2\Bigg)
\nonumber\\[5pt]
    & \quad \;\; - 6 z \sinh a \ d b  \Bigg(d x_3 + \frac{36 \alpha'^2}{L^4} x_3 \Big(\rho d\rho + x_3 d x_3 \Big) \Bigg) \Bigg] \ ,
\nonumber\\[5pt]
    & \widetilde{B}_2 =\frac{\alpha'^3}{\Delta_T}     \Bigg[  e^{2B} \cosh^2 a   \Bigg( 36 x_3 \Big[ \sigma_{3}\wedge \big(\rho d \rho + x_3 dx_3 \big) + \rho d \xi \wedge d \rho \Big] + \Big( \frac{L^4}{\alpha'^2} \sigma_{3} - 54 \rho^2 d\xi \Big)\wedge dx_3 \Bigg)
\nonumber\\[5pt]
& \quad \; - 18 z \rho^2 \sinh a \Big(  d\xi \wedge A_3 + d b \wedge \sigma_{3}  \Big) \Bigg] \ .
\nonumber
\end{align}
%
Finally, the RR sector of the solution reads,
\begin{equation}
\begin{aligned}
& \widetilde{F}_1 =\frac{z L^4}{18 \alpha'^2} \cosh a \ d a \ ,
\\[5pt]
& \widetilde{F}_3 =\frac{L^4}{54\alpha'} \left(  2 A_3 + 3 z x_3 \cosh a d a  \right)    \wedge         \sigma_{1}\wedge \sigma_{2} 
\\[5pt]
& \quad \; +\frac{ z L^4 \alpha'  \cosh a }{\Delta_T}\rho (\sigma_{3}+d\xi)\wedge \Big( e^{2B} \cosh^2 a \ \big(2 x_3 d \rho -3 \rho d x_3 \big)- z \rho \sinh a A_3 \Big) \wedge d a \ ,
\\[5pt]
& \widetilde{F}_5 = \frac{L^4}{6} e^{-2B}  \Big( 24 \ e^{4B} \cosh a d a \wedge \big( \rho \ d \rho+x_3 dx_3 \big) + z dx_3 \wedge d b  \Big) \wedge \textrm{Vol}_{AdS_3}
\\[5pt]
 & \quad \; + \frac{L^4 \alpha'^2 \cosh a}{18 \Delta_T} \Bigg[   e^{2B} \cosh a \rho d\xi \wedge \Bigg(   \Big[ z \Big( \frac{L^4}{\alpha'^2} + 54 \rho^2 \Big) \cosh a d a  - 24 x_3 A_3 \Big] \wedge d\rho
\\[5pt]
 & \quad \; + 18  \rho \Big( 3 z x_3 \cosh a d a -2 db \Big) \wedge dx_3 \Bigg)
\\[5pt]
 & \quad \; - 18 z^2 \sinh a \rho^3 d \xi\wedge \Big( 3 z \sinh a d\rho + x_3 \cos \chi A_3 \Big) \wedge d a  \Bigg] \wedge  \sigma_{1}\wedge \sigma_{2} \ .
\end{aligned}
\label{NN3TTb}
\end{equation}

\section{Central charges in Other Interesting Geometries}
\label{AppCentralC}
In this appendix, we will consider two types of geometries that have attracted attention 
due to their interesting dual QFTs. These geometries can be thought of as dual to anisotropic QFTs, or flows between QFTs in different dimensions. The goal of this appendix is to discuss the application of the formulas \eqref{centralanisotropic} that we developed to calculate central charges in anisotropic theories.
  \subsection*{An $AdS_3\times M_7$ geometry}
The first geometry we will deal with, is dual to  a flow between $\mathcal{N}=4$ SYM in $3+1$ dimensions and 
a $1+1$ CFT with $\mathcal{N}=(4,4)$ SUSY. The solution was written in Section 3.1 of \cite{Maldacena:2000mw}.
The geometry reads,
\begin{equation}
\label{maba}
\begin{aligned}
& ds_{IIB}^2= L^2\sqrt{\Delta}\Big[   e^{2f}(dx_{1,1}^2+ dr^2) +  e^{2g}ds^2_{H_2} \Big] +\frac{L^2}{\sqrt{\Delta}}\Big[\frac{\Delta}{X_1X_3}d\alpha^2 +\frac{\cos^2\alpha}{X_1}d\beta^2
\\[5pt]
& \qquad \;\; + \frac{\cos^2\alpha \cos^2\beta}{X_1}d\phi_1^2 +\frac{\cos^2\alpha \sin^2\beta}{X_1} d\phi_2^2+\frac{\sin^2\alpha}{X_3}(d\phi_3 +\frac{dx}{y})^2\Big] \ ,
\\[5pt]
& \Delta= X_1 \cos^2\alpha + X_3\sin^2\alpha \ , \;\;\; X_1^2=X_3^{-1}=e^{2\varphi}.
\end{aligned}
\end{equation}
There is a five form excited---we will not quote it-- and the dilation is vanishing.
First, we will calculate the central charge of the dual QFT considering as a field theory in $1+1$ dimensions using eq. \eqref{kkmcentral}. Then, we will use our formula for anisotropic backgrounds eq.  \eqref{centralanisotropic}.

If we consider the field theory to be in 2-d, we have that the relevant quantities are,
\begin{equation}
\begin{aligned}
&  d=1,\;\;\; \alpha_0= L^2\sqrt{\Delta} \ e^{2f}, \;\;\;\beta_0=1 \ , \;\;\; e^\Phi=1 \ ,
\\[5pt]
&  \sqrt{\det[g_{int}] \ \alpha_0} \sim L^8e^{f+2g} (X_1^2 X_3)^{-1} \ , \qquad 
\widehat{H} \sim L^{16}e^{2f+4g} \ , 
\label{papapapa}
\end{aligned}
\end{equation}
and we obtain,
\begin{equation}
c_{1+1}\sim L^8\frac{e^{f+2g}}{f'+2g'}.
\label{VAVA}
\end{equation}
The solution asymptotes to the IR and UV as,
\bea
(IR): f\sim -\log r, \;\;\; g\sim 1, \;\;\; (UV):f\sim g\sim -\log r.
\eea
By definition, at the IR fixed point we recover the central charge. At the UV the function is not a constant. Indeed, using that $L^8\sim N^2$,
\beq
 (IR): c_{1+1}\sim N^2, \;\;\; (UV): c_{1+1} \sim r^{-2} .
\eeq
If we consider the theory as an anisotropic 4d theory with
\beq
d=3,\;\;\;\alpha_1= L^2 \ \sqrt{\Delta} \ e^{2f},\;\;\; \alpha_2= L^2 \ \sqrt{\Delta} \ e^{2g} \ y^{-2},\;\;\; \beta_0 = e^{\frac{4}{3}(f-g)}, \;\; e^\Phi = 1.
\eeq
We then calculate
\beq
\sqrt{\det[G_{ij}]} \sim L^8 \ e^{2f+4g} \ (X_1^4 X_3^2)^{-1}.
\eeq
This gives the same function $\widehat{H}$ as in eq.(\ref{papapapa}) above, but different $\beta_0$. For the central charge we get,
\beq
c_{3+1}\sim {\cal N}^2 L^8 \frac{e^{3f}}{(f'+2g')^3}.
\eeq
Also, we observe that this function is constant and proportional to $N^2$
both in the UV and IR. 

Using explicitly the solution in eq.(9) of the paper \cite{Maldacena:2000mw}, we find that for different values of the integration constant $C_1$ in \cite{Maldacena:2000mw}, the central charge is not a monotonically increasing function.
In this sense, this is similar to our examples in Subsection \ref{CCdifdims}.
 \subsection*{Baryonic branch-like geometry}
Let us know consider metric of the Baryonic Branch 
of the Klebanov-Strassler QFT \cite{Klebanov:2000hb,Butti:2004pk}. In the string frame it has the form,
\begin{equation}
\begin{aligned}
& ds^2= e^{\Phi}\hat{h}^{-1/2}dx_{1,3}^2 + e^{\Phi}\hat{h}^{1/2}\Big[e^{2k}d\r^2 + e^{2h}(d\theta^2+\sin^2\theta \ d\varphi^2)
\\[5pt]
& \quad \;\; +\frac{e^{2g}}{4}[(\omega_1-a \ d\theta)^2 +(\omega_2+a\sin\theta \ d\varphi)^2] +
\frac{e^{2k}}{4}(\omega_3+\cos\theta \ d\varphi)^2    \Big].
\end{aligned}
\end{equation}
The usual way of thinking about this is as a 4d CFT, that flows due to  a quasi-marginal deformation of the Klebanov-Witten QFT.
Nevertheless, for the case in which the function,
\beq
\hat{h}=1-\kappa^2 e^{2\Phi} \ ,
\eeq
is taken to be $\hat{h}=1$ (that is $\kappa=0$), we fall into the background describing 
D5 branes wrapped on a two sphere \cite{Maldacena:2000yy}, in which case, the four-dimensional character of the QFT is less clear. 
The relation between both field theories and dual supergravity backgrounds  have been explained in detail in the papers
\cite{Maldacena:2009mw}.

We will calculate the central charge of the system above
considering it, first as a four dimensional QFT and then as a compactified six dimensional QFT.

Indeed, if we consider the system as four dimensional, we get
\bea
 d=3, \;\; \alpha_0=e^{\Phi}\hat{h}^{-1/2},\;\;\; \beta_0= \hat{h}e^{2k}, \;\; \widehat{H} \sim e^{4\Phi+4h+4g+2k} \ \hat{h}.
\eea
We then calculate
\beq
c_{3+1} \sim \beta_0^{3/2}\frac{\widehat{H}^{7/2}}{(\widehat{H}')^3} \sim \frac{\hat{h}^2 e^{2\Phi+2h+2g+4k}}{
(4\Phi' +4h'+4g'+2k' +\frac{\hat{h}'}{\hat{h}})^3}.
\label{central3+1}
\eeq
If, on the other hand we calculate this as if it were an anisotropic six-dimensional QFT, we would have
\begin{equation}
\begin{aligned}
&  d = 5, \;\; \alpha_1=\alpha_2=\alpha_3=e^{\Phi}\hat{h}^{-1/2},\;\;\; \alpha_4=
e^{\Phi+2h}\hat{h}^{1/2},\;\;\;\beta_0=e^{-4h/5 +2k} \hat{h}^{3/5},
\\[5pt] 
& \widehat{H} \sim e^{4\Phi+4h+4g+2k} \ \hat{h}.
\end{aligned}
\end{equation}
Notice that it is the same function $\widehat{H}$, but what changes is the function $\beta_0$.
The central charge of the anisotropic theory is,
\beq
c_{5+1} \sim \beta_0^{5/2}\frac{\widehat{H}^{11/2}}{(\widehat{H}')^5} \sim \frac{e^{2\Phi+2g+6k} \hat{h}^2}{(4\Phi'+4h'+4g'+2k'+\frac{\hat{h}'}{\hat{h}})^5}.
\label{central5+1}\eeq
We now analyze this, using the relevant asymptotics of the functions, as written in
\cite{Conde:2011aa}. We find that in the IR of the QFT ($\rho\to 0$), both definitions agree. Indeed, the function $e^{2h}\sim \rho^2$ while
$e^{2k}\sim e^{2g}\sim e^{2\Phi}\sim \hat{h}\sim 1$ and both give $c\sim \rho^5\to 0$. This is the expected result for a confining QFT.
We will be interested in the UV behavior of the central charge.
Indeed, for the functions we have---see Section 4 in \cite{Conde:2011aa},
\bea
e^{2h}\sim e^{2g}\sim e^{2k}\sim e^{4\rho/3},\;\;\;e^{2\Phi}\sim 1,\;\;\; \hat{h}\sim e^{-8\rho/3} \rho.\nonumber
\eea
We then have that both definitions behave similarly 
\beq
c_{3+1}\sim c_{5+1}\sim \rho.
\eeq
In the 'correct radial variable' $r=e^{2\rho/3}$, this is the logarithmic growth of the degrees of freedom typical of the KS-solution.

Something similar occurs for the particular case $\kappa=0$ or $\hat{h}=1$. In that case, there is yet another solution (this time an exact solution--see \cite{HoyosBadajoz:2008fw}), such that is the same in the IR as the one discussed above, but in the UV behaves as,
\bea
 h\sim \log\rho,\;\;\; \Phi\sim \rho -\log\rho, \;\;\;k=g=0.
\eea
In this case, the definition in eq. \eqref{central3+1} grows unbounded in the UV $c\sim e^{2\rho }\rho$, while the calculation for the anisotropic QFT in eq. \eqref{central5+1} gives a similar growth, both driven by the dilation factor. But at the subleading order, the four dimensional expression grows faster. This indicates that the second definition (the one for the anisotropic QFT) is closer to the true behavior, where the QFT has---from a 4d perspective-- an infinite tower of KK-modes, then becomes a 6d-QFT. The Little string theory is needed to UV-complete the system, this is reflected by the unbounded growth of the central charges.
This situation is similar to what happened in Section \ref{centralchargessectionxx} in computing the central charge of the $AdS_5\to AdS_3$ flow considering it as  2-d QFT.



\providecommand{\href}[2]{#2}\begingroup\raggedright\endgroup

\end{document}